\documentclass[aps,prb,floatfix,superscriptaddress,
amsmath,amssymb,twocolumn]{revtex4}

\usepackage[latin1]{inputenc}
\usepackage{amsmath}
\usepackage{amssymb}
\usepackage{graphicx}
\usepackage{graphics}
\usepackage{color}
\usepackage{multirow}
\DeclareMathAlphabet{\mathit}{OT1}{ptm}{m}{it}
\usepackage{url}
\usepackage{hyperref}
\usepackage{subfigure}

\def \ba {\begin{eqnarray}}
\def \ea {\end{eqnarray}}

\newcommand{\nn}{\nonumber}
\newcommand{\ket}[1]{| #1 \rangle}
\newcommand{\bra}[1]{\langle #1 |}

\newcommand{\op}[1]{\mathbf{#1}}

\newcommand{\ha}{\mathcal{H}}
\newcommand{\pa}{\mathcal{P}_{AF}}
\newcommand{\ta}{\Theta_{AF}}

\newcommand{\tah}{\hat{\Theta}_{AF}}

\newcommand{\ketb}[3]{\ket{#1^{\text{ #2}}_{#3}}}
\newcommand{\brab}[3]{\bra{#1^{\text{ #2}}_{#3}}}

\newcommand{\sew}{\mathit{w}}

\begin{document}

\title{Identifying two-dimensional $Z_2$ antiferromagnetic topological 
insulators}

\author{Fr\'{e}d\'{e}ric B\`{e}gue} \affiliation{Laboratoire de Physique 
Th\'{e}orique, IRSAMC, Universit\'{e} de Toulouse, CNRS, UPS, France }

\author{Pierre Pujol} \affiliation{Laboratoire de Physique 
Th\'{e}orique, IRSAMC, Universit\'{e} de Toulouse, CNRS, UPS, France }

\author{Revaz Ramazashvili} \affiliation{Laboratoire de Physique 
Th\'{e}orique, IRSAMC, Universit\'{e} de Toulouse, CNRS, UPS, France }

\begin{abstract} 
We revisit the question of whether a two-dimensional topological insulator 
may arise in a commensurate N\'eel antiferromagnet, where staggered 
magnetization breaks both the elementary translation and time reversal, 
but retains their product as a symmetry. In contrast to the so-called $Z_2$ 
topological insulators, an exhaustive characterization of antiferromagnetic 
topological phases with the help of a topological invariant has been missing. 
We analyze a simple model of an antiferromagnetic topological insulator 
and chart its phase diagram based on a recently proposed criterion for 
centrosymmetric systems [Fang {\it et al.}, Phys. Rev. B {\bf 88}, 085406 (2013)]. 
We then adapt two methods, originally designed for paramagnetic systems, 
and make antiferromagnetic topological phases manifest. The proposed 
methods apply far beyond the particular example treated in this work, 
and admit straightforward generalization. We illustrate this by considering 
a non-centrosymmetric system, where there are no simple criteria to identify 
topological phases. We also present an explicit construction of edge states 
in an antiferromagnetic topological insulator. 
\end{abstract}


\maketitle

\section{Introduction}
\label{sec.introduction}

Topological states of matter have become a focus of much theoretical 
and experimental effort (see [\onlinecite{ReviewHasanKane,ReviewQiZhang}] 
for detailed reviews, and [\onlinecite{KaneMele1,KaneMele2,Xu06,Wu06}] for 
some of the earlier work on the subject). 
Such an interest is dictated by a remarkable stability of topological 
phases and of their physical properties with respect to perturbation. 

Topological insulators (TIs) provide a simple example of such phases 
in a non-interacting system. From a band structure textbook perspective, 
these are ordinary band insulators with a spectral gap. However, 
in addition to the mere presence of the gap, the bands may have 
a non-trivial reciprocal-space topology, that manifests itself, 
{\em inter alia}, 
by the presence of surface states, stable against moderate bulk 
and surface perturbations. This topology may be encapsulated in a special 
quantum number, assigned to the occupied bulk bands: insulators with an 
odd value of this number have protected gapless surface states 
within the bulk gap, and are called ``topological". For an even value of this 
number, surface states within the bulk gap are not protected; such 
insulators are called ``topologically trivial". Switching between an even 
and an odd value requires closing of the bulk gap and a phase transition; 
otherwise, a smooth variation of the Hamiltonian leaves this number intact. 
Hence it is called a $Z_2$ (``even-odd") topological invariant. 

Symmetry tends to facilitate description, thus it is not surprising 
that some of
the pioneering work on topological insulators studied rather 
symmetric systems:
notably those symmetric with respect to both the time reversal $\theta$ 
and inversion $I$.
Simultaneous presence of these two symmetries guarantees double degeneracy 
of bulk Bloch eigenstates at any momentum in the Brillouin zone. For 
two-dimensional models, one of the better-known examples of this kind is the 
so-called Bernevig-Hughes-Zhang (BHZ) model~\cite{BHZ}, that we briefly review 
in this article. Bulk Bloch eigenstates may also be degenerate in systems with 
more delicate symmetries. A convenient example is provided by a collinear 
N\'eel antiferromagnet, sketched in the Fig. \ref{Fig.Lattice}, where the time 
reversal symmetry $\theta$ is explicitly broken by magnetic order. Black and 
white circles in the figure depict the two sublattices of a square-lattice 
N\'eel 
antiferromagnet, and correspond to the opposite directions of local 
magnetization. 
Both the time reversal $\theta$ and a translation $T_{\bf a}$ by half a period 
${\bf a}$ invert the local magnetization (interchange black and white circles 
in the Fig. \ref{Fig.Lattice}, and thus none of them is a symmetry of the 
system. 
However, the product $\theta T_{\bf a}$ remains a symmetry and conspires with 
inversion $I$ to ensure the double degeneracy of Bloch eigenstates at any 
momentum in the Brillouin zone -- the same way as, in  paramagnetic insulator, 
such a degeneracy appears due to a conspiracy of $\theta$ and $I$. 

With this similarity in mind, one may inquire whether an antiferromagnet 
may host a topologically non-trivial state of matter such as a topological 
insulator and, if so, whether the latter may have properties distinguishing 
it from its non-magnetic counterparts. Several pioneering studies have 
already addressed this issue. In particular, Mong {\em et 
al.}~\cite{MongEssinMoore} 
have posed the question of whether an antiferromagnetic insulator may be 
topologically non-trivial, and whether or not it may be characterized 
by a topological invariant. The authors concluded that, in contrast to 
non-magnetic insulators, the presence of surface states in an antiferromagnet 
is sensitive to whether the surface is symmetric under the same combination 
of a translation and time reversal as the bulk. This result was obtained for 
three-dimensional materials and was confirmed for a special
 choice of boundary. 
The investigation of three-dimensional antiferromagnetic topological 
insulators has, since then, become an extensive subject of 
study.~\cite{Liu13AF,Zhang15,Liu14,Fang15,Fang_Gilbert_Bernevig} 

In a subsequent publication, Fang {\em et al.}~\cite{Fang_Gilbert_Bernevig} 
studied a class of systems that, beyond symmetries involving time reversal, 
also possess an inversion center. They argued that 
antiferromagnetic insulators 
with such properties may  be characterized by a $Z_2$ topological invariant 
that, similarly to a paramagnet, involves the product of parity eigenvalues 
over a set of special points in the Brillouin zone. However, in an 
antiferromagnet 
this set of special points would comprise only a half of the points that are 
relevant 
in the paramagnetic case. 

The systems mentioned above can be viewed as a particular example of 
the so-called crystalline topological insulators, which have been studied in 
3 dimensions \cite{Fu11,Kargarian13,Liu13,Serbyn14,Liu14,Zhang15,Fang15}
and also in the case of the two-dimensional honeycomb lattice 
\cite{Priyamvada13}.
The key ingredient of all these studies is the presence of time reversal 
symmetry 
supplemented by a crystal symmetry. 

Here, we study the precise form of this $Z_2$ invariant 
in an antiferromagnetic 
insulator for a particular generalization of the BHZ model~\cite{BHZ} in two 
dimensions. In the Sec.\ref{sec.BHZ} we briefly review the BHZ model and 
its symmetries. We then extend the model by turning on 
staggered magnetization, 
identify its symmetries and ask whether its topology may be characterized 
similarly to how it is done for the paramagnetic BHZ model. 
In the Sec.\ref{sec.Methods}, we review some of the methods 
used to study topological insulators: the Fu-Kane topological 
invariant~\cite{Fu_Kane2007}, the parallel transport 
method~\cite{Soluyanov_Vanderbilt_2012}, the method of Wannier 
Charge Centers~\cite{Vanderbilt92,Yu11} -- and, finally, an explicit 
construction of edge states. We show that these methods do not 
apply to an antiferromagnet verbatim, and show how they must be adapted. 
The model we consider is centrosymmetric, hence we test our results 
against the $Z_2$ topological invariant in the form proposed by Fang 
{\em et al.}, and find perfect agreement. However, our adaptation of the 
Wannier Charge Center method applies perfectly well even when the 
parity-based criterion no longer does. To illustrate this, in the Sec. 
(\ref{sec.P_broken}) we consider a non-centrosymmetric perturbation 
of the initial Hamiltonian: the method successfully identifies the 
topological phases. Finally, the Sec. \ref{sec.CDO} contains 
concluding remarks and an outlook. 
 
\section{The BHZ model and its extension to antiferromagnetism}
\label{sec.BHZ}

Bernevig, Hughes and Zhang proposed a model to describe topological insulating 
phases in HgTe/CdTe quantum wells.~\cite{BHZ} In this section, 
we first present 
the Hamiltonian they introduced. We then see how it is modified 
in the presence 
of 
antiferromagnetism, and study how magnetic order affects its topological 
properties.

\subsection{The BHZ model}

We consider a square lattice, defined by lattice vectors 
$\op{R}=pa\hat{\op{X}}+qa\hat{\op{Y}}$, 
with $p,q \in \mathbb{Z}$ (see the Fig. \ref{Fig.Lattice}). A unit cell 
labelled by 
$\op{R}$ hosts four single-electron states, $\ket{\op{R},n}$: two 
$s$-type orbitals, 
$\ket{\uparrow,s}$ and $\ket{\downarrow, s}$, and two $p$-type orbitals, 
$\ket{\uparrow, p_x+ip_y}$ and $\ket{\downarrow, p_x-ip_y}$.  

We will use the index $\nu$ to denote the $s$-state ($\nu=+$) or a 
$p$-state ($\nu=-$), and $\sigma$ for the spin. In this basis, the BHZ 
Hamiltonian may be written as:

\begin{figure}[t]
\includegraphics[scale=0.4]{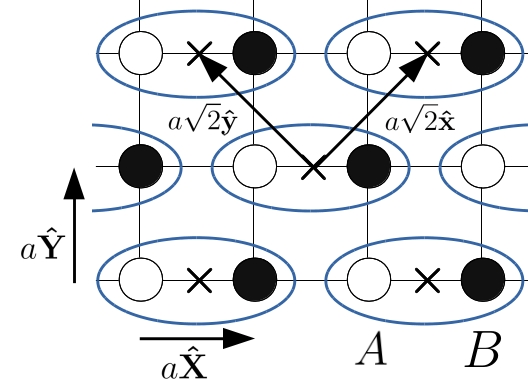}
\caption{(color online) Square lattice, on which the Hamiltonian is defined. 
In the absence of a staggered magnetic field, the primitive Bravais 
lattice vectors are $a\op{\hat{X}}$ and $a\op{\hat{Y}}$. In the presence 
of staggered magnetization, the dimerized lattice is defined by the 
primitive vectors $a\sqrt{2}\op{\hat{x}}$ and $a\sqrt{2}\op{\hat{y}}$. 
In this case, a unit 
cell (in blue) comprises two sites, A and B (white and black 
dots, respectively). The 
center of a unit cell (denoted $\op{r}$ in the main text) is represented 
by a cross.} 
\label{Fig.Lattice}
\end{figure}

\begin{align}\label{eq.BHZ real space }
\ha_{BHZ}&=\sum_{\op{R},\sigma,\nu} (\mu+ \nu\Delta\mu) 
c_{\sigma,\nu}^\dagger(\op{R}) c_{\sigma,\nu}(\op{R})  \nn\\
& +\sum_{<\op{R},\op{R'}>,\sigma,\nu} (t+ \nu \Delta t)
(c_{\sigma,\nu}^\dagger(\op{R}) c_{\sigma,\nu}(\op{R'}) + h.c) \nn\\
& +\sum_{\op{R},\sigma,\nu} 
-i\alpha\sigma(c_{\sigma,\nu}^\dagger(\op{R}+a{\bf \hat X}) 
c_{\sigma,-\nu}(\op{R}) - h.c)\nn\\
& +\sum_{\op{R},\sigma,\nu} \alpha\nu(c_{\sigma,\nu}^\dagger
(\op{R}+a{\bf \hat Y})c_{\sigma,-\nu}(\op{R}) + h.c).
\end{align}
The first term ($\mu_\pm=\mu\pm\Delta\mu$) originates from the energy 
difference of the the s- and p-symmetric orbitals. The second term corresponds 
to the nearest-neighbor hopping between the same orbitals, 
with different hopping 
amplitudes, $t_\pm=t\pm\Delta t$. Finally, the remaining two terms hybridize 
the two species via the amplitude $\alpha$, and are of spin-orbital 
nature. 

The Hamiltonian is non-interacting, and its ground state is built 
by filling the 
single-electron states up to the Fermi energy. At half-filling, 
the bulk spectrum 
has an insulating gap. However, depending on the Hamiltonian parameters, 
the system may be a trivial or a topological insulator, in the sense described 
in the Introduction --  as we argue below. 
It turns out that the trivial phase is realized 
for $| \Delta\mu |> 4|\Delta t|$: a boundary 
of such a system hosts an even number of pairs of edge states. By contrast, 
when $| \Delta\mu |< 4|\Delta t|$, the system is in a 
topological phase, and has an 
odd number of pairs of states at the boundary. Thus, in the topological phase, 
at least a single pair of chiral edge states is guaranteed to exist. 
Transition from a topological to the trivial state implies 
closing the band gap 
(here, this occurs at $| \Delta\mu |=4|\Delta t|$) and is 
a quantum phase transition. 
Hence the term ``topologically protected edge states". Topology of either of 
the phases can be characterized by the parity of the number 
of pairs of edge states, 
which, as we will see below, is related to the $Z_2$ topological invariant. 

It is interesting to note that the strength $\alpha$ 
of the spin-orbit coupling 
does not appear in the above inequalities despite being responsible for the 
existence of the insulating gap.

Translational invariance of the crystal lattice 
allows one to define the Bloch Hamiltonian
\begin{equation}\label{Bloch transfo}
H_{BHZ}(\op{k})=
e^{-i\op{k}\cdot\hat{\op{R}}}\ha_{BHZ}e^{i\op{k}\cdot\hat{\op{R}}}
\end{equation}
and obtain
\begin{align}\label{eq.BHZ k space}
H_{BHZ}(\op{k})&=\mu+2t \cos(ak_X)+2t \cos(ak_Y)\nn\\
&\ +(\Delta\mu +2\Delta t \cos(ak_X)+2\Delta t \cos(ak_Y))\tau^z\nn\\
&\ -2\alpha \sin(ak_X) s^z\tau^x+2\alpha \sin(ak_Y) \tau^y,
\end{align}
where the $s^a$ and $\tau^b$, with $a$ and $b$ standing for $x,~y$ and $z$, 
are the Pauli matrices acting in the spin and orbital spaces, respectively. 
The identity operator in either of the two spaces is omitted for brevity: for 
example, the last term in the Eq.(\ref{eq.BHZ k space}) acts as the identity 
operator in the spin space. 

The BHZ Hamiltonian is invariant under the time reversal $\Theta=i s^yK$, 
where $K$ is complex conjugation. The operator $\Theta$ is anti-unitary, 
and $\Theta^2 = -1$. 

Using the Kramers theorem arguments, one can show that the eigenstates of the 
BHZ Hamiltonian come in Kramers pairs (see appendix \ref{app.degeneracy} for 
a general proof), a state at momentum $\op{k}$ being related to its degenerate 
counterpart at momentum $\op{-k}$. This has an important consequence for a 
set of special points $\Gamma_i=(\Gamma^x,\Gamma^y)$ in the Brillouin zone, 
called time reversal invariant momenta (TRIM). These special 
points satisfy the 
equality $-\Gamma_i=\Gamma_i+\op{G}$, where $\op{G}$ is a reciprocal lattice 
vector, and $\Gamma^x,\Gamma^y$ take values $0$ or $\pi /a$. As $\Gamma_i$ 
is equivalent to $-\Gamma_i$, the TRIM states are doubly degenerate, 
an important property that we will use later.

The Hamiltonian is also invariant under inversion 
$\mathcal{P}=\sum_\op{R} \tau_z\ket{-\op{R}}\bra{\op{R}}$ (the $\tau_z$ comes 
from the fact that the s-orbital remains invariant under inversion, while the 
p-orbital 
acquires a minus sign). Note that 
$\mathcal{P}H(\op{k})\mathcal{P}^{-1}=H(\op{-k})$ 
and $\mathcal{P}^2=1$.

Combining the $\Theta$ and $\mathcal{P}$, one can show that each 
Bloch eigenstate also has a degenerate partner at the same momentum.

\subsection{ Generalization to a staggered magnetization}

Looking to study the effect of anti-ferromagnetism on topological 
insulators, we introduce a staggered magnetization, following the Refs. 
[\onlinecite{Kulikov:1984,Revaz:PRL.2008,Revaz:PRB.2009,
Fang_Gilbert_Bernevig,Guo_Feng_Shen}], via the term: 
\begin{equation}
\label{eq.term mag}
\sum_{\op{R},\sigma,\nu} (-1)^\op{R}\sigma m \ 
c_{\sigma,\nu}^\dagger(\op{R}) c_{\sigma,\nu}(\op{R}) , 
\end{equation}
where the $m$ is the product of the antiferromagnetic ordered moment 
and the constant that couples it to the conduction electron spin. 
The Eq.(\ref{eq.term mag}) introduces antiferromagnetic order 
phenomenologically, 
independently of the precise form of the interaction that gives rise to 
magnetism. 
At the same time, it entirely neglects both thermal and quantum fluctuations 
of the magnetic order. This may be justified for the ordered moment that is 
noticeable on the scale of the Bohr magneton, and at temperatures well below 
the N\'eel temperature. 

Staggered magnetization in the Eq.(\ref{eq.term mag}) doubles the unit 
cell of the paramagnetic state and reduces its translation symmetry. The 
new Bravais lattice is now defined by the vectors 
$\op{r}=pa\sqrt{2}\hat{\op{x}}+qa\sqrt{2}\hat{\op{y}}, p,q \in \mathbb{Z}$,
and contains two sites per unit cell, A and B, positioned at 
$\op{R}_A=\op{r}-\frac{a}{2}\hat{\op{X}}$ and 
$\op{R}_B=\op{r}+\frac{a}{2}\hat{\op{X}}$ respectively
(see the Fig. \ref{Fig.Lattice}). Hereafter we set $a=\frac{1}{\sqrt{2}}$. The 
unitcell of lattice vector $\op{r}$ now possesses eight states, 
$\ket{\op{r},n}$.

We define the Bloch Hamiltonian with staggered magnetization as:
\begin{equation}
\label{Bloch transfo AF}
H(\op{k})=e^{-i\op{k}\cdot\hat{\op{r}}}\ha e^{i\op{k}\cdot\hat{\op{r}}}.
\end{equation}

One may note the difference between the Eqs.(\ref{Bloch transfo}) and
(\ref{Bloch transfo AF}). Here, in contrast to the choice made by Guo 
{\it et al.}, 
the real positions of the sites, $\op{R}_A$ and $\op{R}_B$, have been replaced 
by the position $\op{r}$ of the center of the unit cell to which the sites 
belong. 
This choice in the definition of the Bloch Hamiltonian ensures 
$H(\op{k}+\op{G})=H(\op{k})$ for any reciprocal lattice vector 
$\op{G}$\cite{Fruchart14},
a property that will prove useful later, as we wish to define quantities 
that are continuous over the BZ torus. We obtain:

\begin{align}\label{eq.BHZ+mag k space}
H(\op{k})&=\mu+\Delta\mu \tau^z +m s^z \sigma^z \nn\\
&+(2C_-^2+2C_-C_+) (t\sigma^x + \Delta t \tau^z \sigma^x)\nn\\
&+(2C_-S_- +2C_+S_-)  (t\sigma^y + \Delta t \tau^z \sigma^y)\nn\\
&-2\alpha S_- (C_-  s^z\tau^x \sigma^x + S_- s^z\tau^x \sigma^y)\nn\\
&+2 \alpha S_+   (C_- \tau^y \sigma^x+ S_-\tau^y \sigma^y)\nn\\
\end{align}
where $C_\pm \equiv \cos\left[(k_x \pm k_y)/2\right]$ and 
$S_\pm \equiv \sin\left[(k_x \pm k_y)/2\right]$, while $\sigma$, $s$ 
and $\tau$ are the Pauli matrices acting in the sub-lattice (A and B), 
spin and orbital spaces, respectively.

The time reversal and the elementary translation both invert the local 
magnetization, 
thus none of them is a symmetry of the antiferromagnetic state. However, their 
product $\ta=T\Theta$ ($T$ being the translation by a vector $a\hat{\op{X}}$) 
remains a symmetry. 

\begin{figure}
\includegraphics[scale=0.35]{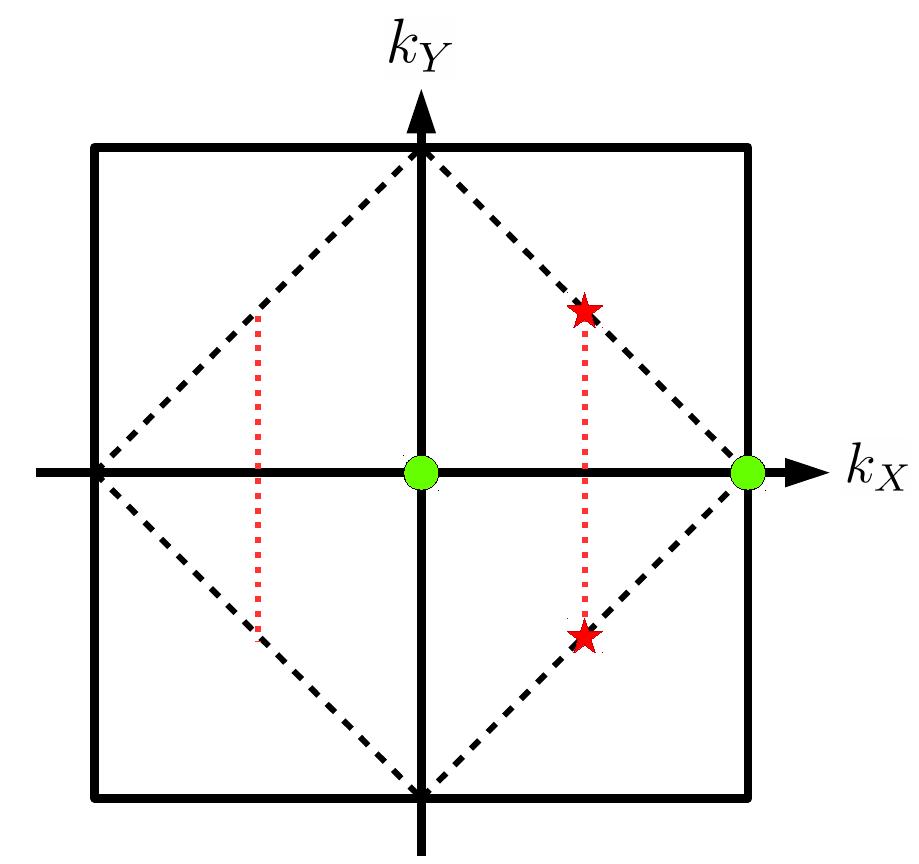}
\caption{(color online) The big solid black square depicts the paramagnetic 
Brillouin Zone. It is reduced to the antiferromagnetic Brillouin Zone (black 
dashed square) in the presence of the staggered magnetization. The red 
dotted lines correspond to points in the antiferromagnetic Brillouin Zone, 
where $\ta^2=1$. They have no horizontal counterparts since the chosen 
$\ta$ involves translation only along the $x$-axis. The two red stars denote 
the two A-TRIM, where $\ta^2=1$, while the green circles correspond 
to the B-TRIM, where $\ta^2=-1$. 
} 
\label{fig.BZ}
\end{figure}

The $\ta$ is also anti-unitary but does not square to $-1$ for an arbitrary 
momentum 
in the Brillouin zone. Indeed, as the time reversal operator commutes with any 
purely 
spatial transformation, it commutes with any translation, and thus $\ta^2=-T^2 
$. One 
may note that $T^2$ is a symmetry of the antiferromagnetic state and acts on a 
state 
$\ket{\Psi_{n,\op{k}}}$ as per 
$T^2 \ket{\Psi_{n,\op{k}}} = e^{-i2 a \op{k}\cdot \hat{\op{X}}} 
\ket{\Psi_{n,\op{k}}}$.
As a result, at certain points in the BZ ($k_X=\pi/(2a)~ mod~ \pi/a$) the 
$\ta^2$ 
acts as the identity, which is an obstacle to the definition of Kramers pairs 
(see 
the Fig. \ref{fig.BZ}). 

However, the system is still invariant under the inversion symmetry, 
represented by the operator $\pa$, whose expression is given in the Appendix 
\ref{app.degeneracy}.

Thanks to both symmetries, one may show that it is still possible to separate 
the states in Kramers-like pairs relating a state of momentum $\op{k}$ to a 
state 
of momentum $\op{-k}$ (see Appendix \ref{app.degeneracy}). This will prove 
indispensable in the next section, when computing the $Z_2$ invariant in the 
antiferromagnetic case. Moreover, one may show that the combined symmetry 
$\pa\ta$ is anti-unitary and squares to $-1$. 
Thus, each Bloch eigenstate possesses an orthogonal degenerate 
partner at the same momentum, and the bulk bands are doubly 
degenerate.~\cite{Revaz:PRL.2008,Revaz:PRB.2009}

As before, the system has a insulating gap at half-filling, and two questions 
arise:
\begin{itemize}
\item Does a topological phase survive for a finite value of the staggered 
magnetization?

\item Starting from a trivial phase with $m=0$, can we obtain a 
topological phase by switching on a staggered magnetization?
\end{itemize} 

Fang and collaborators~\cite{Fang_Gilbert_Bernevig} proposed an expression
for the $Z_2$ invariant for a two-dimensional N\'eel 
antiferromagnet, symmetric 
under both the $\ta$  and $\pa$. We will discuss this criterion in more 
detail in the next section (see the Eq.(\ref{eq.def inv topo with xi AF})).

\begin{figure}[t]
\subfigure[]{\includegraphics[height=6cm]{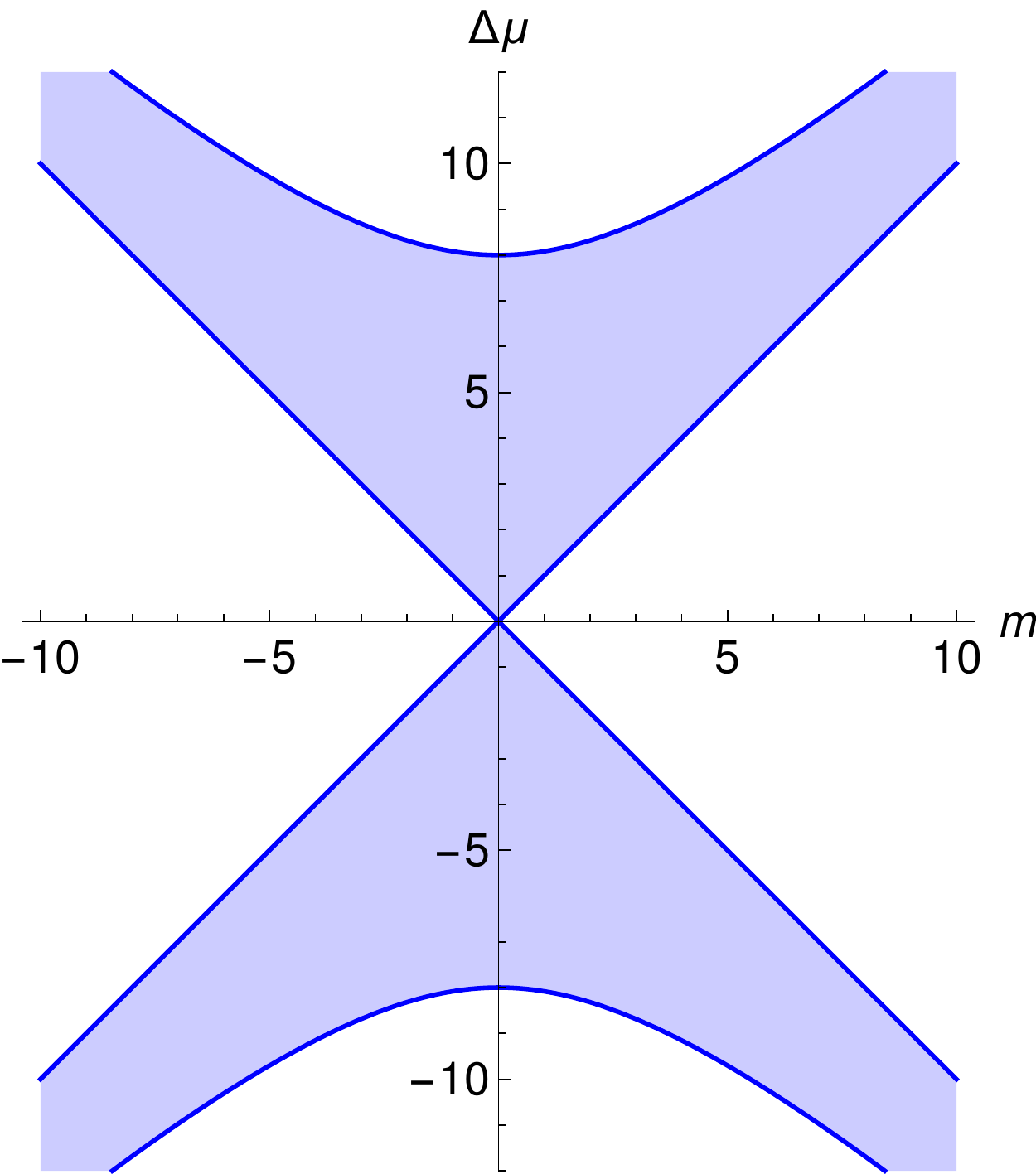}}

\subfigure[]{\includegraphics[height=3cm]{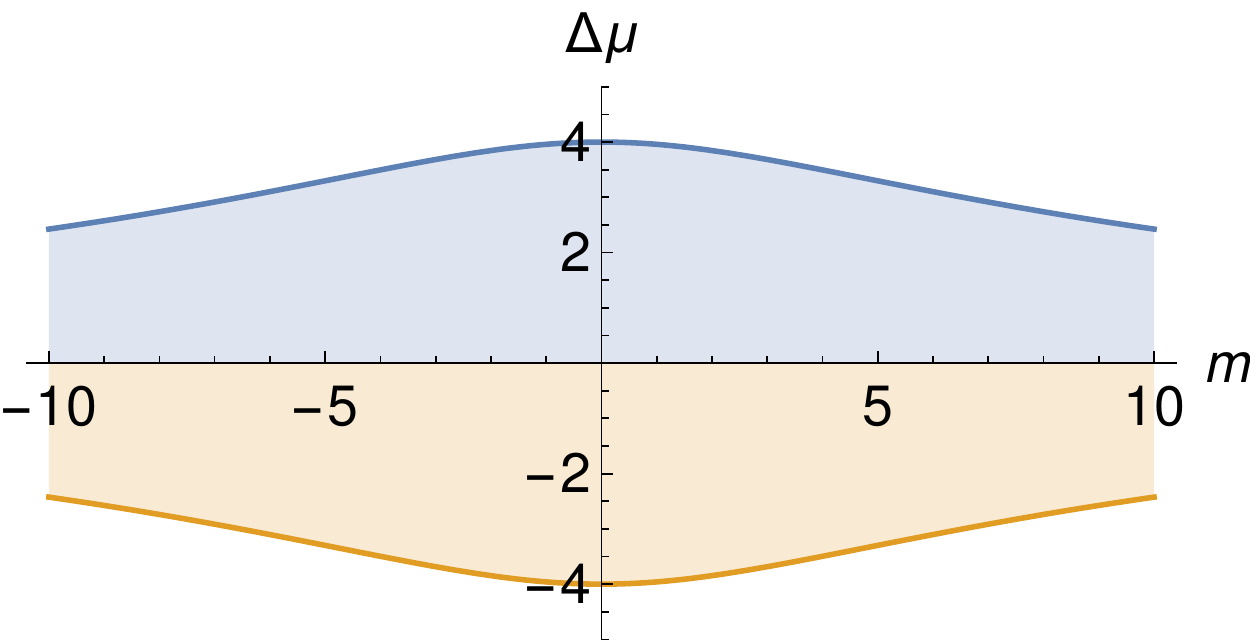}}
\caption{(color online) Phase diagram of the BHZ Hamiltonian of the Eq. 
(\ref{eq.BHZ k space}) 
in the presence of staggered magnetization, for $t_+ = 3$ and $t_- =1$. 
(a) The phase diagram at half-filling. The blue region corresponds to the 
topological 
phase. The $m=0$ vertical line corresponds to the phase diagram of the BHZ 
Hamiltonian. One may notice that a topological phase at $m=0$ survives until 
a finite threshold value of $m$, and becomes trivial for larger values. 
Moreover, from the trivial phase at $m=0$, the system may be driven 
into the topological phase by turning on a staggered magnetization. 
(b) Overlay of the phase diagrams at 1/4 and 3/4 filling. 
The blue region corresponds to a topological phase at 1/4 filling, while 
the yellow region corresponds to a topological phase at 3/4 filling.
}
\label{PhaseDiagram}
\end{figure}

Applied to the case at hand, the criterion due to Fang {\em et al.} 
as expressed 
by the Eq. (\ref{eq.def inv topo with xi AF}) suggests the positive answer to 
both 
questions. Indeed, e.g. for $\alpha=2$, $t_+=3$ and $t_-=1$, we obtain the phase diagram 
in the Fig. \ref{PhaseDiagram}, where the filled region corresponds to values 
of $m$ and $\Delta\mu$ where we have a topological insulator 
(all of the numerical results presented in this article will assume those 
values for $t_+$ and $t_-$). The phase diagram shows that, starting in the 
topological phase at $m=0$ and upon increasing $m$, the system remains 
non-trivial until $m=\Delta\mu$. On the other hand, if one starts in a trivial 
phase at $m=0$, there is a range of $m$, where the system becomes 
topologically non-trivial, corresponding to regions where $|m|< |\Delta\mu |$ 
and $2|\Delta\mu |<\sqrt{m^2+16 t_+^2}+\sqrt{m^2+16 t_-^2} $. 
We verified this by applying the method detailed in the next 
section to several 
sets of parameters. The results were in perfect agreement with the prediction 
of Fang and collaborators. 

Because of the staggered magnetization, we are now working with a 
8-band model,
and once again, the symmetries protect doubly-degenerate bands. Thus, 
it is now
possible to open insulating gaps at $1/4$ and $3/4$ filling, that have no 
equivalent
in the BHZ model. It is also interesting to see if we may have topological 
phases at 
such fillings. Using the expression (\ref{eq.def inv topo with xi AF}) due to 
Fang 
{\it et al.} for the $Z_2$ invariant, we obtain the phase diagram in the 
Fig. \ref{PhaseDiagram}, which is once again in perfect agreement 
with our numerical results.  

\section{Methods for characterizing topological insulators}
\label{sec.Methods}

In this section we discuss some of the techniques used to characterize 
time reversal-symmetric topological insulators, using, as an illustration, 
the BHZ Hamiltonian with and without staggered magnetization. 

The first method, due to Fu and Kane~\cite{Fu_Kane2006} yields an expression
for the $Z_2$ invariant that can be computed analytically,
 knowing continuously 
defined Bloch functions in the bulk. The second method is an adaptation of the 
Fu-Kane approach to cases where the Bloch states may be computed only 
numerically, by studying either the phase variation of Bloch functions across 
the Brillouin zone~\cite{Soluyanov_Vanderbilt_2012}, or the so-called Wannier 
charge center trajectories~\cite{Vanderbilt92,Yu11}. Finally, the third method 
amounts to an explicit construction of edge states, as the number 
of edge states 
may distinguish between the trivial and the topological phase.

\subsection{Computation of the Fu-Kane topological invariant}
\label{subsec.Fu and Kane invariant}

We first present the main steps in the derivation of the $Z_2$ topological 
invariant 
due to Fu and Kane, and then the modifications necessary to account for the 
presence of staggered magnetization. The technical details are given in 
the Appendices \ref{app.Fu and Kane} and \ref{app.Fu and Kane AF}.

As the BHZ Hamiltonian enjoys translational invariance, its eigenstates can be 
written as:
\begin{equation}\label{eq.lien psi et u}
\ket{\Psi_{n,\op{k}}}=e^{i\op{k}\cdot\op{r}}\ket{u_{n,\op{k}}}
\end{equation}
with $n$ from 1 to 4 for BHZ, and with $\ket{u_{n,\op{k}}}$ being the 
eigenvectors 
of the Bloch Hamiltonian of the Eq.(\ref{eq.BHZ k space}). We choose the 
$\ket{\Psi_{n,\op{k}}}$ to be continuous over the BZ.

As stated previously, the BHZ Hamiltonian is time reversal-invariant, so the 
eigenstates 
of the BHZ Hamiltonian come in Kramers pairs. We can thus separate the 
eigenstates 
in $(\{\ket{\Psi_{\op{k},\alpha}^I},\ket{\Psi_{\op{k},\alpha}^{II}}\},
\op{k}\in BZ, \alpha=1,2)$, 
in such a way that 
\begin{eqnarray}\label{eq.defchi}
\ket{\Psi^{I}_{\alpha,\op{-k}}}&=&-e^{i\chi_{\op{k},\alpha}}
\Theta \ket{\Psi^{II}_{\alpha,\op{k}}} \nn \\
\ket{\Psi^{II}_{\alpha,\op{-k}}}&=&e^{i\chi_{\op{-k},\alpha}}
\Theta \ket{\Psi^{I}_{\alpha,\op{k}}}.
\end{eqnarray}
The value of the topological invariant will depend on how the phase $\chi$ 
varies across the BZ.

The topological invariant is defined via the time reversal polarization 
$P^\Theta_{k_y}$, introduced by Fu and Kane~\cite{Fu_Kane2006} in terms of the 
center of mass position 
along $\op{x}$ of hybrid Wannier functions, computed at a fixed $k_y$. 
We define 
and 
calculate the $P^\Theta_{k_y}$ in the Appendix \ref{app.Fu and Kane}. 
The $Z_2$ 
topological invariant may be defined as per 
\begin{equation}\label{eq.def inv topo with P}
\Delta = \sum_{\Gamma^y\in \{0,\pi /a\}} P^\Theta_{\Gamma^y}\ mod\ 2 .
\end{equation}
As explained in the Ref.~[\onlinecite{Fu_Kane2006}], this quantity is a 
topological invariant, independent of the gauge chosen for the Bloch states. 

The topological invariant may be recast in terms of the sewing matrix
\begin{equation}
\sew(\op{k})_{mn}=\bra{\Psi_{m,\op{-k}}} \Theta\ket{\Psi_{n,\op{k}}} ~, 
\end{equation}
evaluated at the four TRIM as per 
\begin{equation}\label{eq.def inv topo with sew}
(-1)^\Delta = \prod_{i} \frac{\sqrt{\det[\sew (\Gamma_i)]}}{Pf [\sew 
(\Gamma_i)]}, 
\end{equation}
where the index $i$ labels the different TRIM. Even though the 
above expression 
appears to rely only on the TRIM states, it has to be computed in a 
gauge where 
the eigenstates are defined continuously on the BZ torus.

For a time reversal-symmetric system in the presence of inversion symmetry 
(which is the case for BHZ Hamiltonian), it turns out that the 
Eq.(\ref{eq.def inv topo with sew}) admits the form~\cite{Fu_Kane2007}
\begin{equation}\label{eq.def inv topo with xi}
(-1)^\Delta=\prod_{i=1}^4 \prod_{\alpha=1}^N \xi_{\alpha}(\Gamma_i)
\end{equation}
where $\xi_{\alpha}(\Gamma_i)=\pm 1$ are the parity eigenvalues 
of the filled eigenstates $\ket{\Psi^{I}_{\alpha,\Gamma_i}}$.

Contrary to the Eq.(\ref{eq.def inv topo with sew}), the expression in 
the r.h.s 
of the Eq.(\ref{eq.def inv topo with xi}) has the advantage of relying only on 
the knowledge of the states at the TRIM. Indeed, $\prod_{\alpha=1}^N 
\xi_{\alpha}(\Gamma_i)$ is gauge-invariant, so one can compute the 
$\xi_{\alpha}(\Gamma_i)$ separately for the different TRIM, 
without insisting on a continuous 
definition 
of the states across the BZ. However, such an expression can be obtained 
only for a centrosymmetric system.

In the AF case, we doubled the number of bands in the Bloch Hamiltonian -- and 
thus the index $n$, that was defined in the Eq.(\ref{eq.lien psi et u}) 
and took values from 1 to 4, now varies from 1 to 8. 
As mentioned above, due to the breaking of the time-reversal symmetry, 
we are no longer guaranteed Kramers degeneracy all over the BZ. 
Indeed, there is a line $k_x-k_y=\pm \pi$ of states in the BZ, where $\ta^2=1$ 
and, strictly speaking, the Kramers theorem does not hold. However, thanks 
to the inversion symmetry, we can recover the degeneracy necessary to write 
an equivalent of the Eq.(\ref{eq.defchi}):
\begin{eqnarray}\label{eq.defchiAF}
\ket{\Psi^{I}_{\alpha,\op{-k}}}&=&-e^{i\chi_{\op{k},
\alpha}}e^{i\Phi_{\op{k}}/2}\ta \ket{\Psi^{II}_{\alpha,\op{k}}} \nn \\
\ket{\Psi^{II}_{\alpha,\op{-k}}}&=&e^{i\chi_{\op{-k},\alpha}}
e^{-i\Phi_{-\op{k}}/2}\ta \ket{\Psi^{I}_{\alpha,\op{k}}}
\end{eqnarray}
for $\alpha=1...4$, and where $\Phi_{\op{k}}$ is defined so that 
\begin{equation}
\ta^2=-\sum_\op{k}e^{i\Phi_{\op{k}}}\ket{\op{k}}\bra{\op{k}}.
\end{equation}

As above, we define the topological invariant via the time reversal 
polarization as per the Eq.(\ref{eq.def inv topo with P}). However, 
in an antiferromagnet this expression cannot be reduced to a simple 
form as in the Eq.(\ref{eq.def inv topo with sew}): indeed, the sewing 
matrix 
\begin{equation}
\tilde{\sew}_{k_{b}}(k_a)_{mn}=\bra{\Psi_{m,\op{-k}}} 
\ta\ket{\Psi_{n,\op{k}}}  .
\end{equation}
is no longer anti-symmetric at all the TRIM. In fact, at those TRIM, 
where $\ta^2=1$, that we refer to as A-TRIM in Fig. \ref{fig.BZ}, 
the Pfaffian in the  Eq.(\ref{eq.def inv topo with sew}) cannot even 
be defined. 

However, the topological invariant $\Delta$ of the 
Eq.(\ref{eq.def inv topo with P}) 
may be recast in a form that, albeit less elegant, remains valid in 
the case of the AF 
order, for a continuous gauge and under the convenient choice of axes 
$\op{a}=\op{x}$ 
and $\op{b}=\op{x}-\op{y}$ described in the appendix \ref{app.Fu and Kane AF} :
\begin{eqnarray}\label{eq.def inv topo with chi}
\Delta =\sum\limits_{\Gamma^b\in \{0,\pi\}} &\bigg[&\frac{1}{2\pi i}
\int_{0}^{\pi}dk_a \nabla_{k_a}\text{log\ det}
[\tilde{\sew}_{\Gamma_b}(k_a)] \nn\\
&+&\frac{1}{\pi} \sum_\alpha (\chi_{\alpha,0,\Gamma_b}+
\chi_{\alpha,\pi,\Gamma_b})\bigg]\mathrm{mod}~2
\end{eqnarray}

This expression has the advantage of depending only on $\chi$, and of being 
computable given a continuous set of Bloch states. It justifies 
the computation 
of the following section. 
Fang {\it et al.} proposed an expression for the topological invariant, 
similar to the Eq. (\ref{eq.def inv topo with xi}):
\begin{equation}\label{eq.def inv topo with xi AF}
(-1)^\Delta=\prod_{\Gamma_i \in B-TRIM} 
\prod_{\alpha=1}^N \xi_{\alpha}(\Gamma_i)
\end{equation}
where the B-TRIM are those where $\ta^2=-1$. The 
Eq.(\ref{eq.def inv topo with xi AF}) does not directly follow from 
the definition (\ref{eq.def inv topo with P}) in terms of 
time-reversal polarization, but is equivalent to our result 
(\ref{eq.def inv topo with chi}) 
if there is no band inversion at a single A-TRIM (which is true in 
the case at hand).

\subsection{Parallel Transport Method}
\label{sec.Alexey}

\begin{figure*}
\centering
\includegraphics[height=3cm]{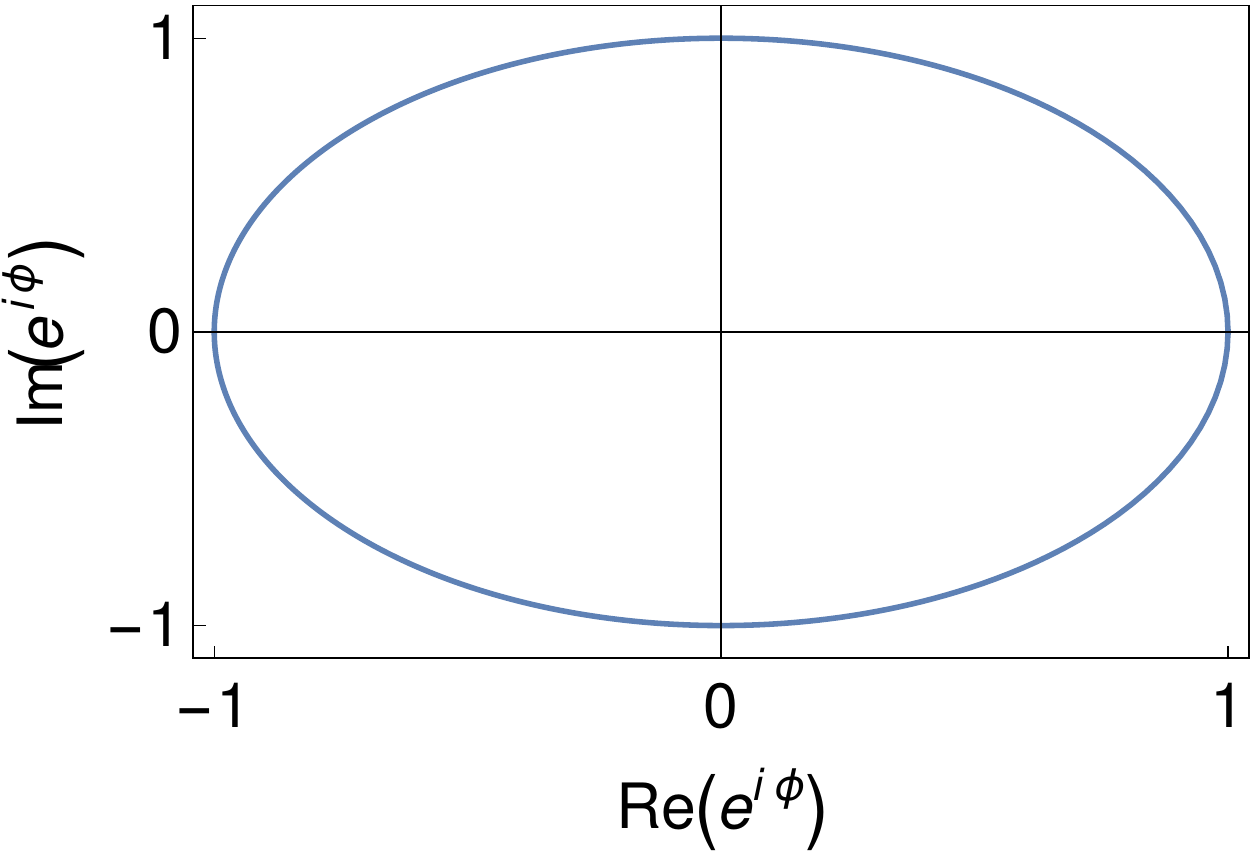} 
\includegraphics[height=3cm]{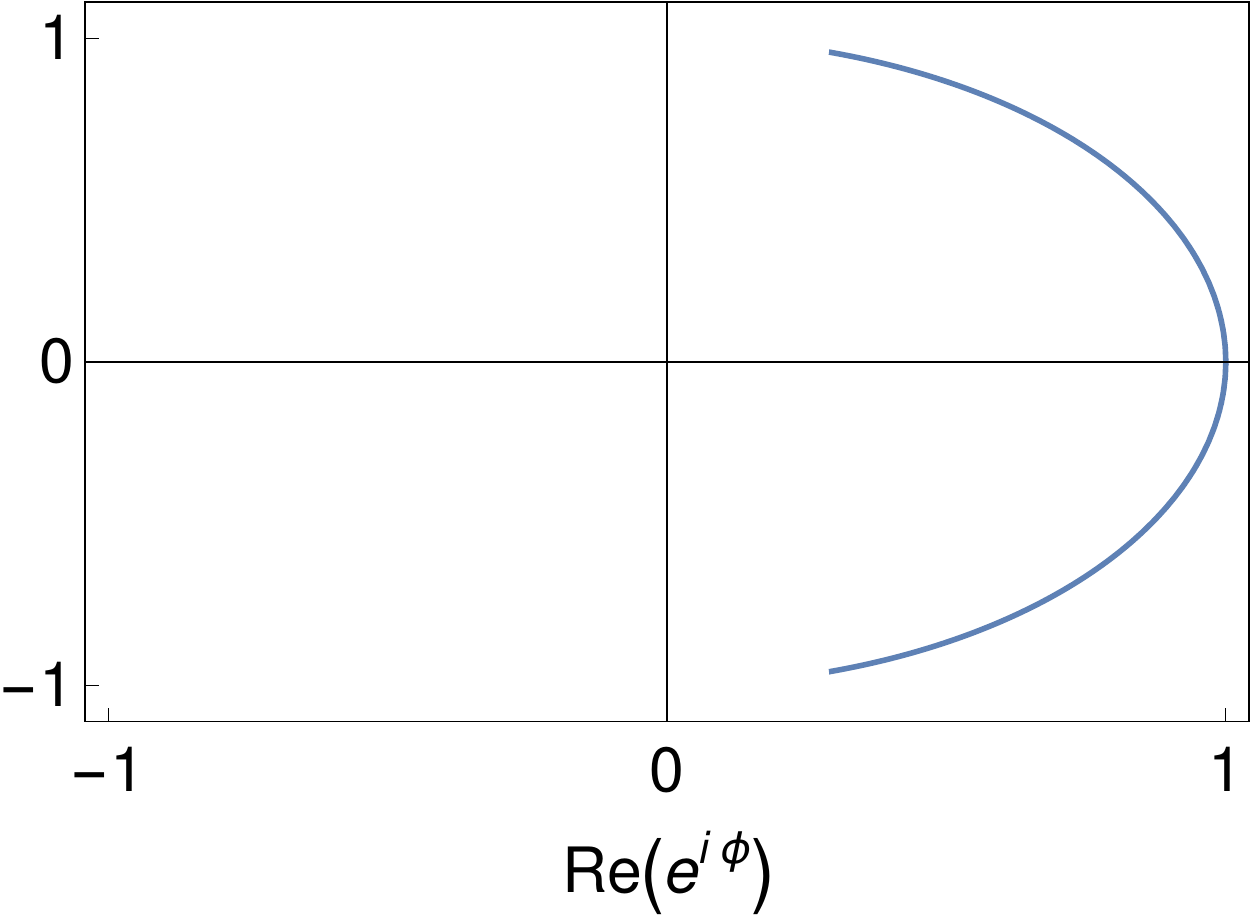} 
\includegraphics[height=3cm]{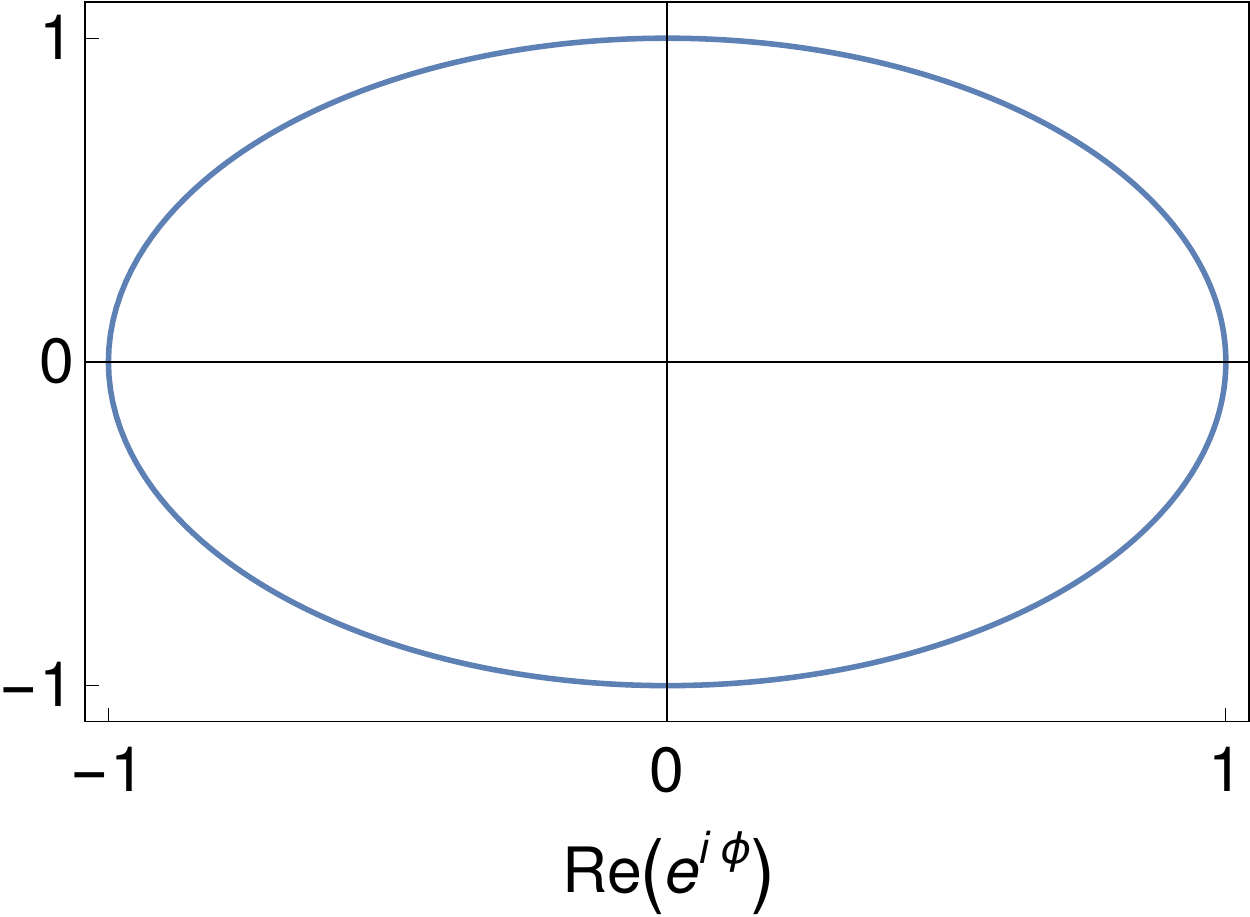} 
\includegraphics[height=3cm]{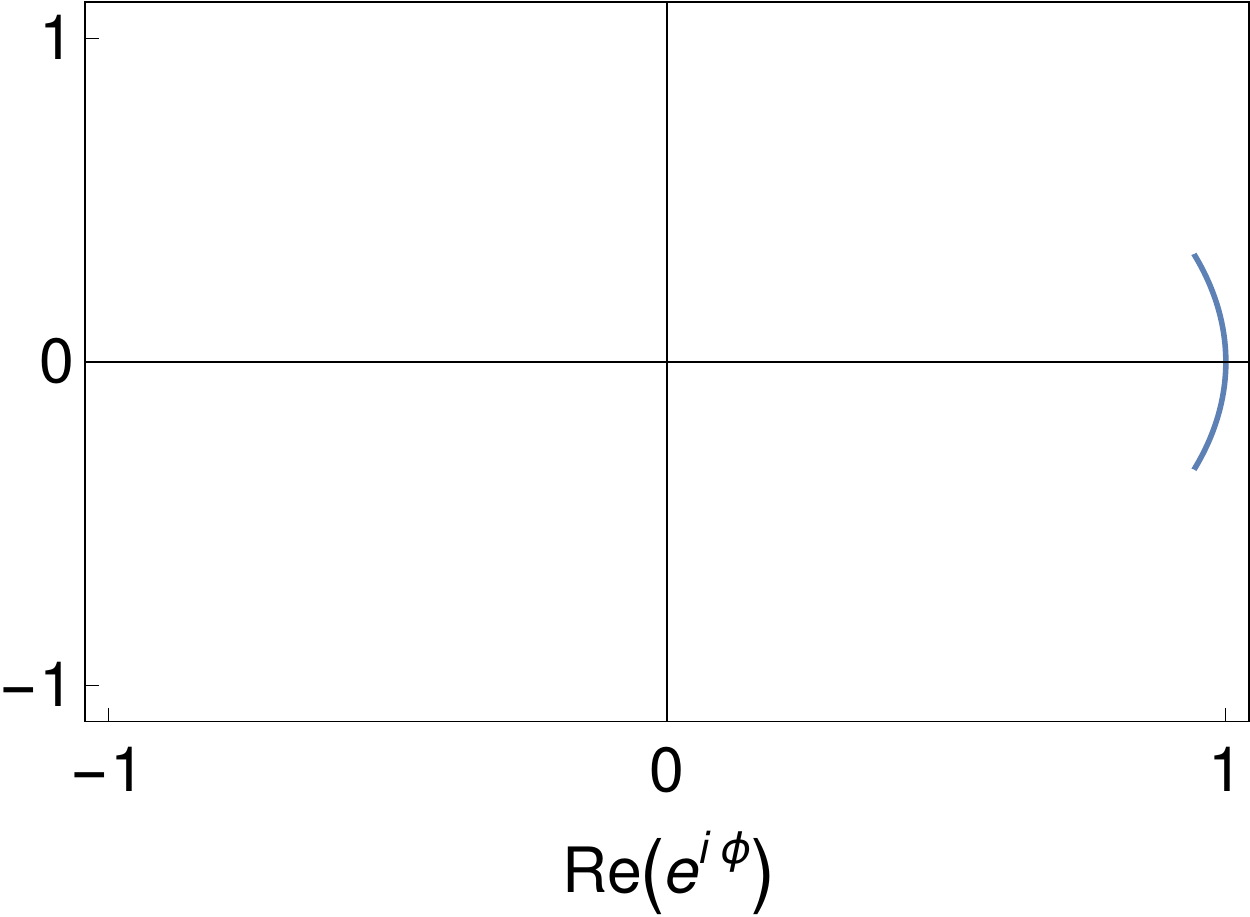}

\ \ \includegraphics[height=2.96cm]{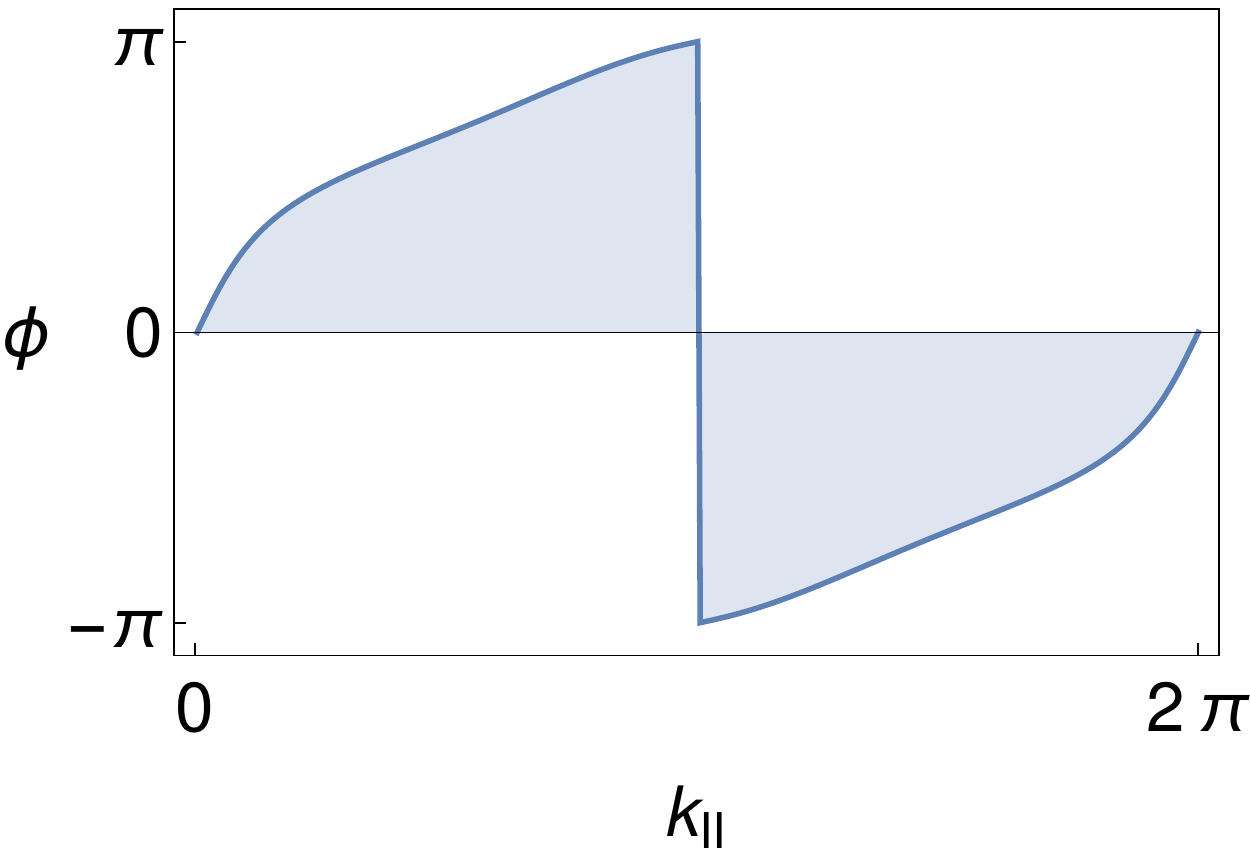} 
\includegraphics[height=2.96cm]{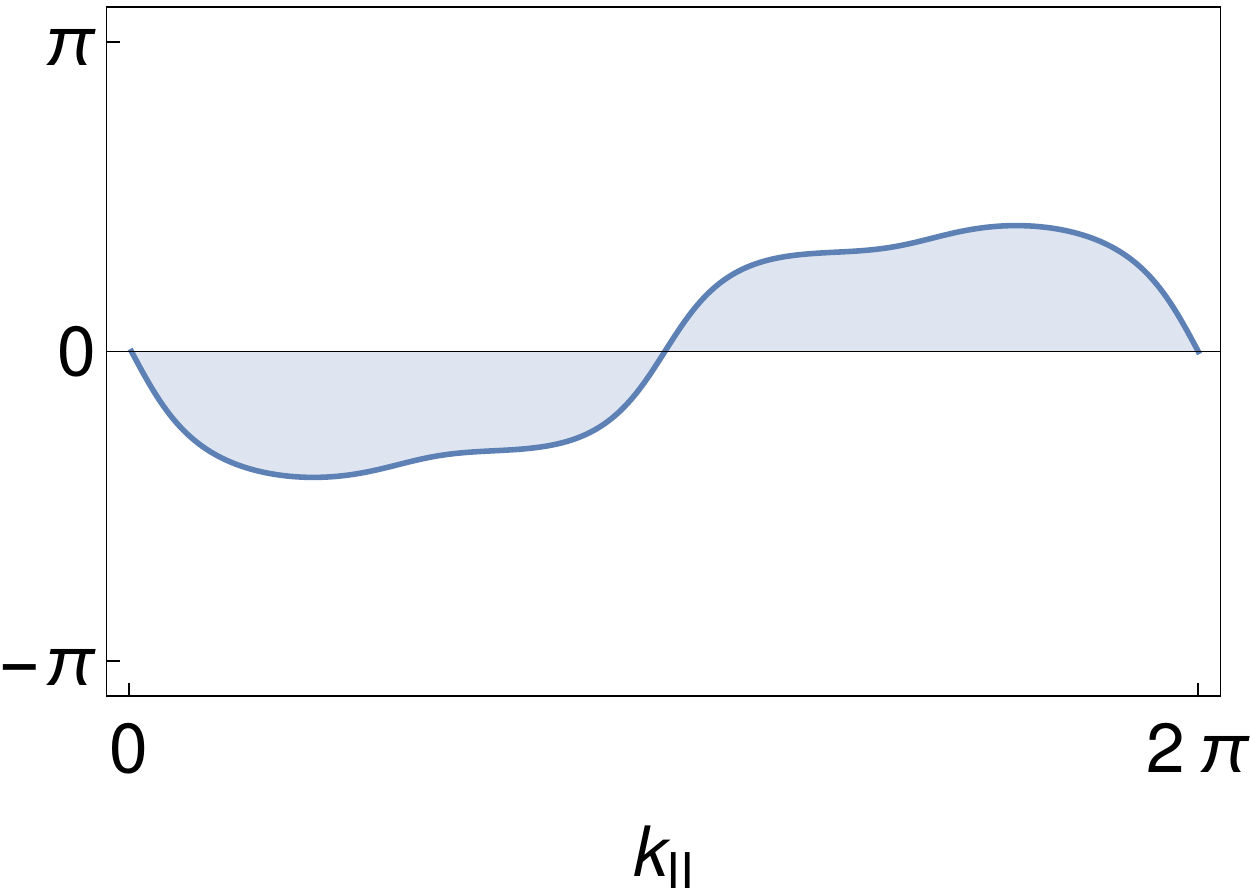} 
\includegraphics[height=2.96cm]{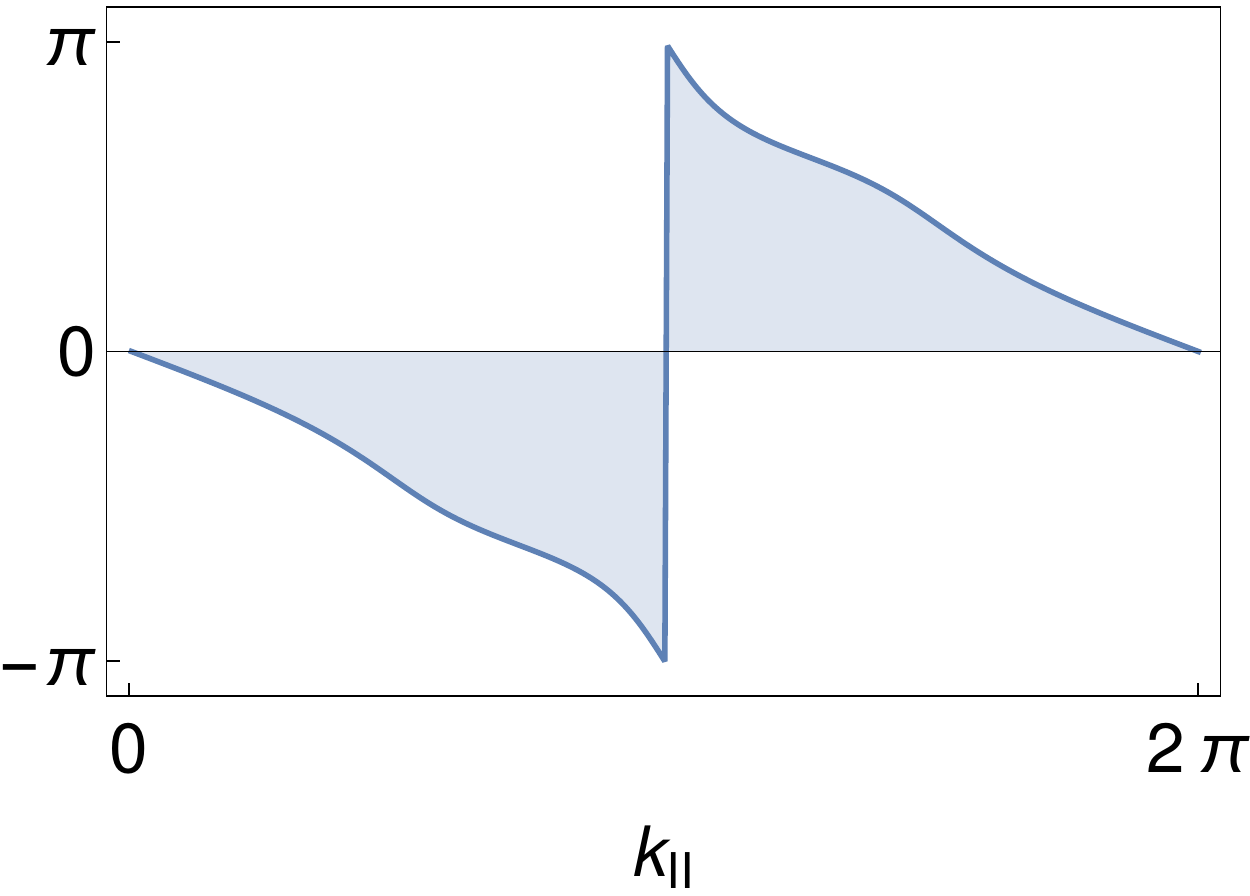} 
\includegraphics[height=2.96cm]{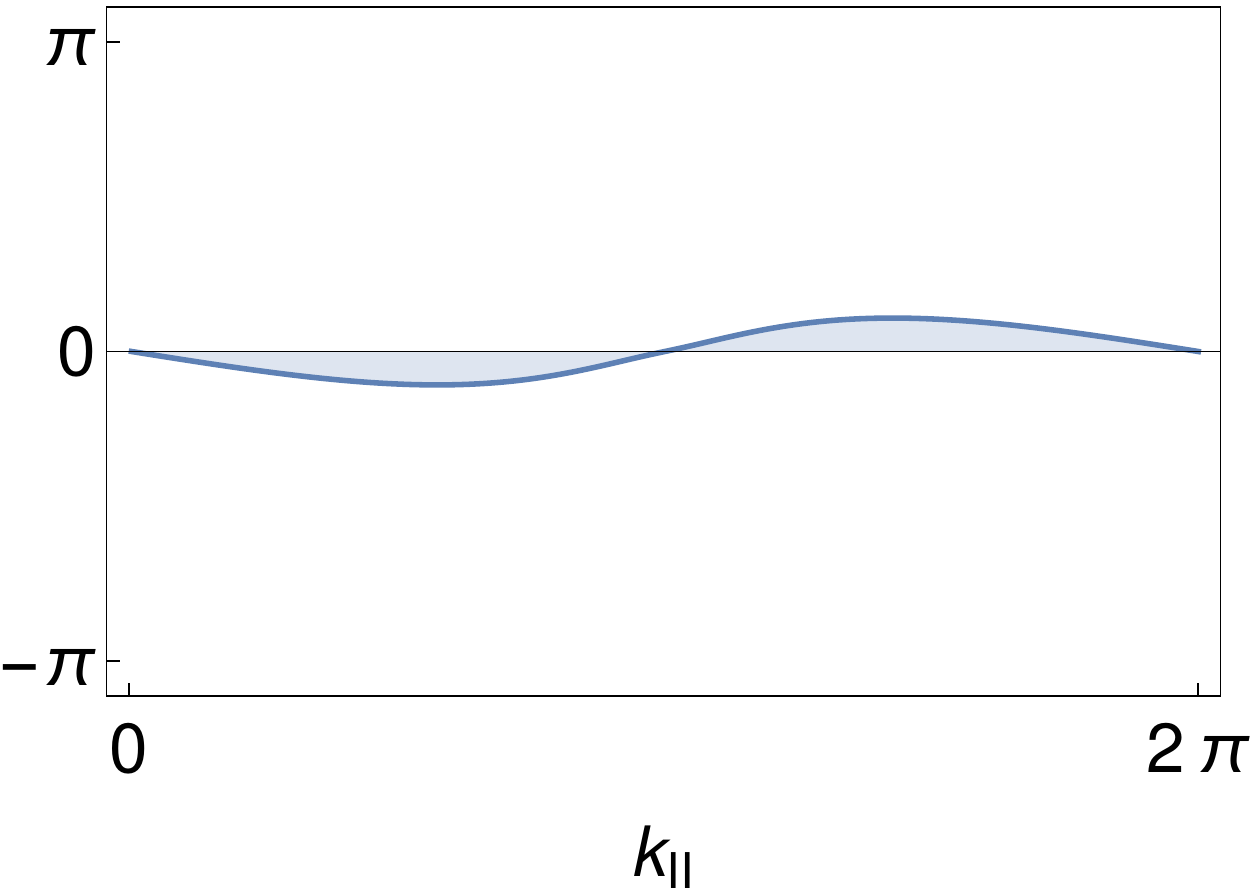}

\includegraphics[height=3cm]{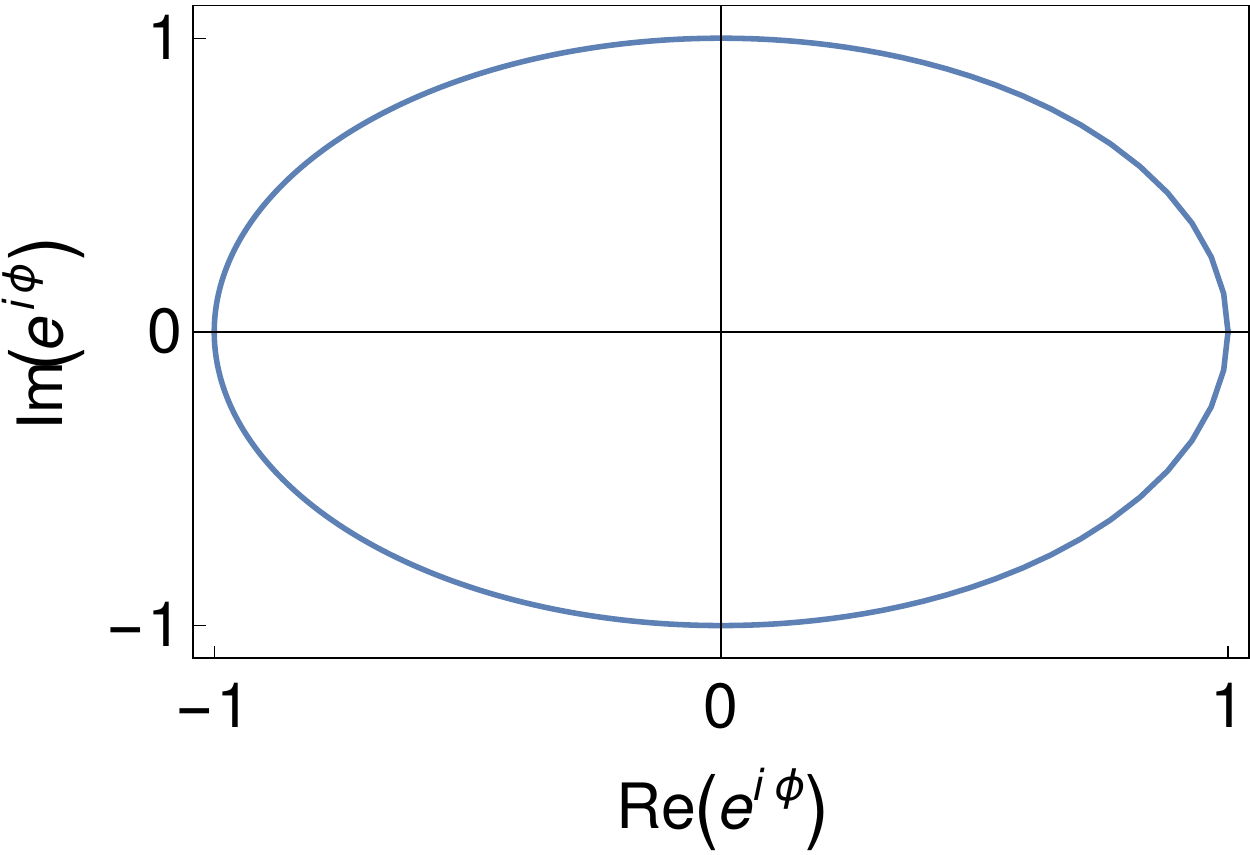} 
\includegraphics[height=3cm]{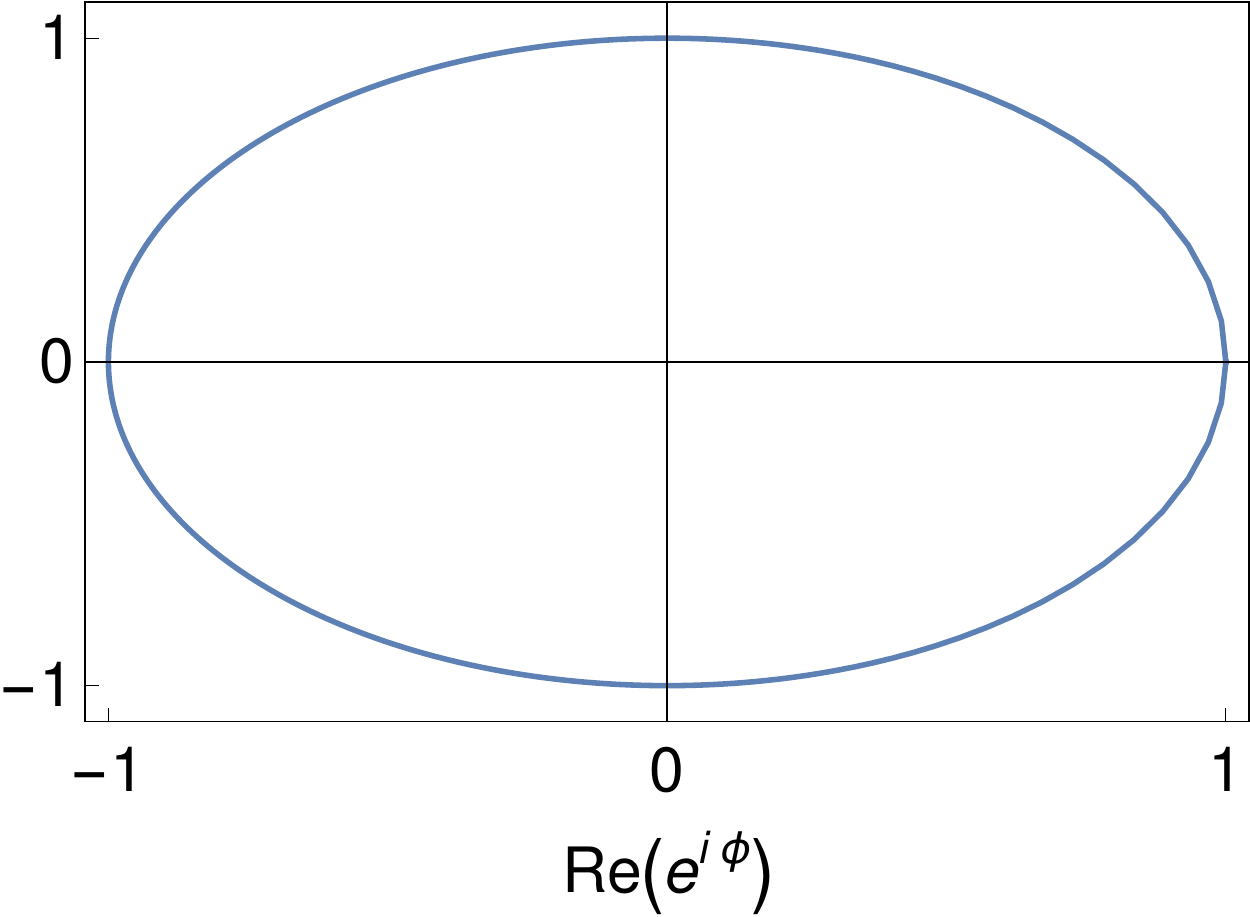} 
\includegraphics[height=3cm]{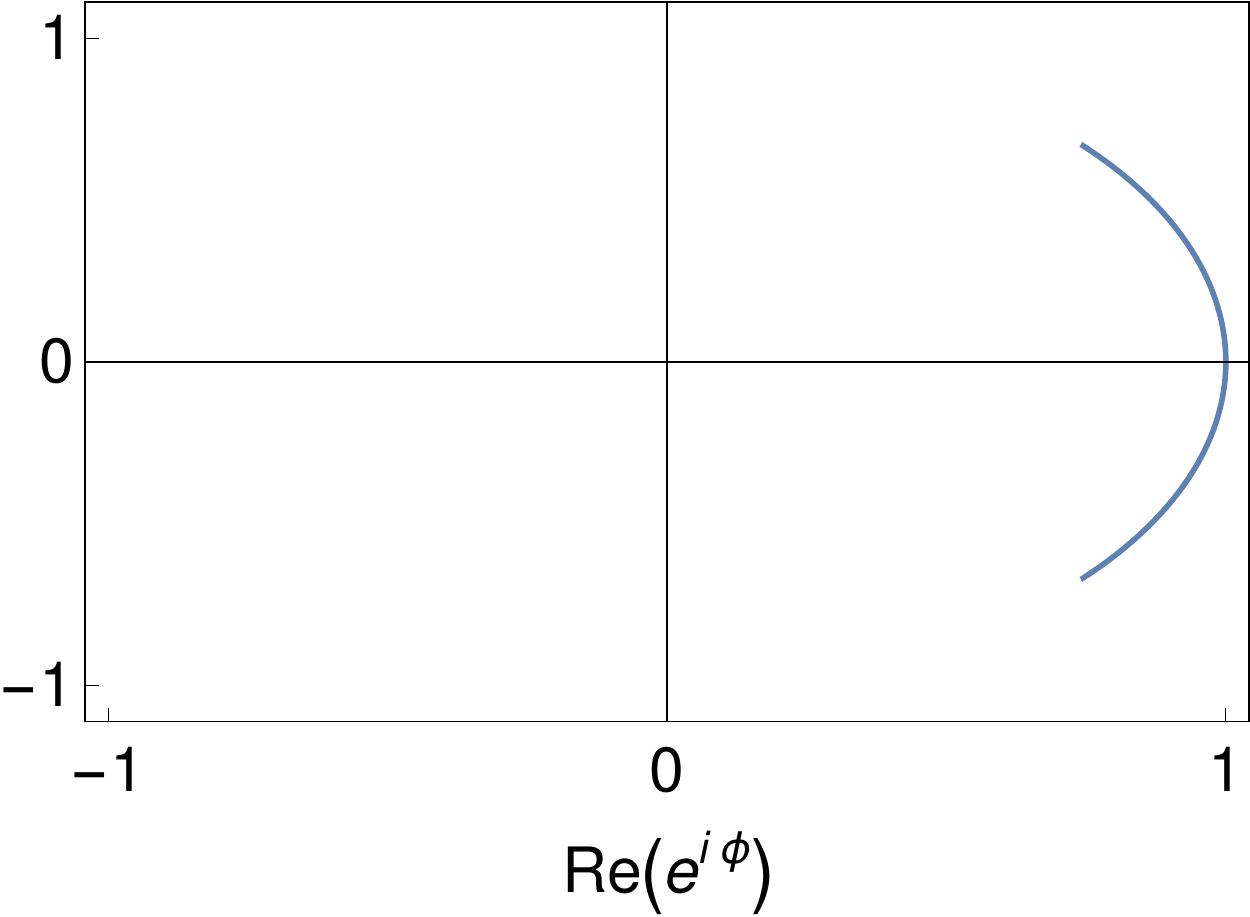} 
\includegraphics[height=3cm]{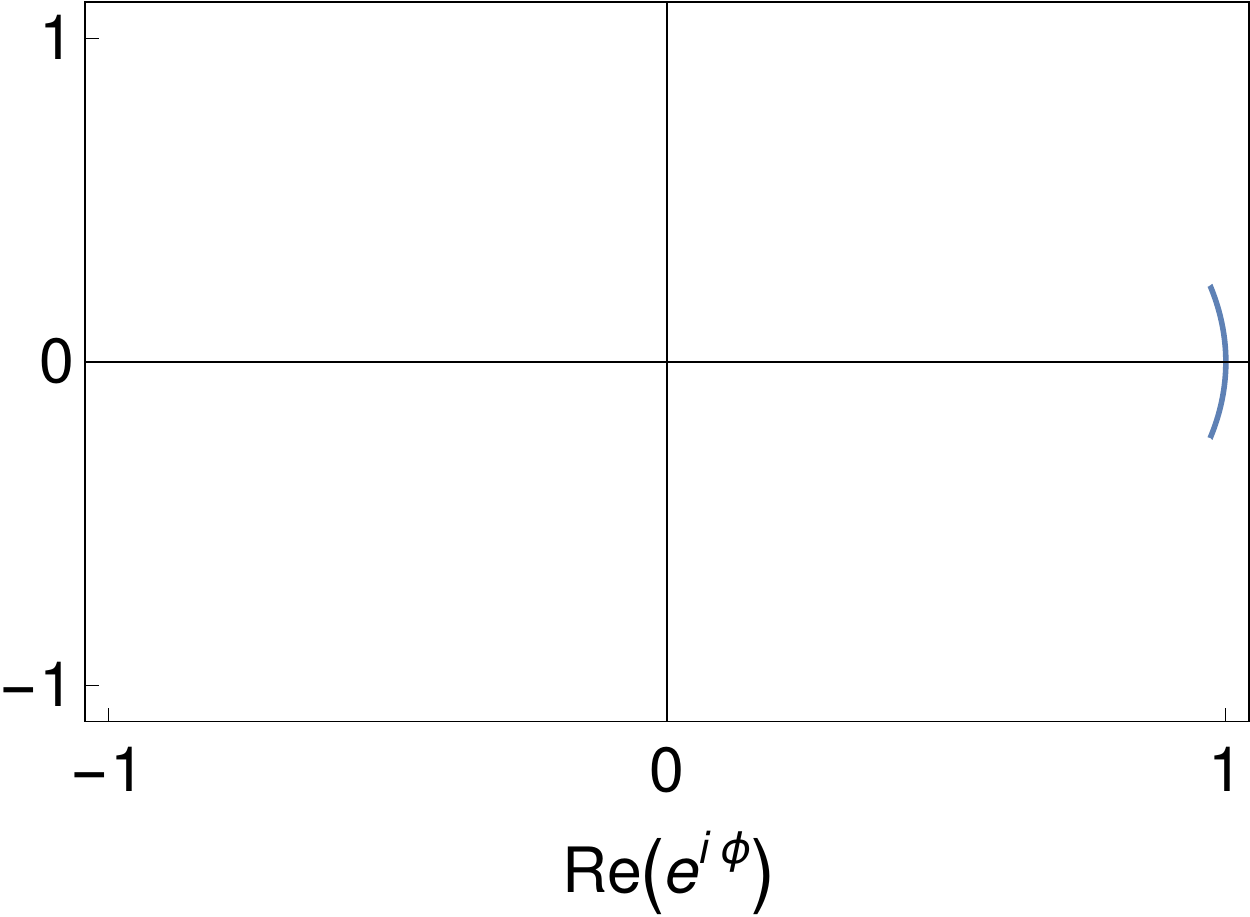}

\ \ \includegraphics[height=2.96cm]{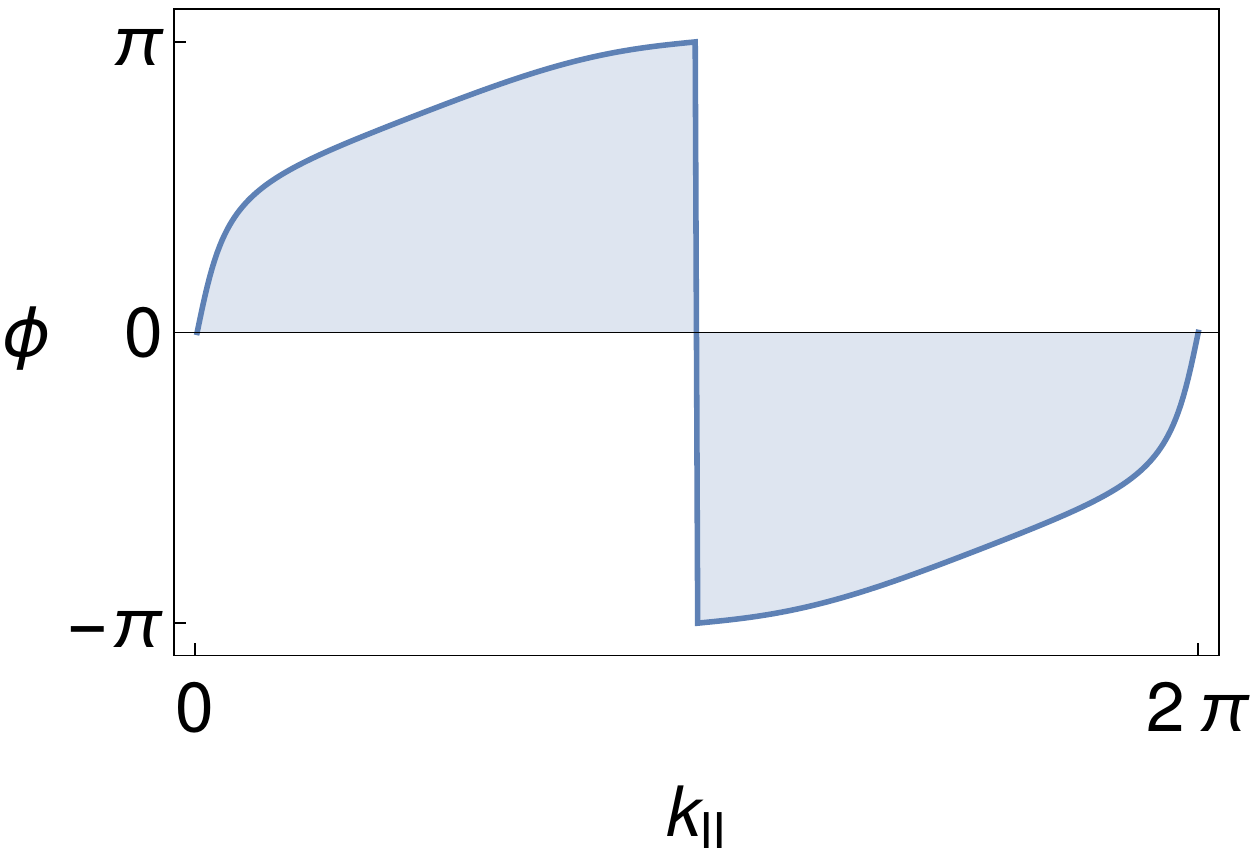} 
\includegraphics[height=2.96cm]{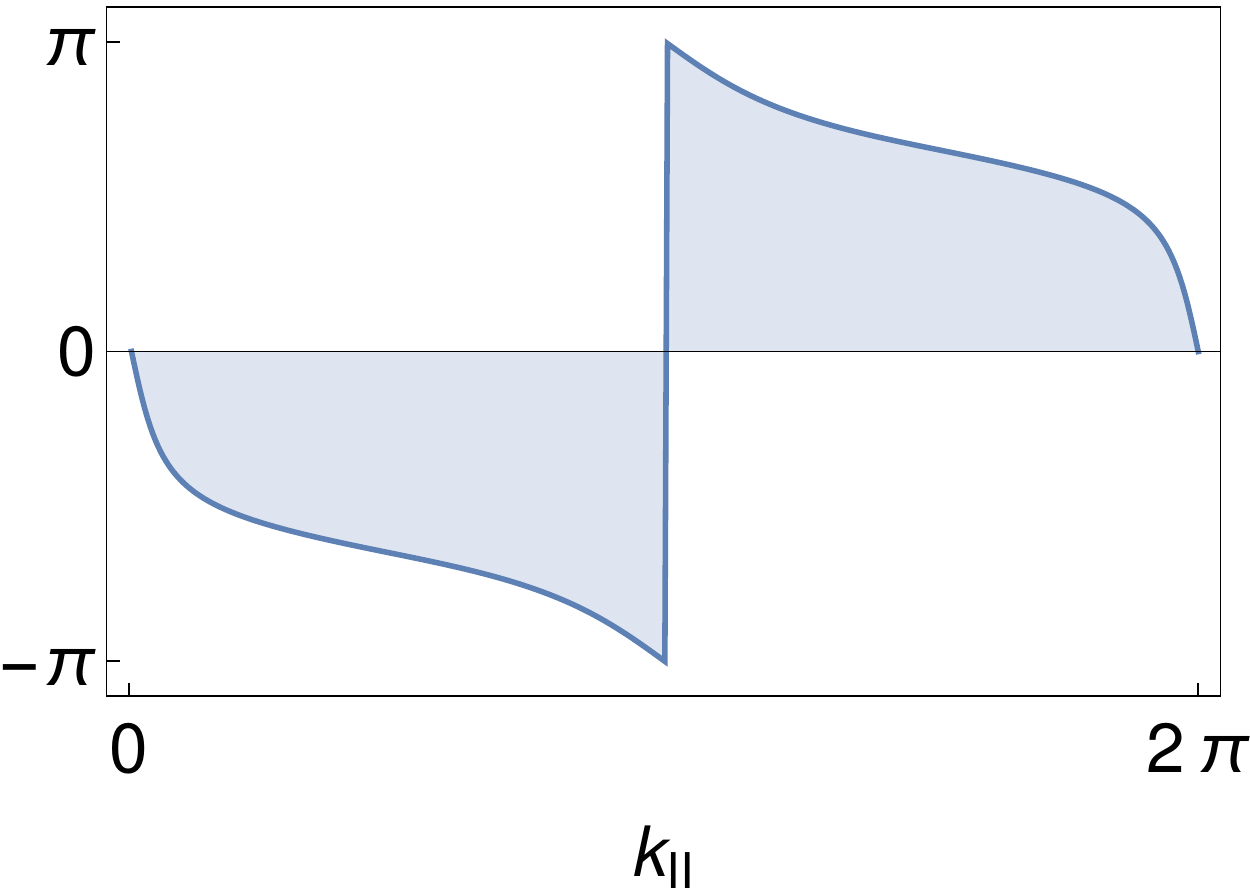} 
\includegraphics[height=2.96cm]{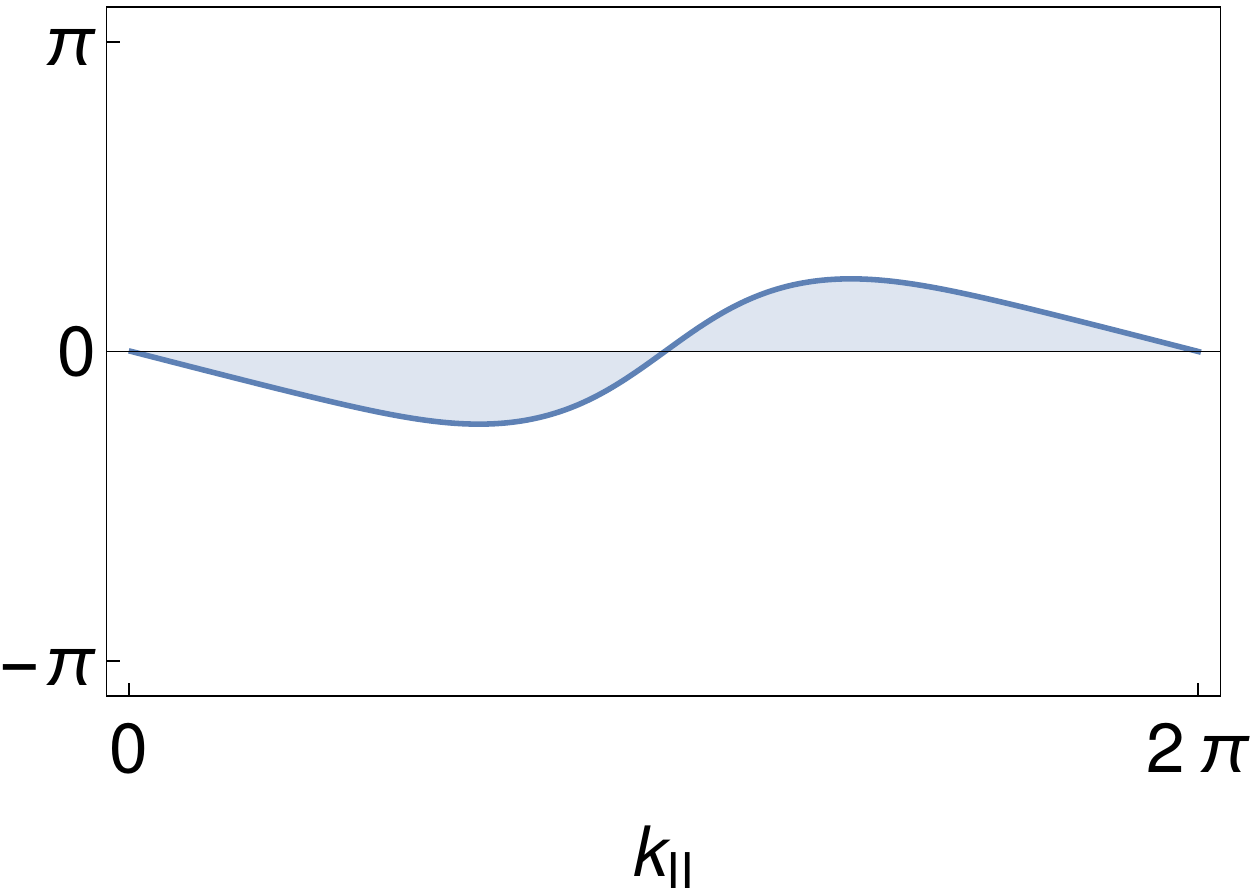} 
\includegraphics[height=2.96cm]{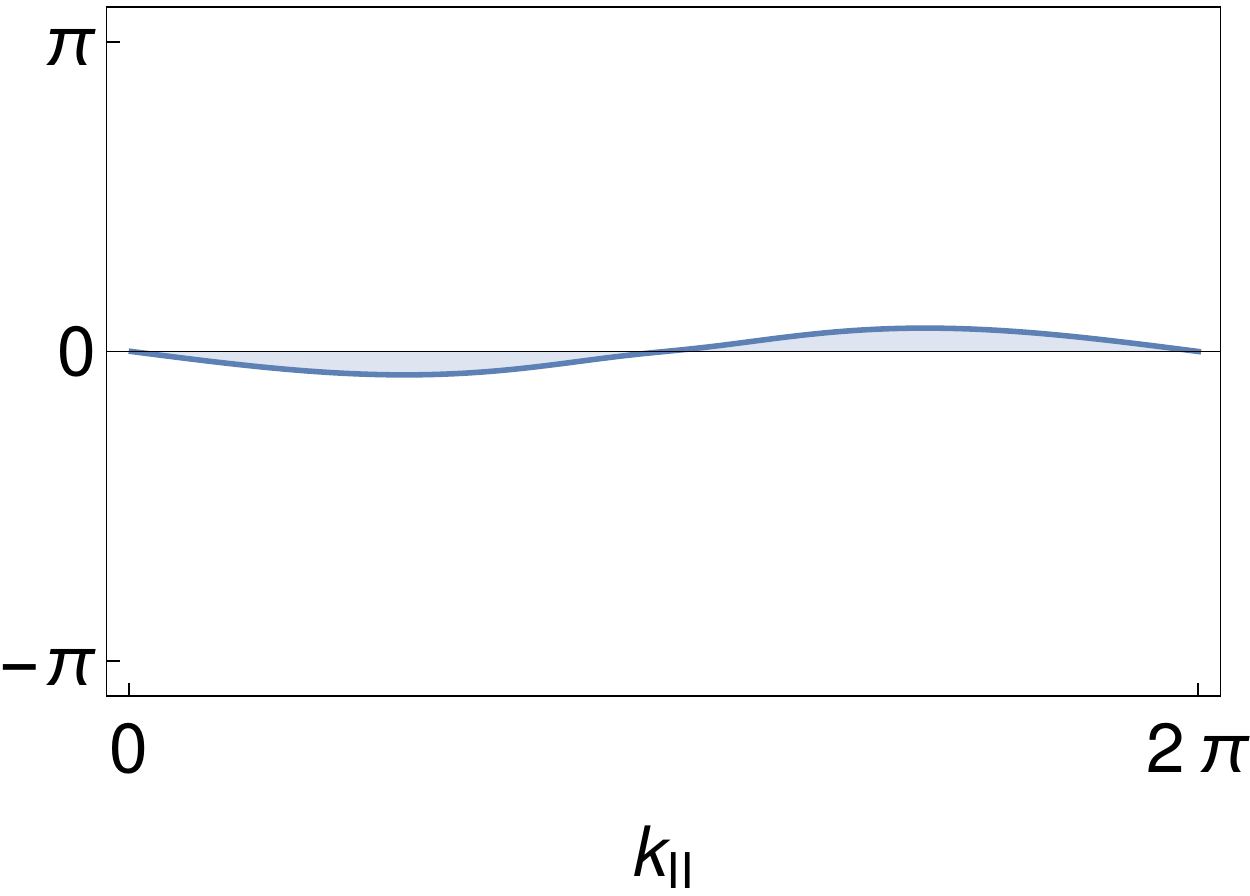}

\caption{(color online) Results of the parallel transport method for 
$\Delta\mu=3$. For the two first lines $m=2$ and for the two last lines 
$m=5$. Each column corresponds to the study of a pair 
of degenerate bands. From left to right we go from the lower to the higher 
energy bands. 
In line 1 and 3, $e^{i\phi}$ is plotted in the complex plane as a 
parametric function of $k_\parallel$. In line 2 and 3, $\phi$ is plotted as 
a function of $k_\parallel$. These plots and 
the Eq.(\ref{eq.def inv topo winding number}) allow us to evaluate the 
$Z_2$ topological invariant. In both cases, the system is a trivial 
insulator at 3/4 filling, and a topological insulator at 1/4 filling. At 
half filling, the system is in the topological phase for  $m=2$ 
and in the trivial phase for  $m=5$. }
\label{Fig.graph.32.35}
\end{figure*}

\begin{figure*}
\centering
\includegraphics[height=3cm]{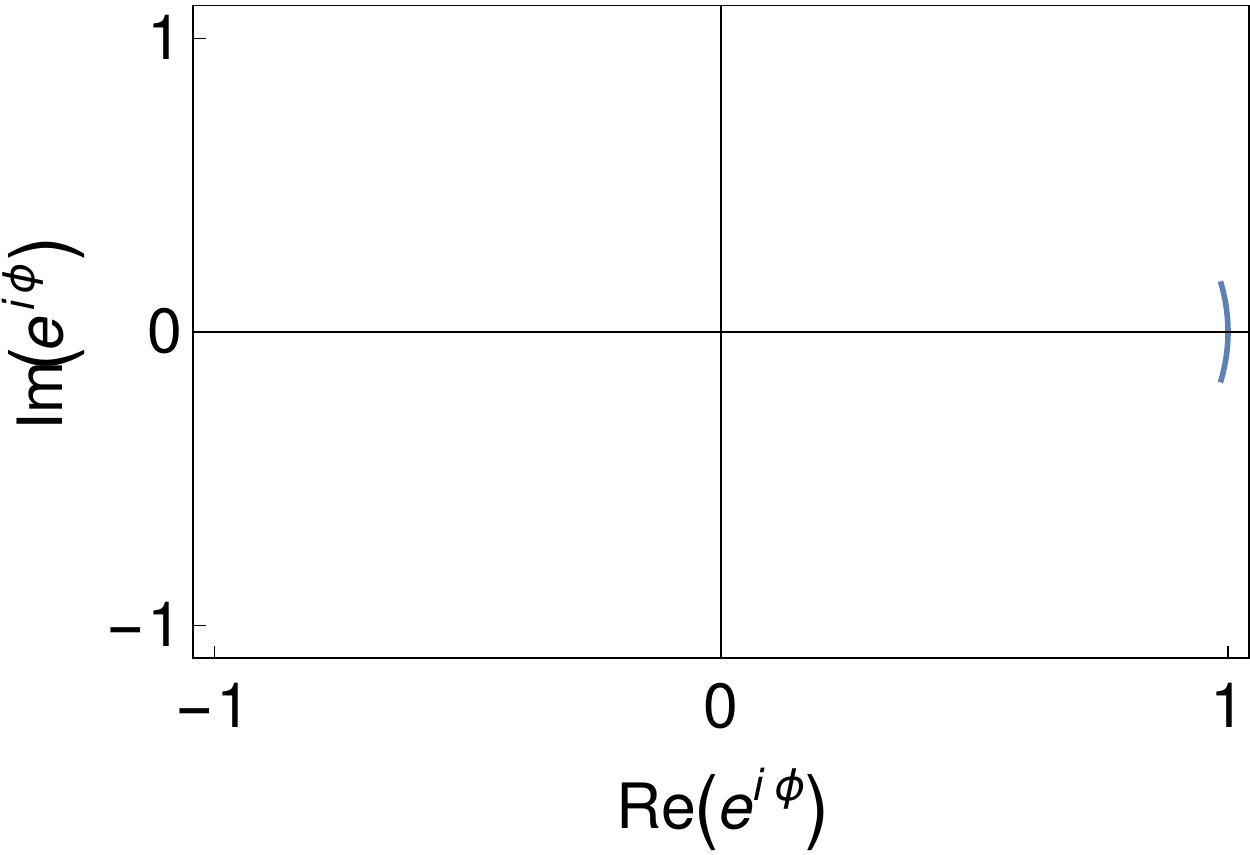} 
\includegraphics[height=3cm]{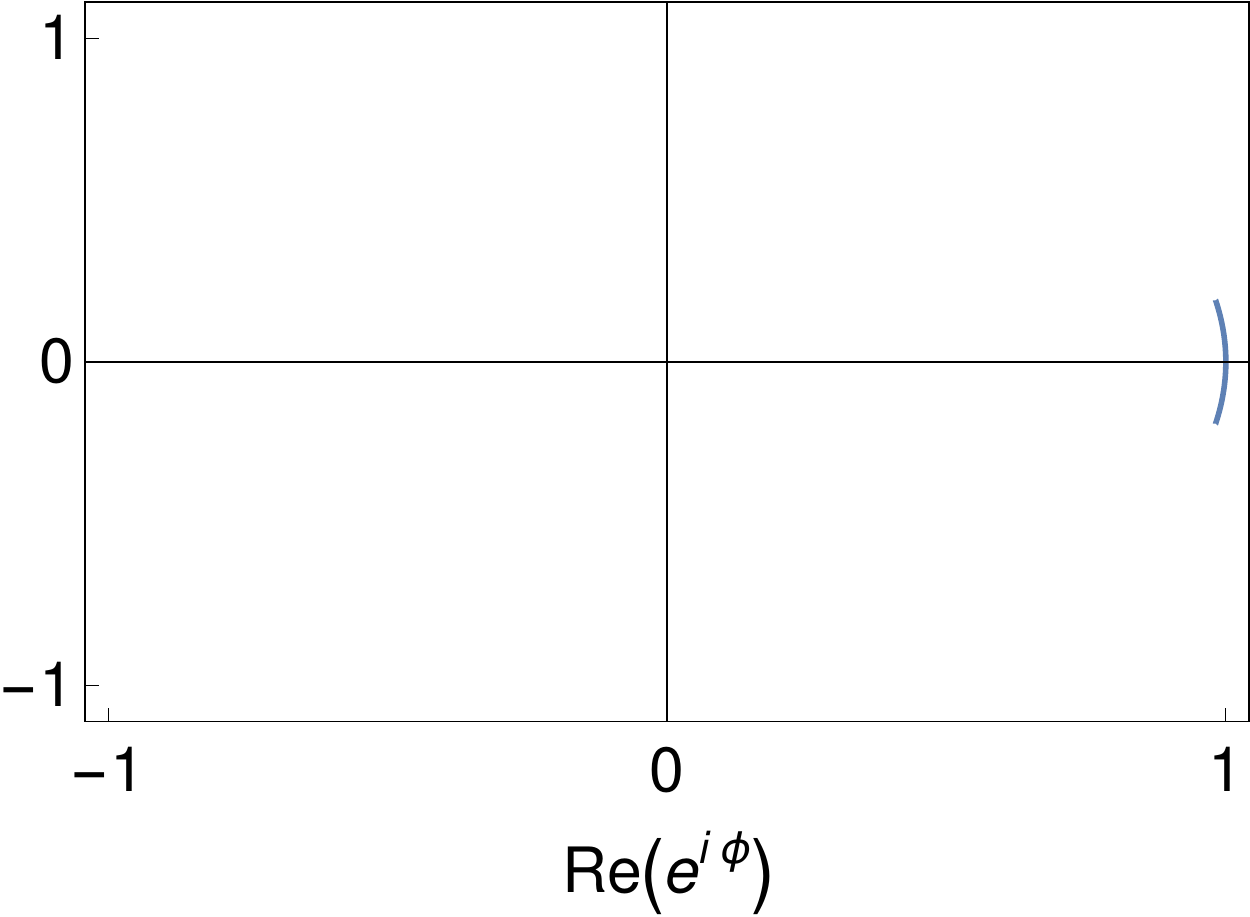} 
\includegraphics[height=3cm]{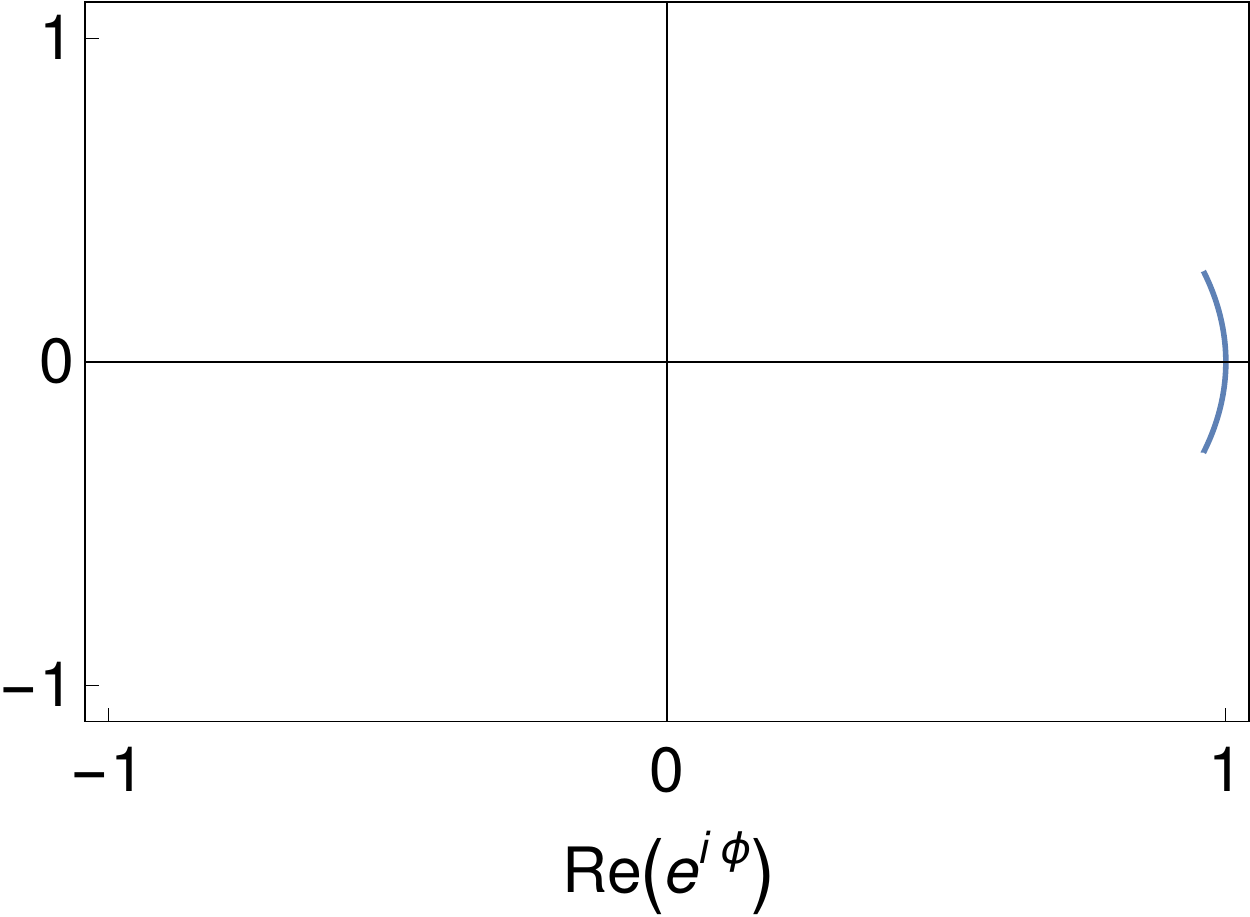} 
\includegraphics[height=3cm]{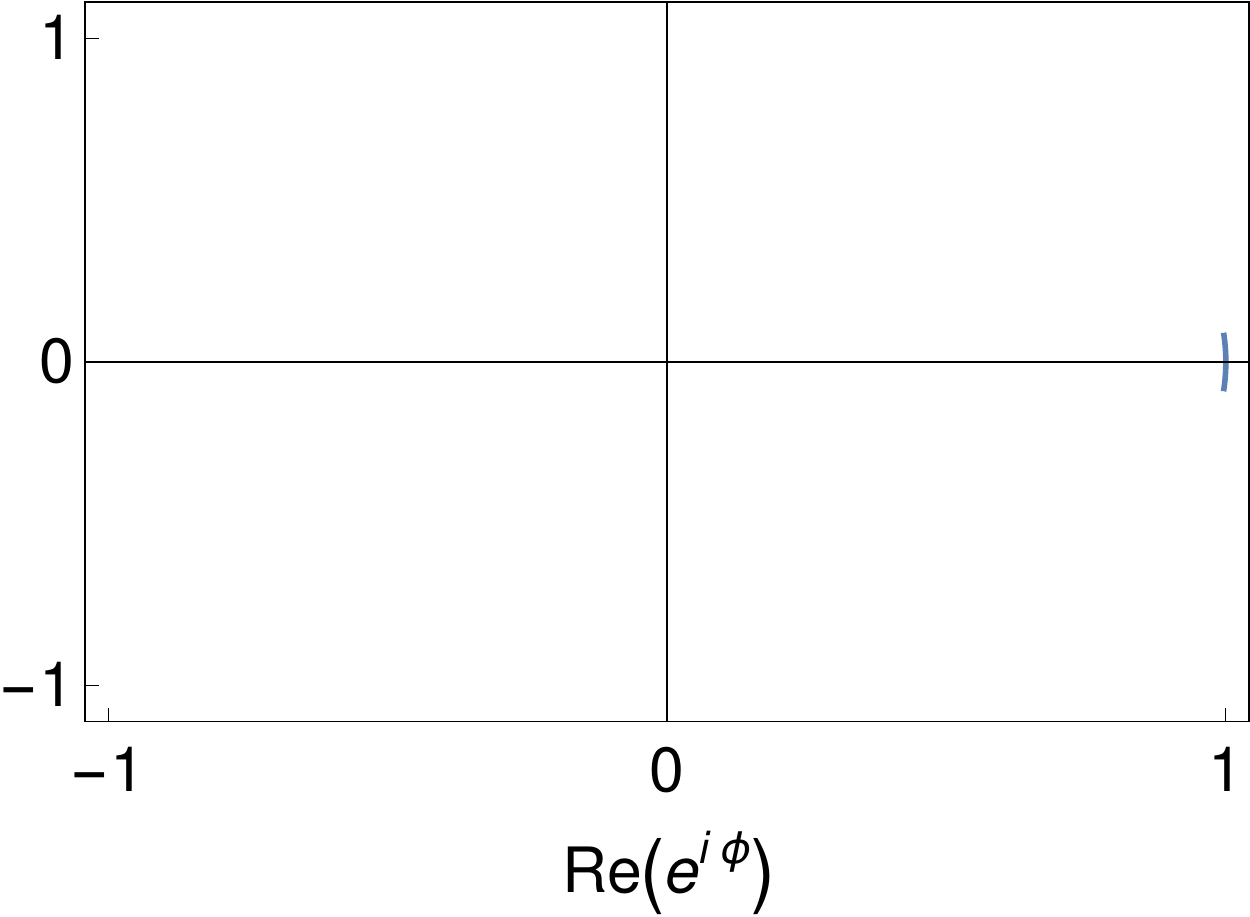}

\ \ \includegraphics[height=2.96cm]{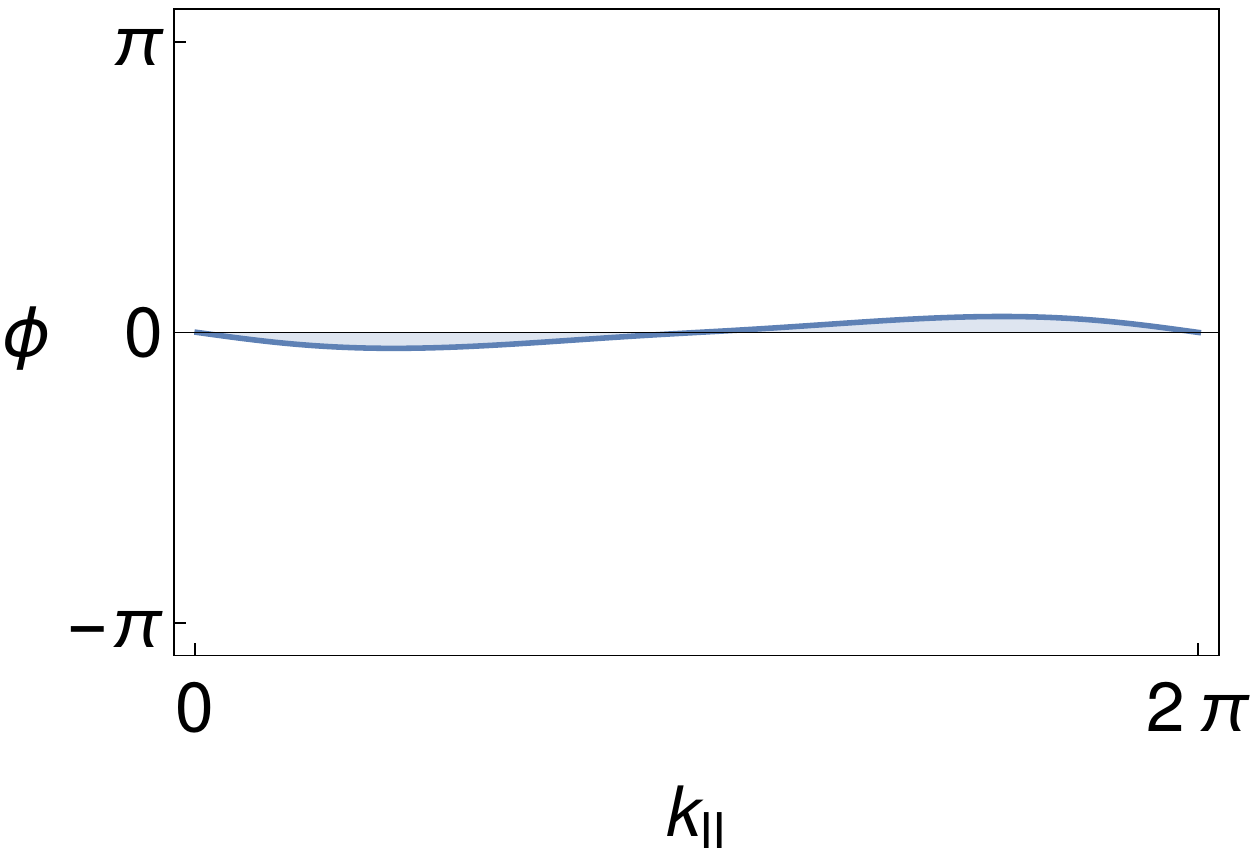} 
\includegraphics[height=2.96cm]{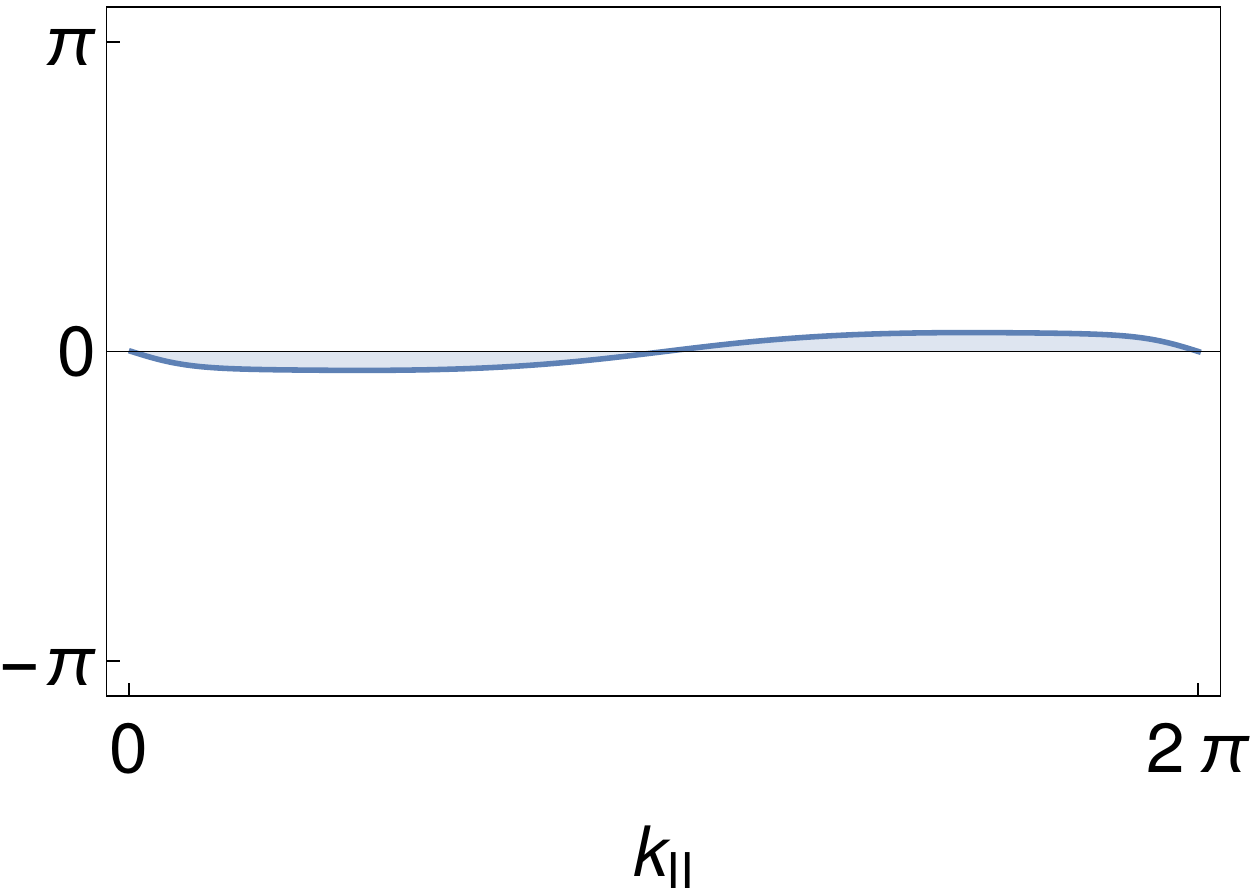} 
\includegraphics[height=2.96cm]{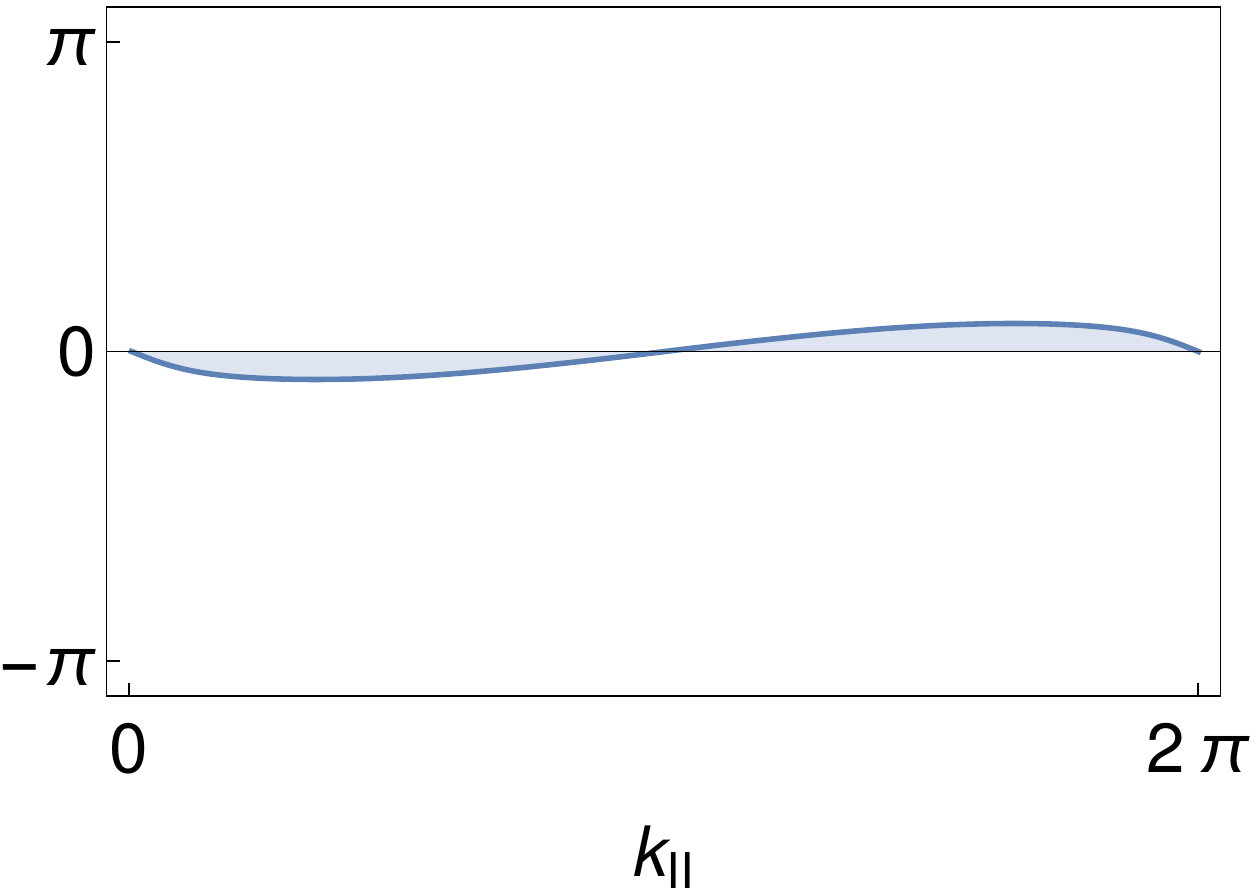} 
\includegraphics[height=2.96cm]{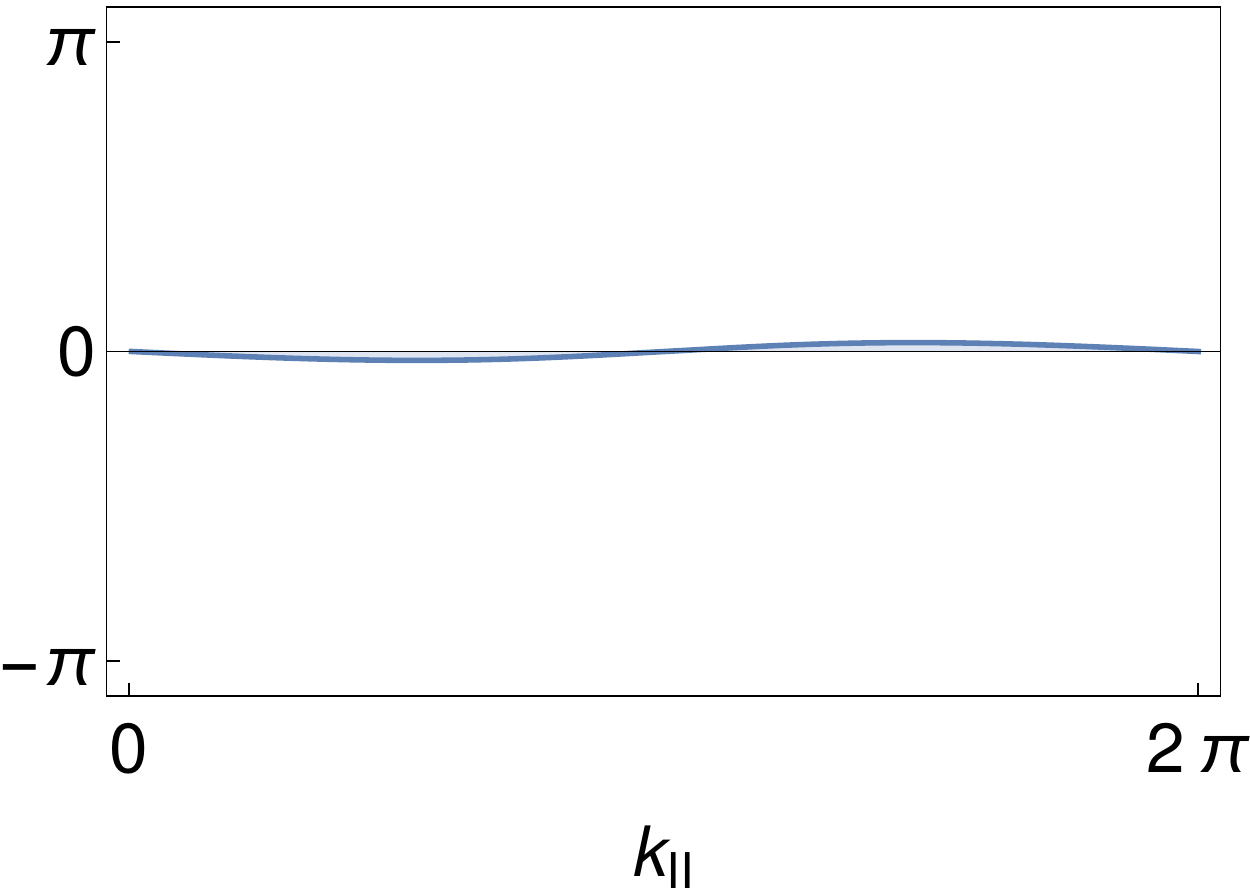}

\includegraphics[height=3cm]{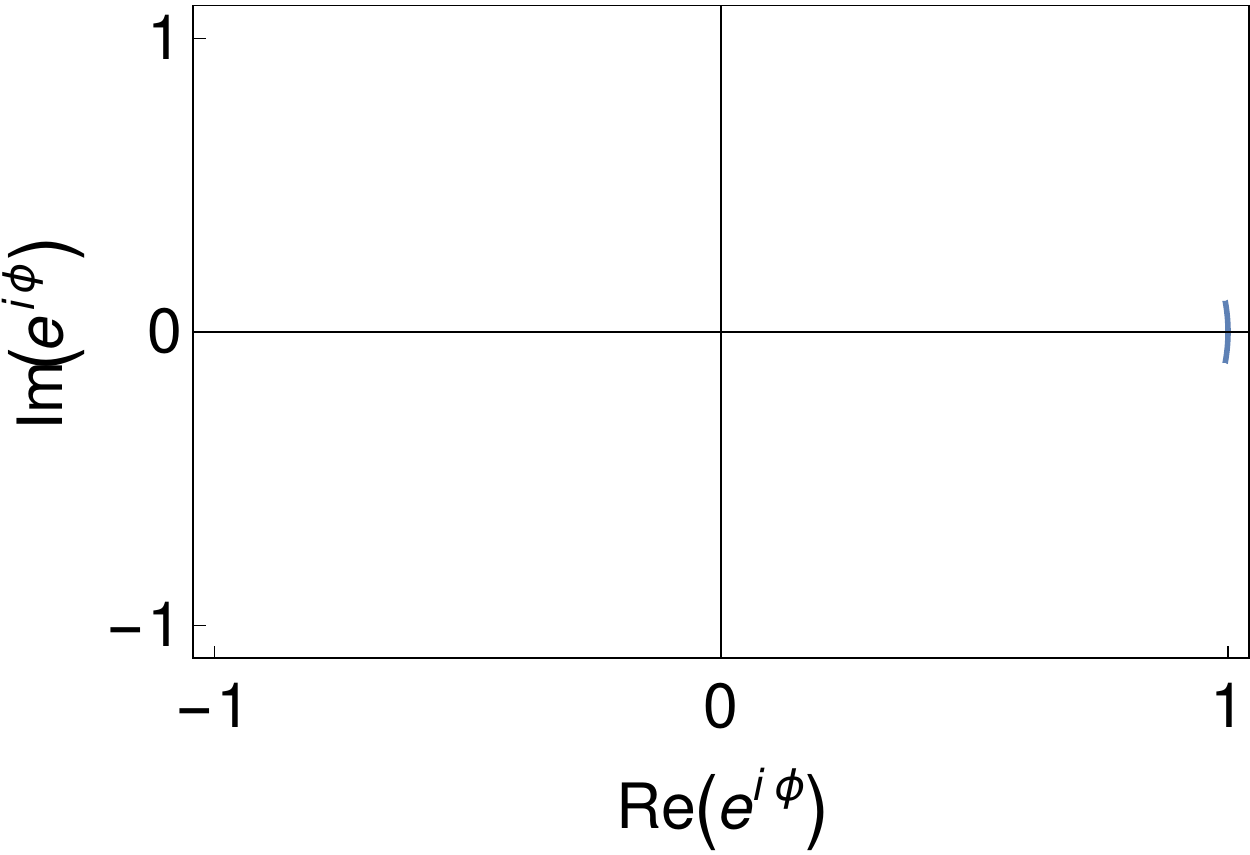} 
\includegraphics[height=3cm]{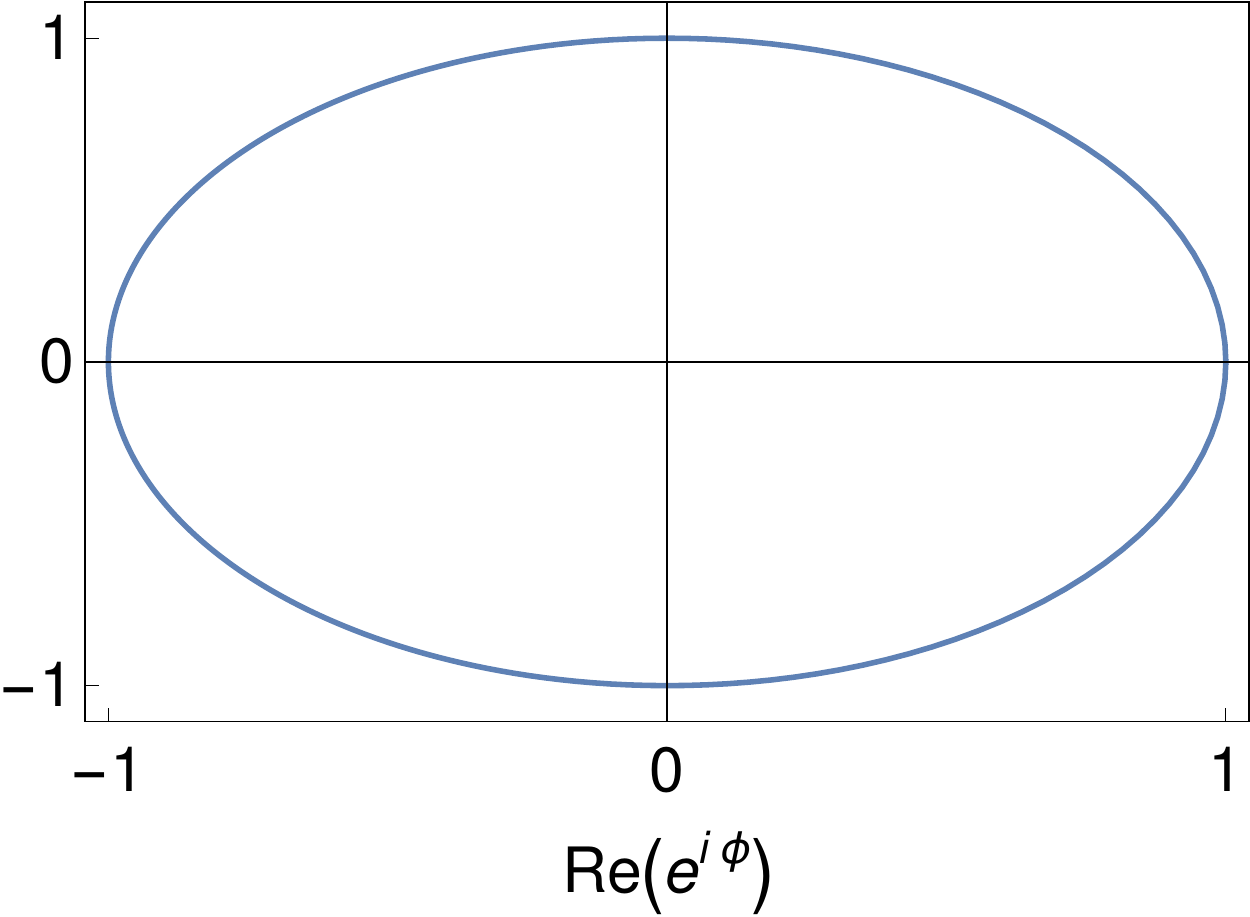} 
\includegraphics[height=3cm]{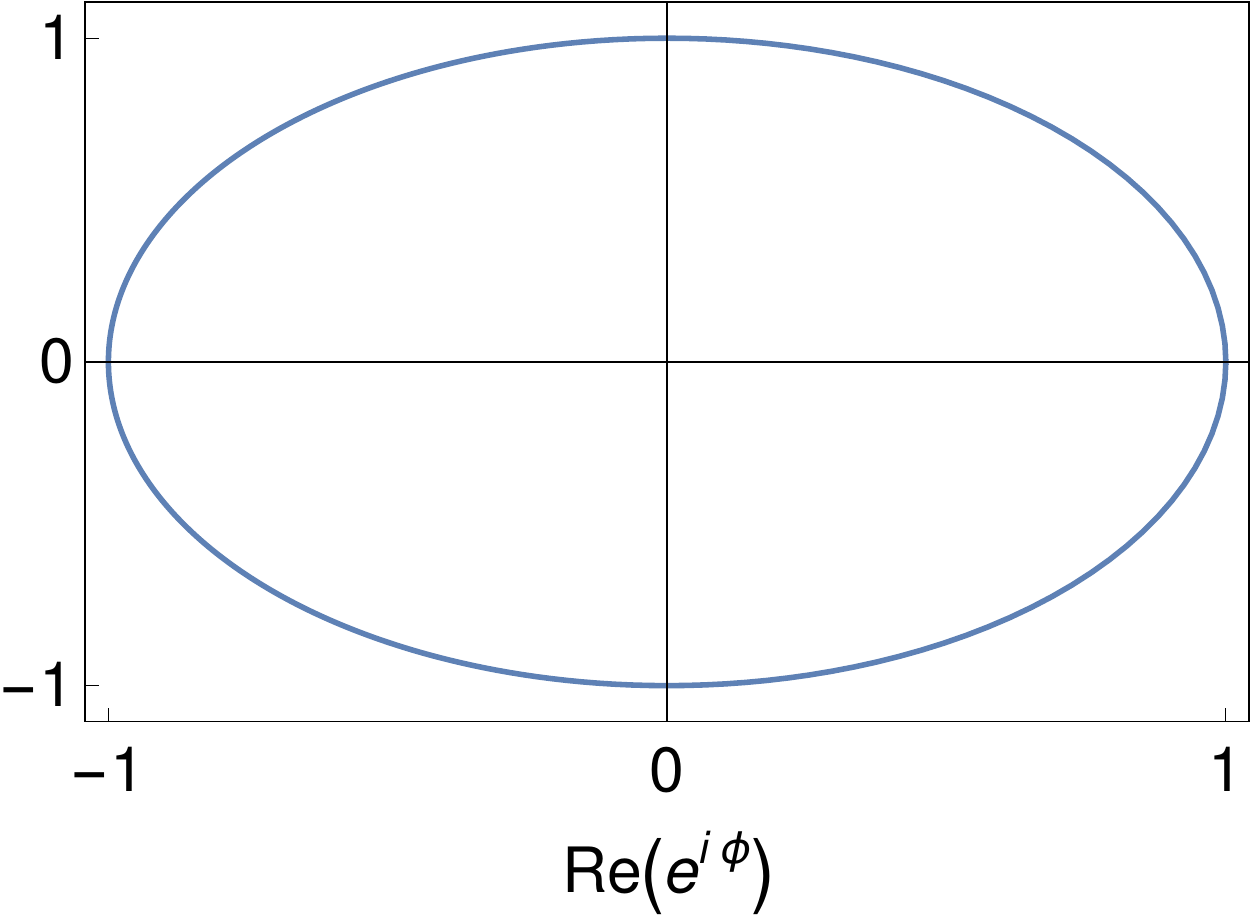} 
\includegraphics[height=3cm]{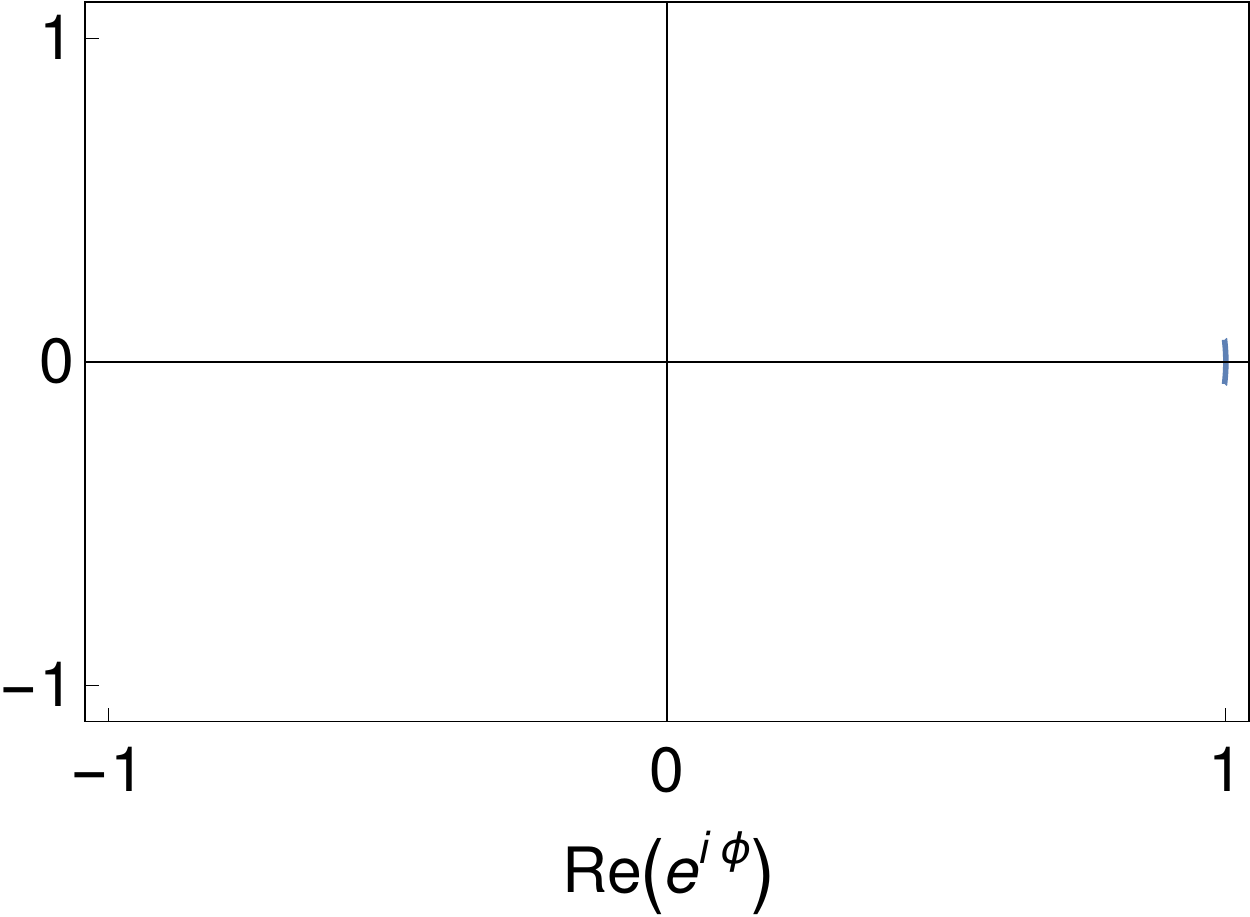}

\ \ \includegraphics[height=2.96cm]{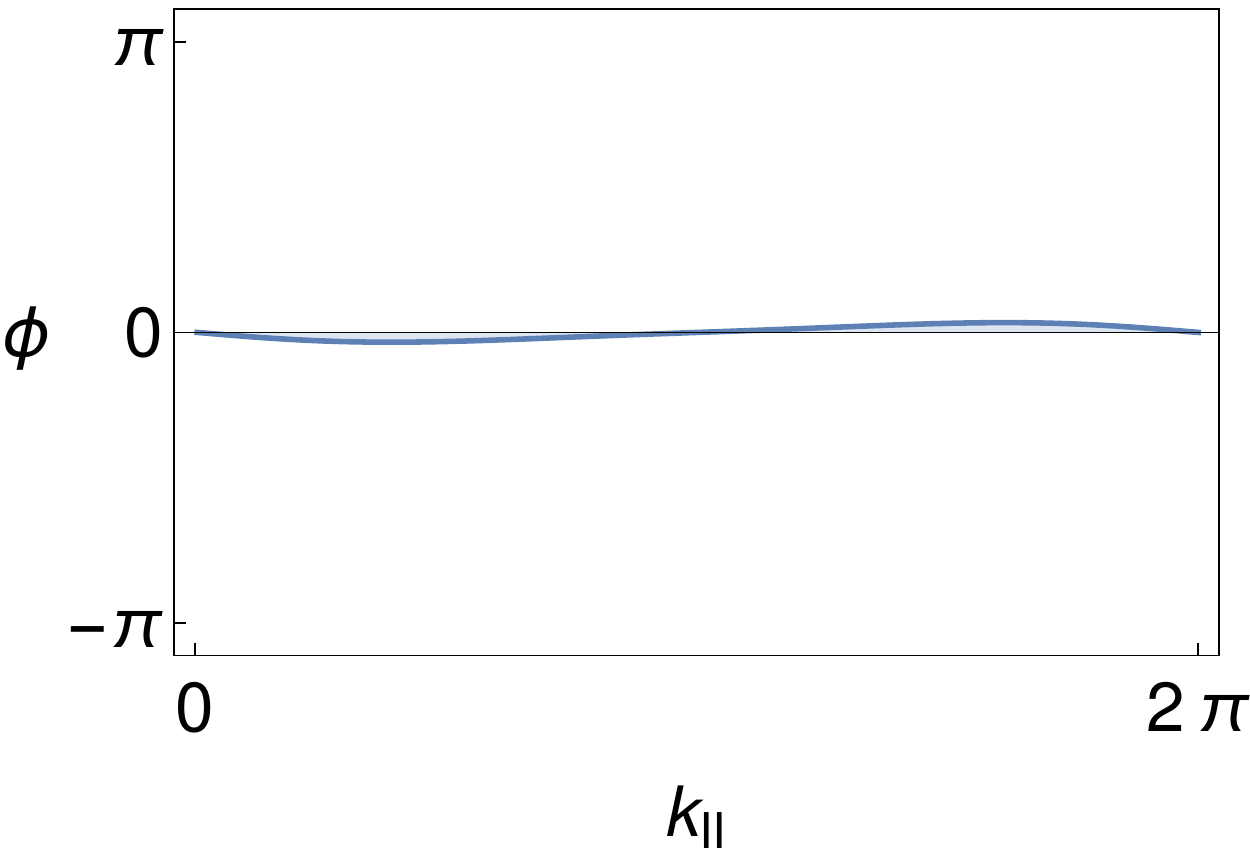} 
\includegraphics[height=2.96cm]{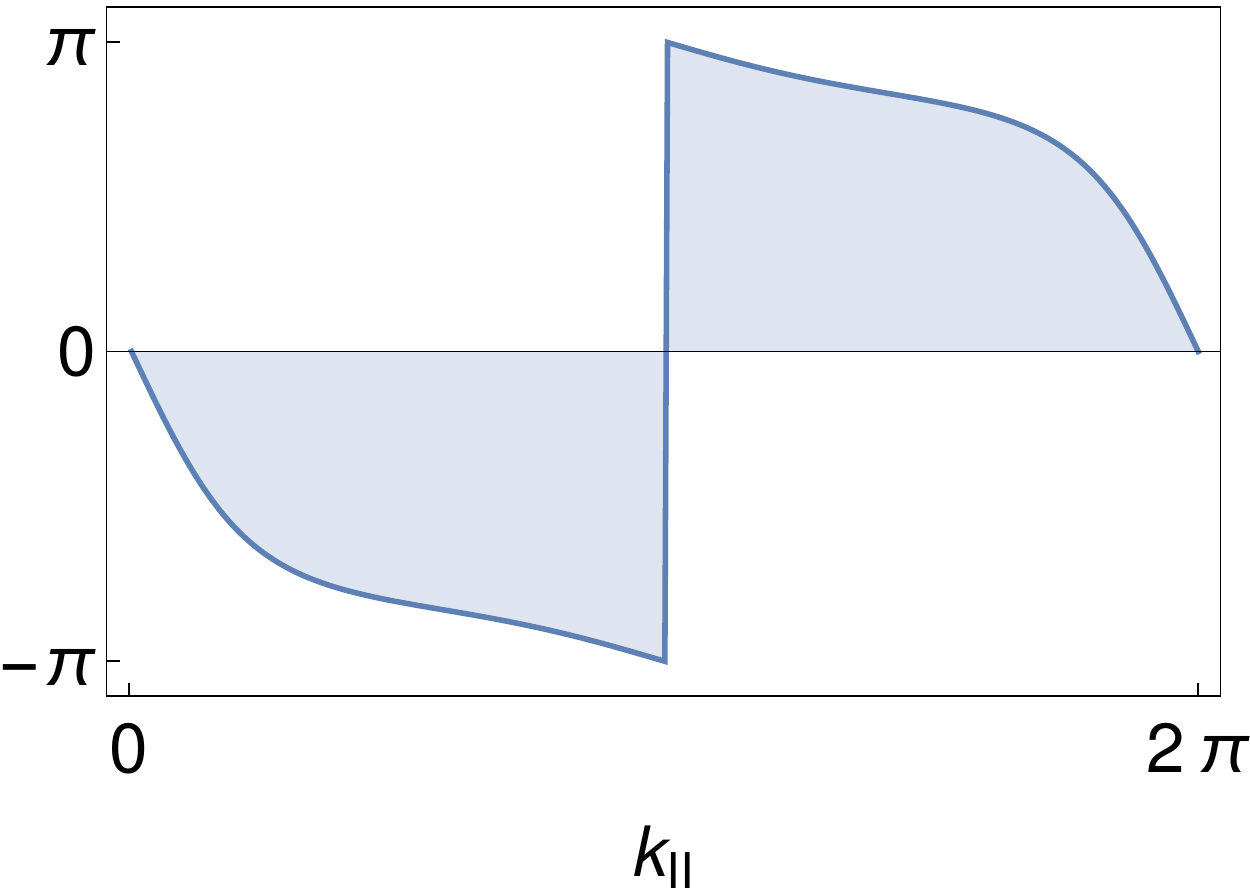} 
\includegraphics[height=2.96cm]{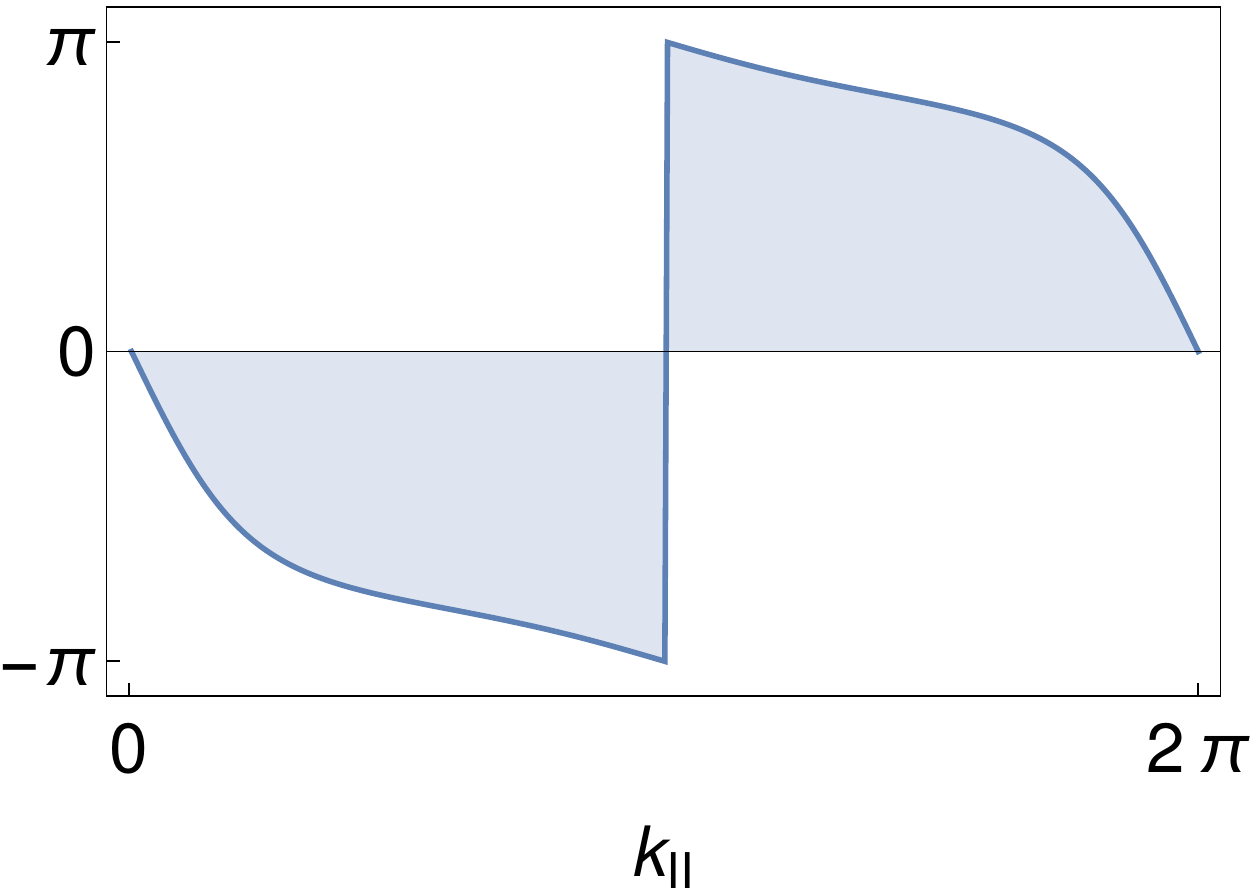} 
\includegraphics[height=2.96cm]{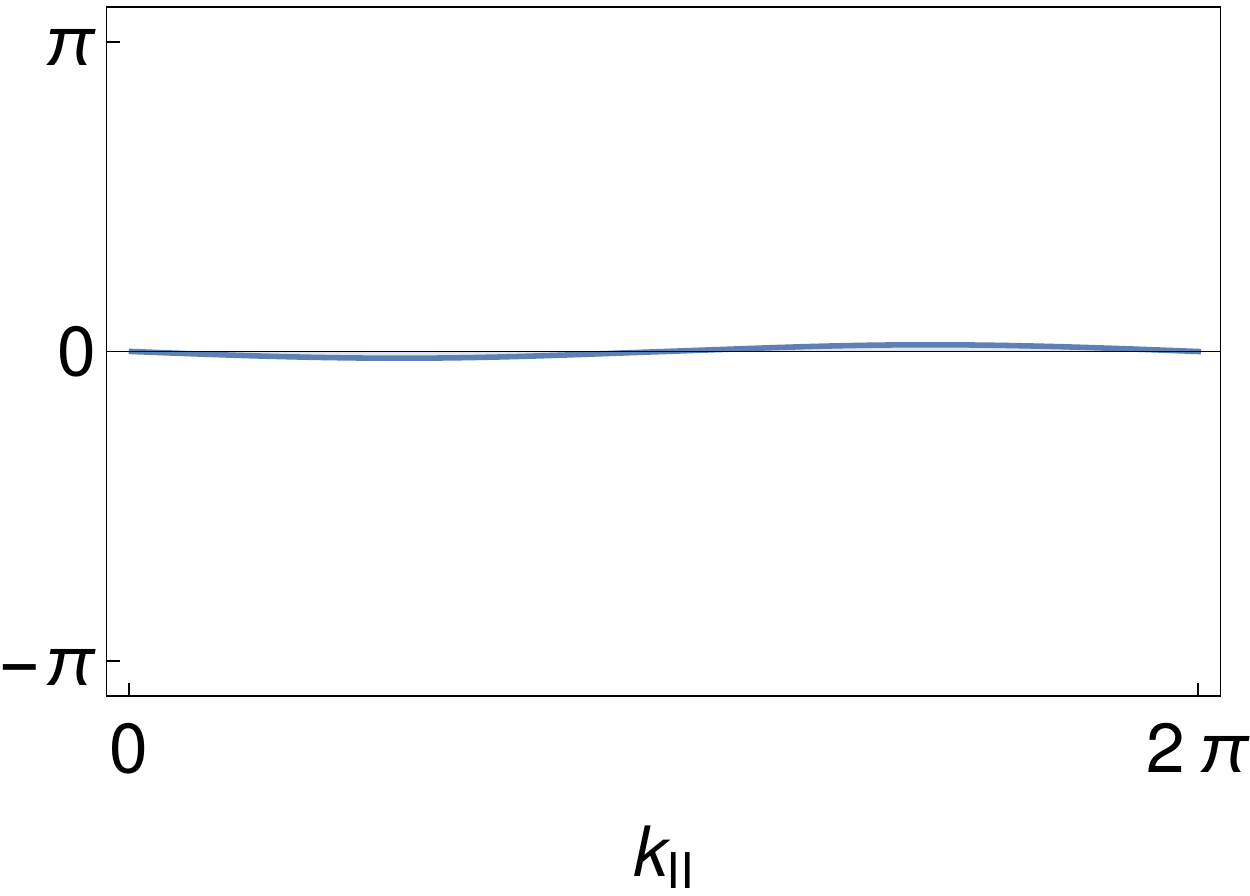}

\caption{(color online) Same as the Fig. \ref{Fig.graph.32.35} 
but for $\Delta\mu=9$. 
We have again $m=2$ for the two first lines and $m=5$ for 
the two last lines. Each 
column corresponds to the study of a pair of degenerate bands. 
In both cases, the 
system is a trivial insulator at 1/4 and 3/4 filling. At half filling, 
the system is in the 
trivial phase for  $m=2$ and in the topological phase for  $m=5$. }
\label{Fig.graph.92.95}
\end{figure*}

In this section we describe another method, proposed by Soluyanov and 
Vanderbildt,~\cite{Soluyanov_Vanderbilt_2012} for computing the topological 
invariant $\Delta$, 
which turns out to be efficient for numerical identification of the 
topological phases 
in an antiferromagnet.

For time reversal symmetric topological insulators, the latter approach 
involves enforcing a constraint, setting the relative phase $\chi$ to zero. 
Eq.(\ref{eq.defchi}) is thus replaced by:
\begin{eqnarray}\label{eq.defchitilde}
\ket{\tilde{\Psi}^{I}_{\alpha,\op{-k}}}&=&
-\Theta \ket{\tilde{\Psi}^{II}_{\alpha,\op{k}}} \nn \\
\ket{\tilde{\Psi}^{II}_{\alpha,\op{-k}}}&=&
\Theta \ket{\tilde{\Psi}^{I}_{\alpha,\op{k}}}
\end{eqnarray}
Then one has to verify if, in this gauge, it is still possible to define 
the eigenstates continuously over the entire Brillouin zone torus: 
an obstruction to do so implies a non-zero value of the topological 
invariant.

The equivalence between the two approaches, that of the 
Eq.(\ref{eq.defchi}) 
and Eq.(\ref{eq.defchitilde})) can be obtained via a singular gauge 
transformation 
similar to the one usually performed in the case of a point flux.

Once again, the Eq. (\ref{eq.defchitilde}) can be adapted to the AF 
case as per 
\begin{eqnarray}\label{eq.defchitildeAF}
\ket{\tilde{\Psi}^{I}_{\alpha,\op{-k}}}&=&
-e^{i\Phi_{\op{k}}/2}\ta \ket{\tilde{\Psi}^{II}_{\alpha,\op{k}}} \nn \\
\ket{\tilde{\Psi}^{II}_{\alpha,\op{-k}}}&=&
e^{-i\Phi_{-\op{k}}/2}\ta \ket{\tilde{\Psi}^{I}_{\alpha,\op{k}}}
\end{eqnarray}
We will use this result to compute numerically the topological invariant, 
adapting 
the method proposed in Ref.~[\onlinecite{Soluyanov_Vanderbilt_2012}] for 
time 
reversal-invariant topological insulators. The idea is the following: we 
have seen previously 
that a non-trivial value of the $Z_2$ topological invariant can be seen as 
an 
obstruction to define the eigenstates continuously over the Brillouin zone 
torus 
in a gauge that enforces the $\ta-$symmetry. The problem is that a 
numerical diagonalization of the Hamiltonian at each ${\bf k}$ point 
typically yields a highly 
discontinuous set of eigenstates.

Indeed, at each momentum value the numerical diagonalization
yields eigenstates that are defined up to an arbitrary phase $\phi$, 
that generally does not vary smoothly upon passing from one 
momentum value to the next, and manifests itself as a spurious 
phase factor $e^{i\phi}$ in the scalar product 
$\langle u(\op{k}) \ket{u(\op{k} + \Delta \op{k})}$ of the two eigenstates 
at the nearby momenta. The problem becomes even more delicate in the 
case of degenerate bands, where the phase factor $e^{i\phi}$ may also arise 
due to a rotation of the basis of degenerate states. Thus, we need to 
redefine 
our eigenstates in order to obtain a continuous gauge. 
In practice, for each pair of degenerate bands, we use parallel transport 
to obtain 
a smooth definition of the eigenstates over the cylindrical BZ with edges 
$k_y = \pi$ 
and $k_y = -\pi$, that respect Eq.(\ref{eq.defchitildeAF}). In this gauge, 
the possible 
discontinuity due to the topological nature of the system is removed to the 
edges of the cylinder (for more details, see A. Soluyanov and D. 
Vanderbilt.~\cite{Soluyanov_Vanderbilt_2012}). To probe this discontinuity,
we compute the "reconnection phase" for the pair of bands labelled 
$\alpha$:
\begin{equation}
e^{i\phi_\alpha(k_x)}=\langle\tilde{\Psi}^{I}_{\alpha,k_x,k_y=-\pi}  
\ket{\tilde{\Psi}^{I}_{\alpha,k_x,k_y=\pi}}
\end{equation}
The winding number $\Delta_\alpha$ of this phase yields the value of the 
topological invariant for the given pair of bands. $\Delta_\alpha=0$ means 
that it is possible 
to find a time reversal-symmetric gauge where the states are defined 
continuously 
over the entire BZ torus. By contrast, $\Delta_\alpha=1$ implies a 
topologically non-trivial phase.

For a given filling, the topological invariant is given by  
\begin{equation}\label{eq.def inv topo winding number}
\Delta=\sum_{\alpha\in FB} \Delta_\alpha ~mod~ 2 ,
\end{equation} 
where $FB$ stands for filled bands.

We applied this method to the BHZ Hamiltonian in the presence of staggered 
magnetization. The Figs. \ref{Fig.graph.32.35} and \ref{Fig.graph.92.95} 
present 
the results for four sets of parameters. For each set, we studied the four 
pairs 
of degenerate bands and plotted the four reconnection phases 
$e^{i\phi_\alpha(k_x)}$. From the figures, 
the values of the $\Delta_\alpha$ can be easily extracted, and conclusions on 
the trivial or topological nature of the system at 1/4, 1/2 and 3/4 filling 
can be drawn. 
All our results, be it presented here or not, are in perfect agreement with 
the phase 
diagram in the Fig. \ref{PhaseDiagram}.

\subsection{Wannier Charge Centers}

\begin{figure}
\centering
\includegraphics[height=2.8cm]{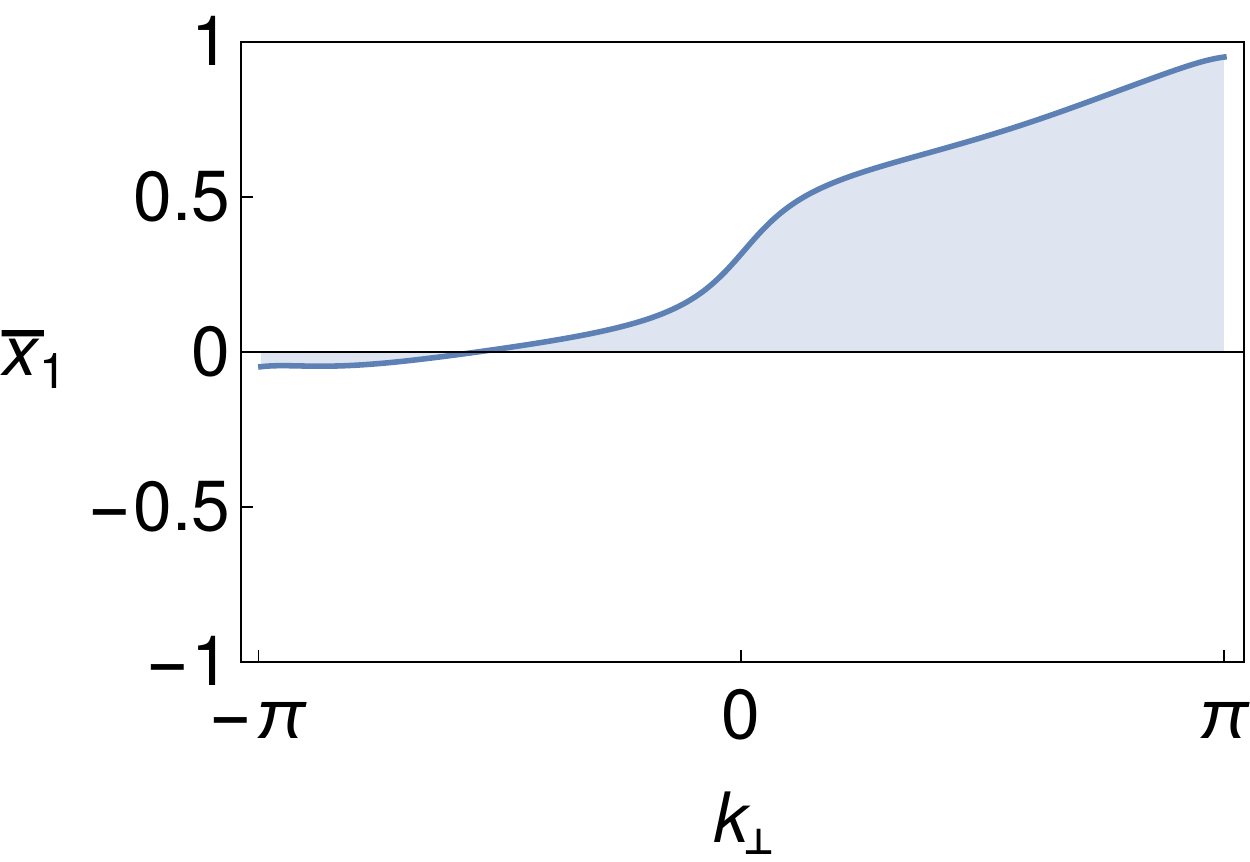} 
\includegraphics[height=2.8cm]{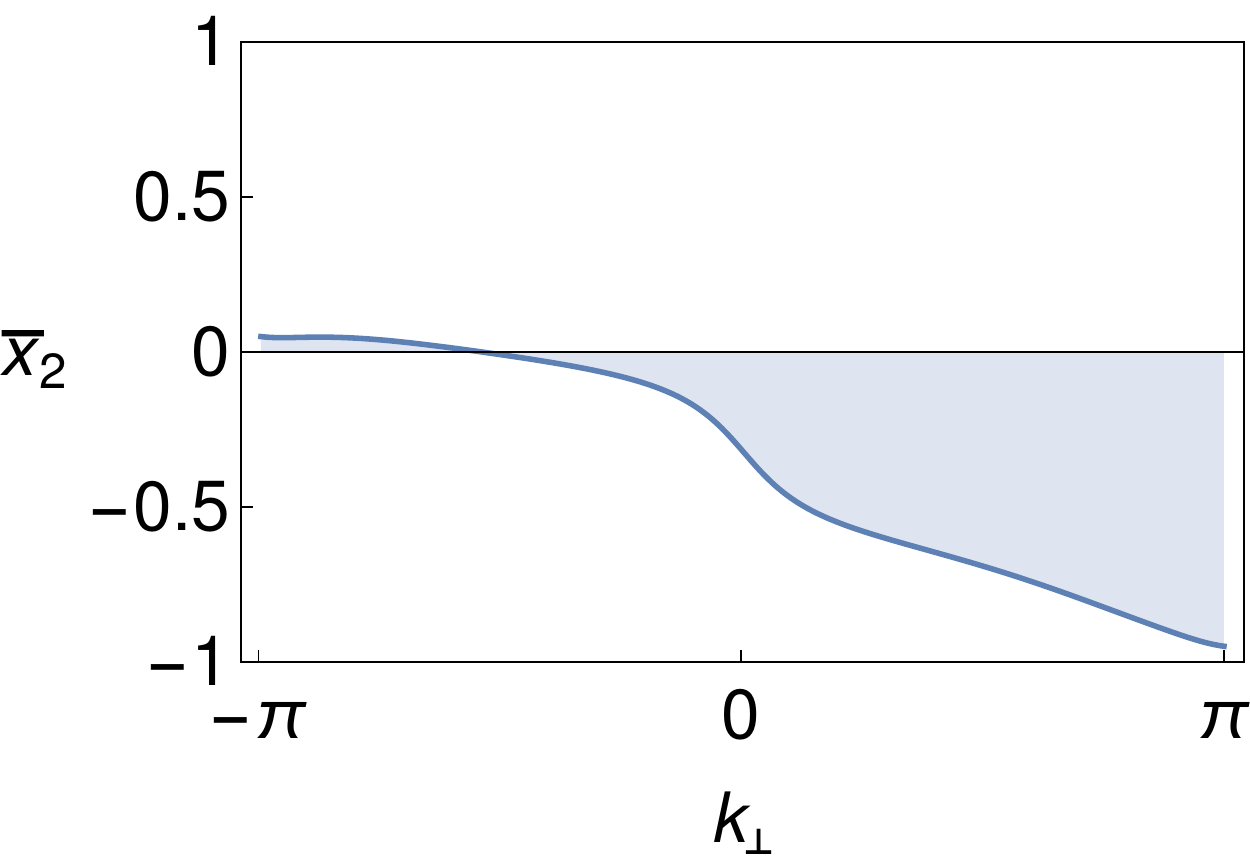} 

\caption{(color online) Position of the WCC for the first pair of Kramers 
degenerated band for a given set of the parameters, as the function of 
$k_y=k_\perp$. Depending on the band, we have $\Delta=\pm 1$, corresponding 
to a topological phase. }
\label{Fig.WCC.ex}
\end{figure}

Another way to probe the topological nature of the system is to study the flow 
of the Wannier Charge Centers (WCC).~\cite{Vanderbilt92,Yu11} Indeed, 
as explained in the Appendix 
\ref{app.Fu and Kane}, one may define the hybrid Wannier functions as per 
\begin{equation}
\ket{X,s,\alpha,k_y}=\frac{1}{2\pi}\int_{-\pi}^{\pi}dk_x
e^{-ik_x( X-\op{\hat{X}} )}\ket{u^{s}_{\alpha,k_x,k_y}}.
\end{equation}
The WCC are then defined as the expectation value 
$\bar{x}^s_{k_y,\alpha}=\bra{0,s,\alpha,k_y}\hat{x}\ket{0,s,\alpha,k_y}$. 
The combination of time reversal and inversion symmetries requires that 
$\bar{x}^I_{k_y,\alpha}+\bar{x}^{II}_{k_y,\alpha} \in N$, 
meaning that the charge center of a pair of Kramers partner states lies at the 
unit 
cell center. However, the center of a single band may flow as $k_y$ varies, 
and 
this flow may characterize the topology of the system. To show this, we need 
to 
modify the definition of the topological invariant from the Eq. 
(\ref{eq.def inv topo with P}) as per 
\begin{equation}\label{eq.new def inv}
\Delta = P^I_{k_y=0}-P^I_{k_y=2\pi}\ mod\ 2 .
\end{equation}
Above, the two definitions were equivalent thanks to the choice of gauge 
in the Eq.(\ref{eq.defchiAF}) and the special choice of basis axes in the 
Appendix \ref{app.Fu and Kane AF}, that yield 
\begin{equation}\label{eq.condition on P}
P^I_{k}=P^{II}_{-k}+N_I,\ N_I\in \mathbb{N}
\end{equation}
and
\begin{equation}
P^I_{k+G}=P^{I}_{k}+N_G,\ N_G\in \mathbb{N}
\end{equation}
Here, for convenience, we choose to treat $\op{x}$ and $\op{y}$ as basis 
vectors, 
Thus, the Eq.(\ref{eq.condition on P}) is not valid anymore. Under this 
condition, the 
Eq.(\ref{eq.new def inv}) is a better definition of the topological invariant 
than the Eq.(\ref{eq.def inv topo with P}), as it is directly related to the 
Chern 
number associated with the band corresponding to one of the two Kramers 
partners. 
The Eq.(\ref{eq.new def inv}) transforms to 
\begin{eqnarray}
\Delta &=&\sum_{\alpha\in FB}  
(\bar{x}^I_{k_y=0,\alpha}-\bar{x}^{I}_{k_y=2\pi,\alpha})\ mod\ 2 \nn\\
&=& \sum_{\alpha\in FB} (\bar{x}^I_{k_y=-\pi,\alpha}-
\bar{x}^{I}_{k_y=\pi,\alpha})\ mod\ 2 .
\end{eqnarray}

\begin{figure*}
\centering
\includegraphics[height=3cm]{WC1v32.pdf} 
\includegraphics[height=3cm]{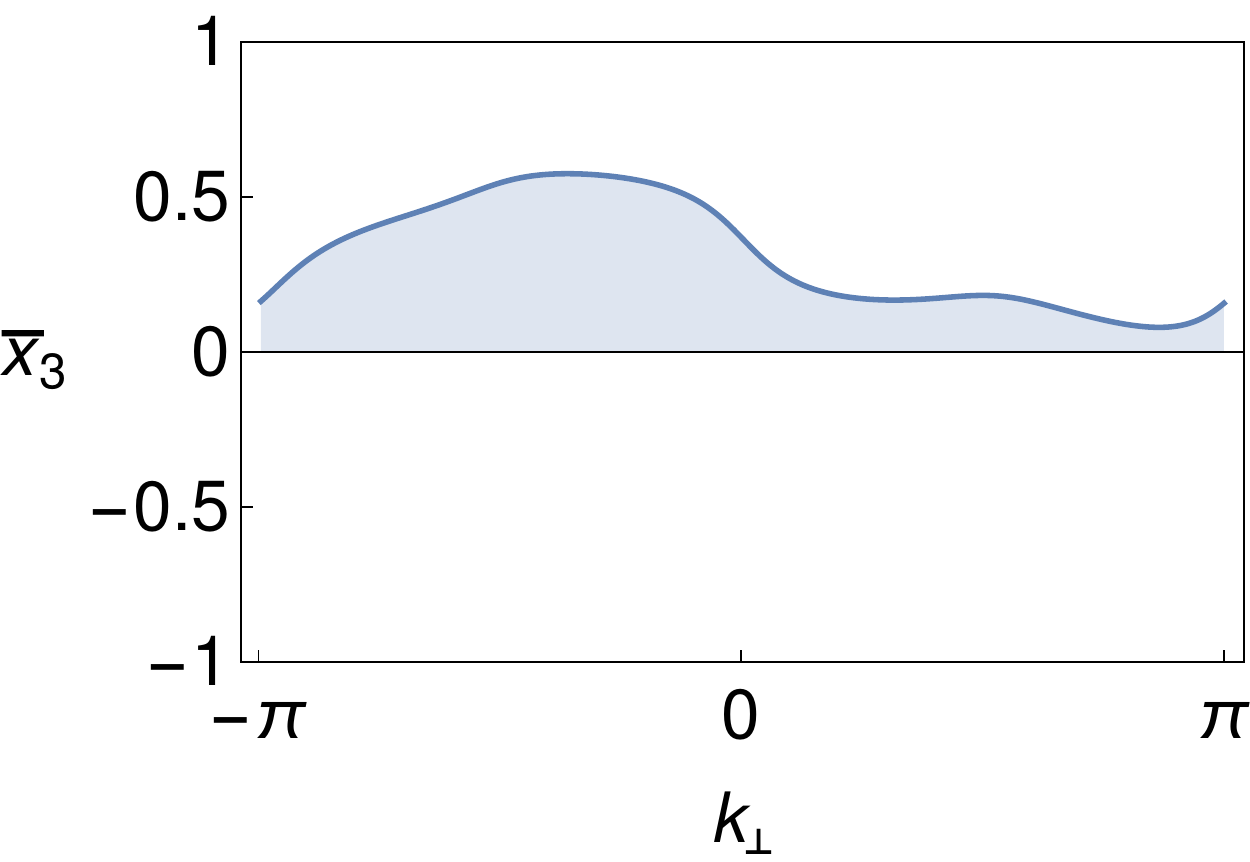} 
\includegraphics[height=3cm]{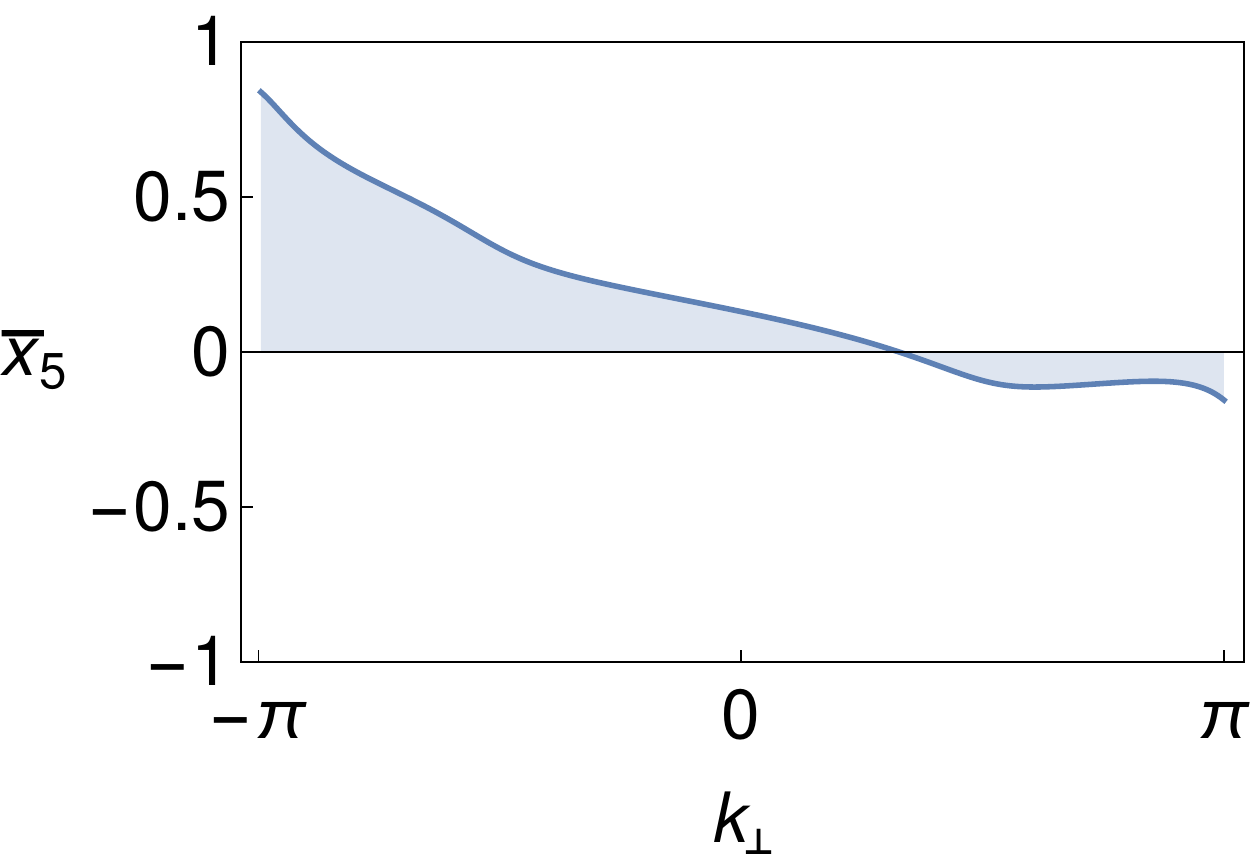} 
\includegraphics[height=3cm]{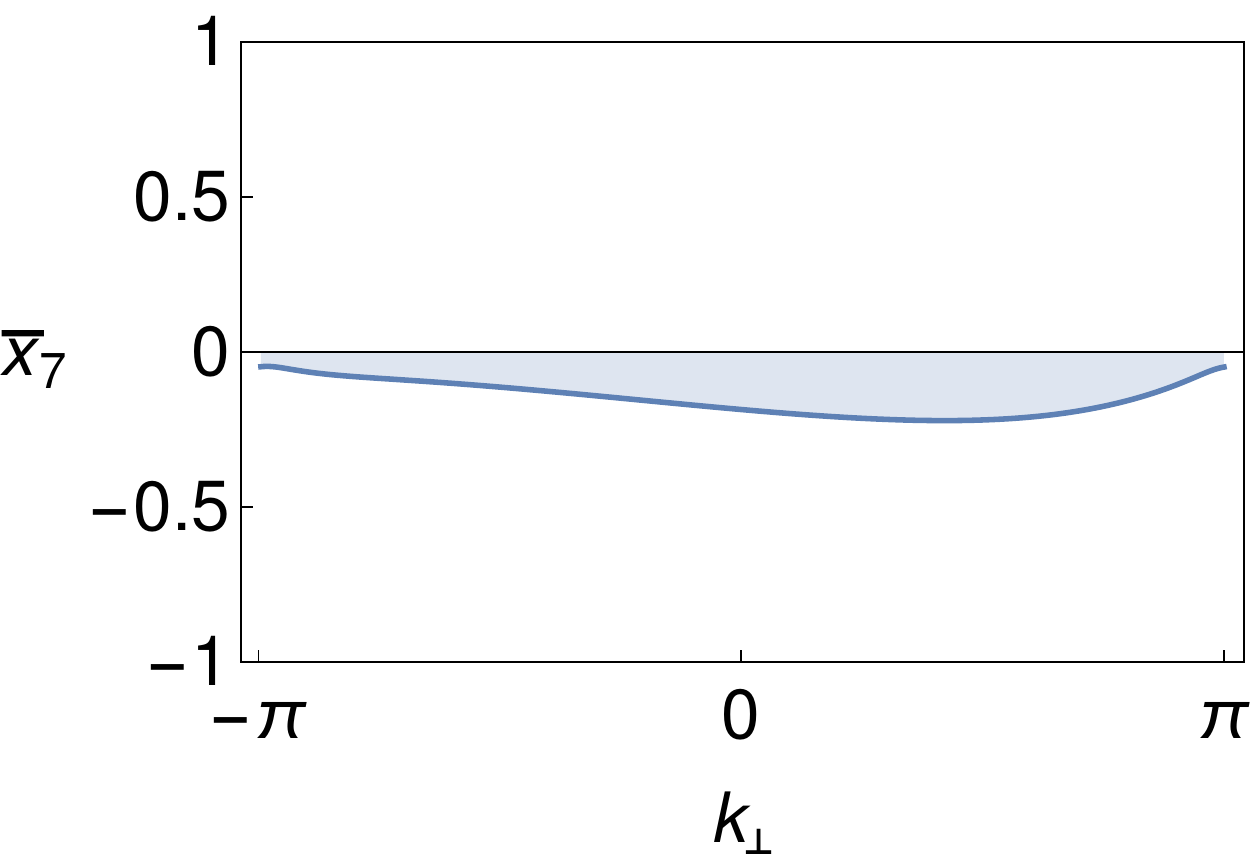}

\includegraphics[height=3cm]{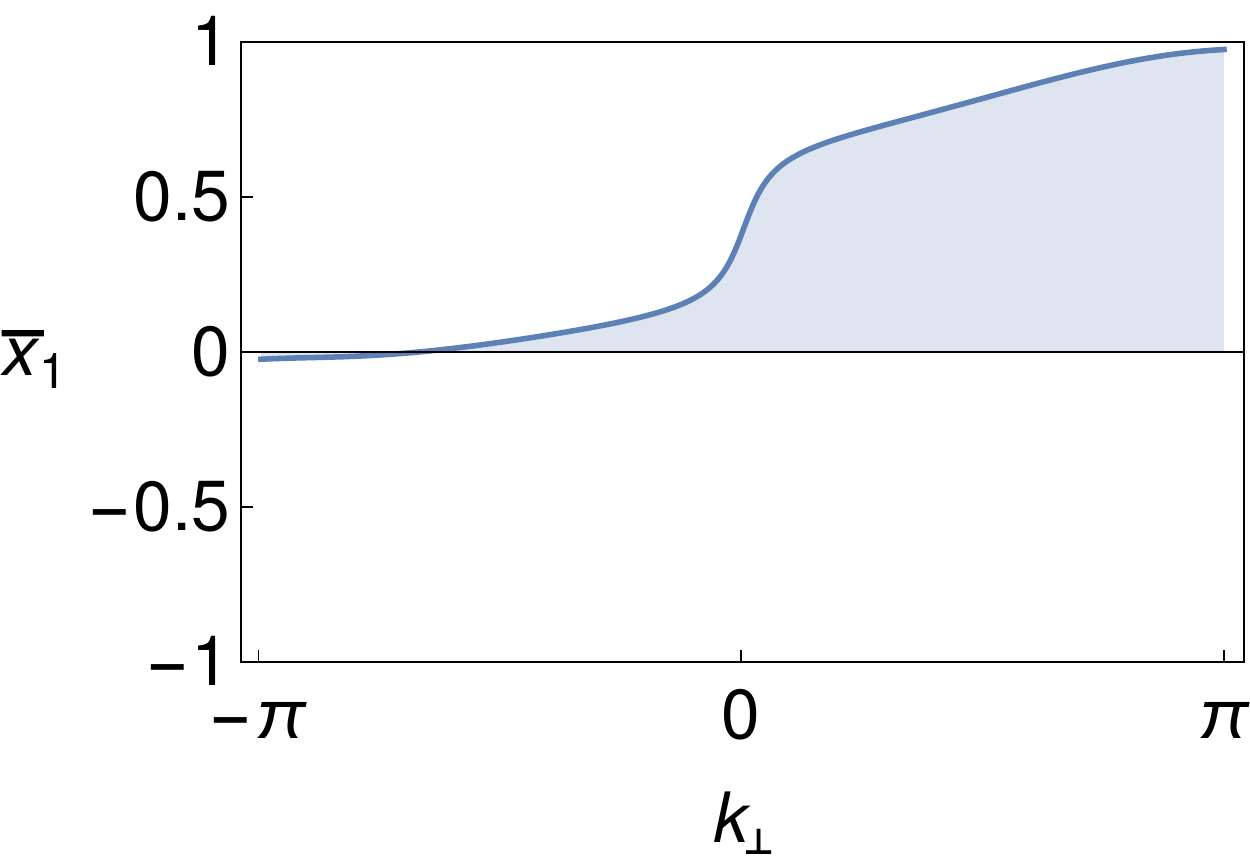} 
\includegraphics[height=3cm]{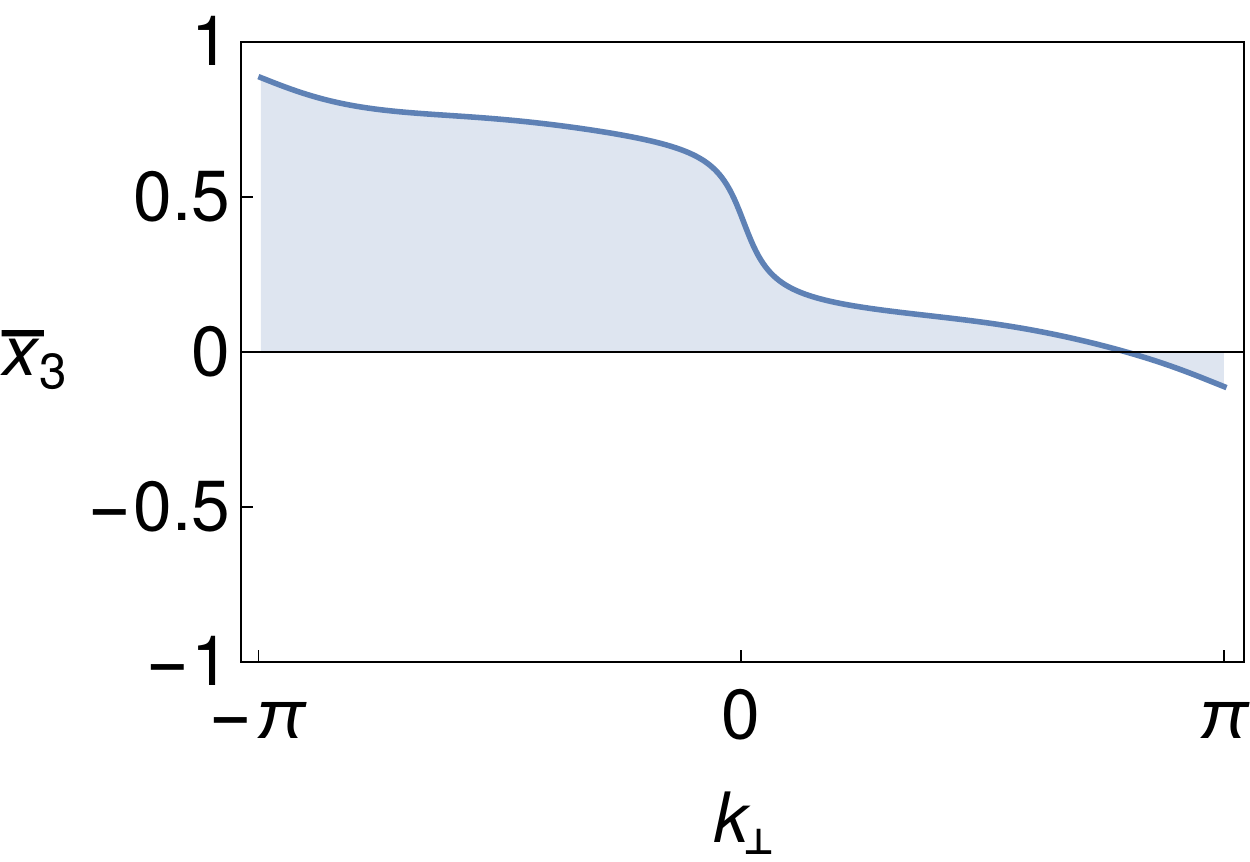} 
\includegraphics[height=3cm]{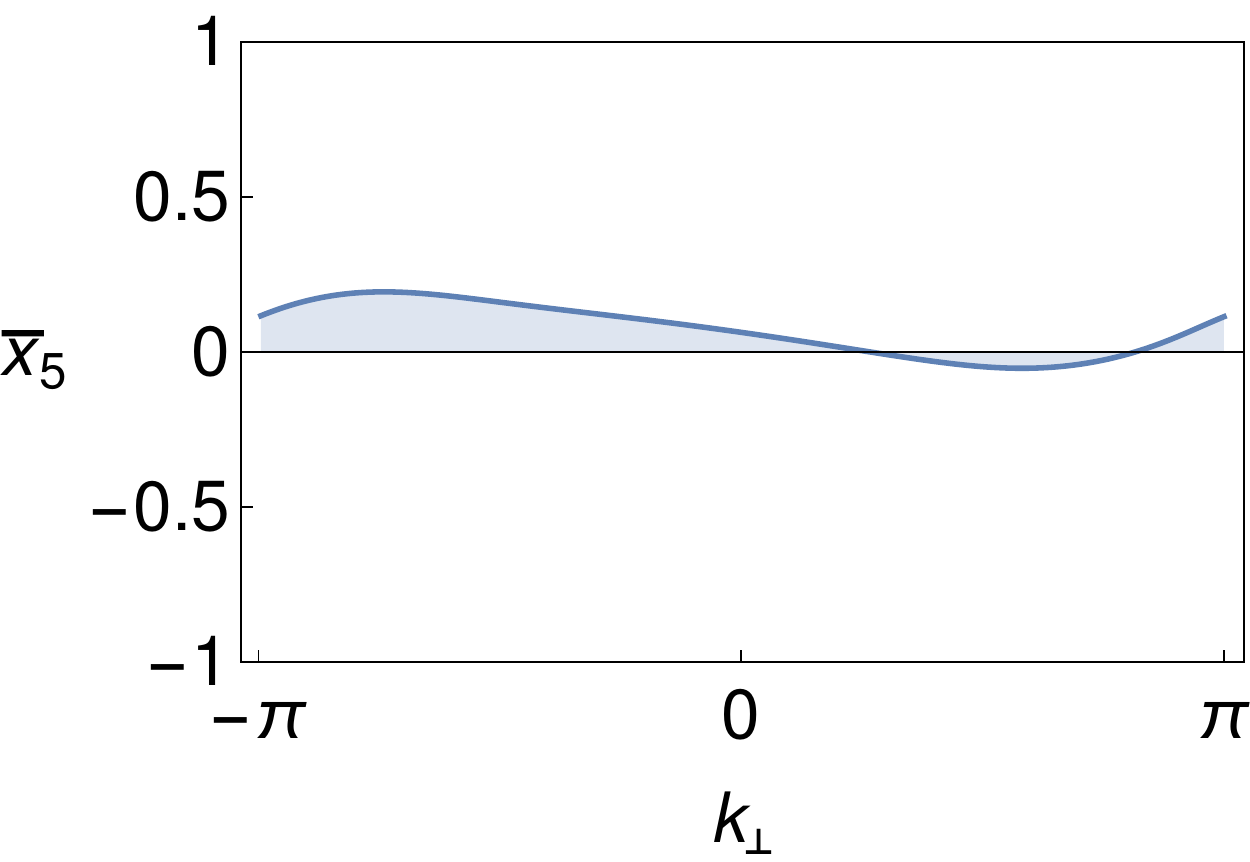} 
\includegraphics[height=3cm]{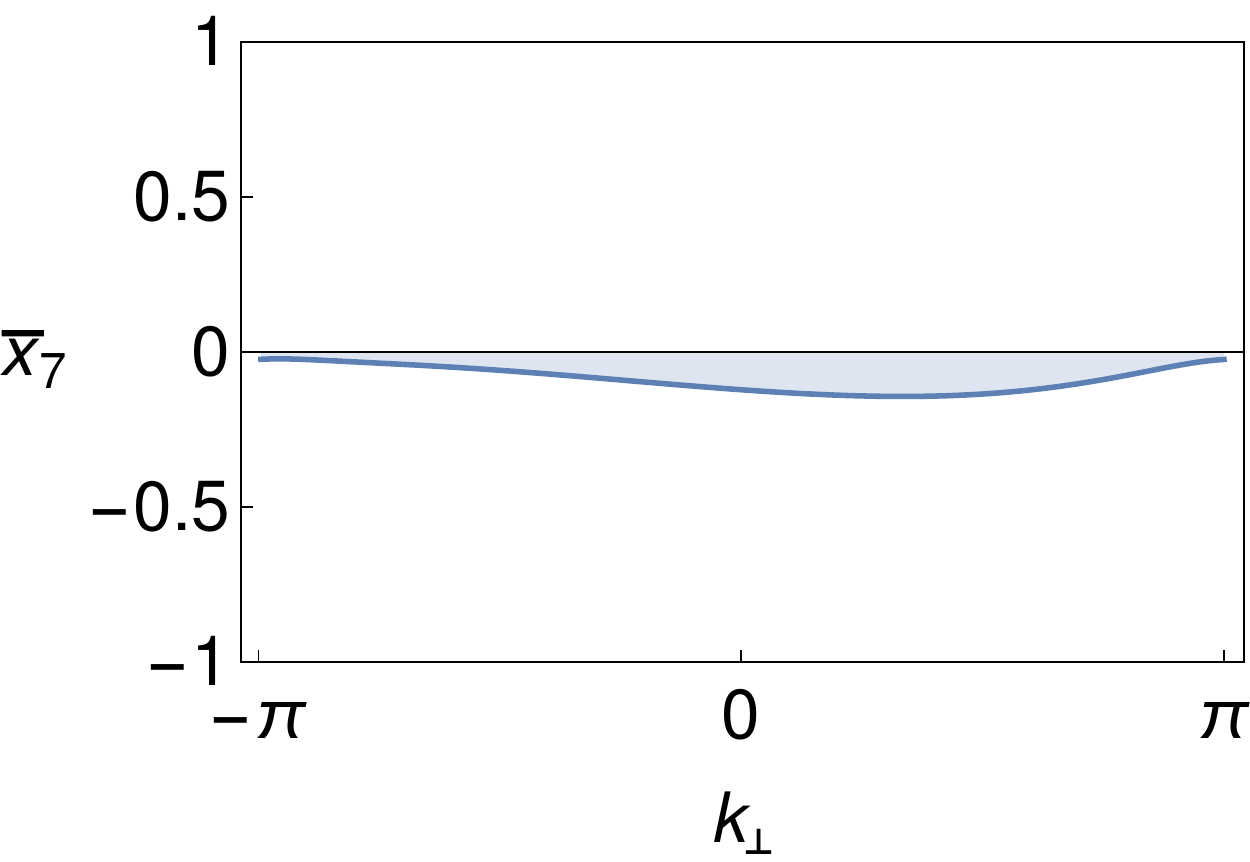}

\includegraphics[height=3cm]{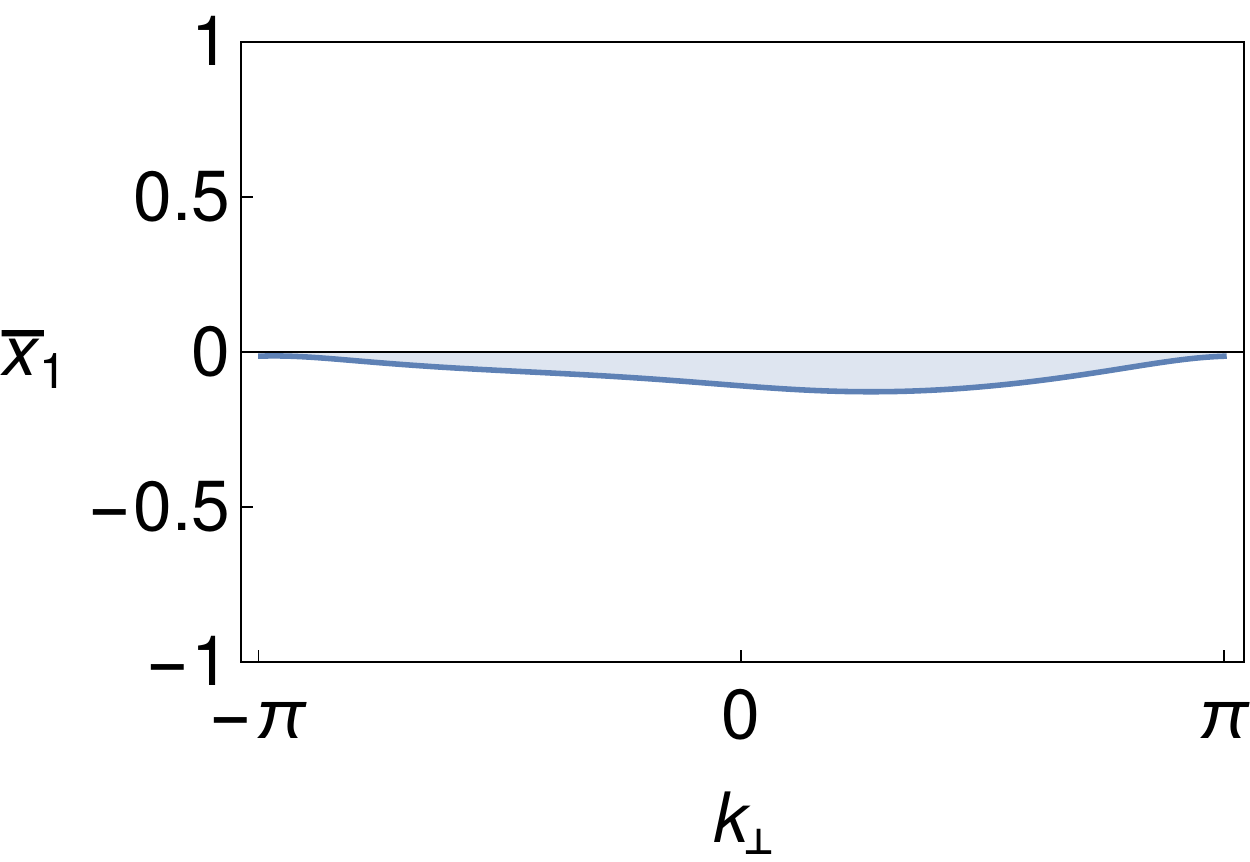} 
\includegraphics[height=3cm]{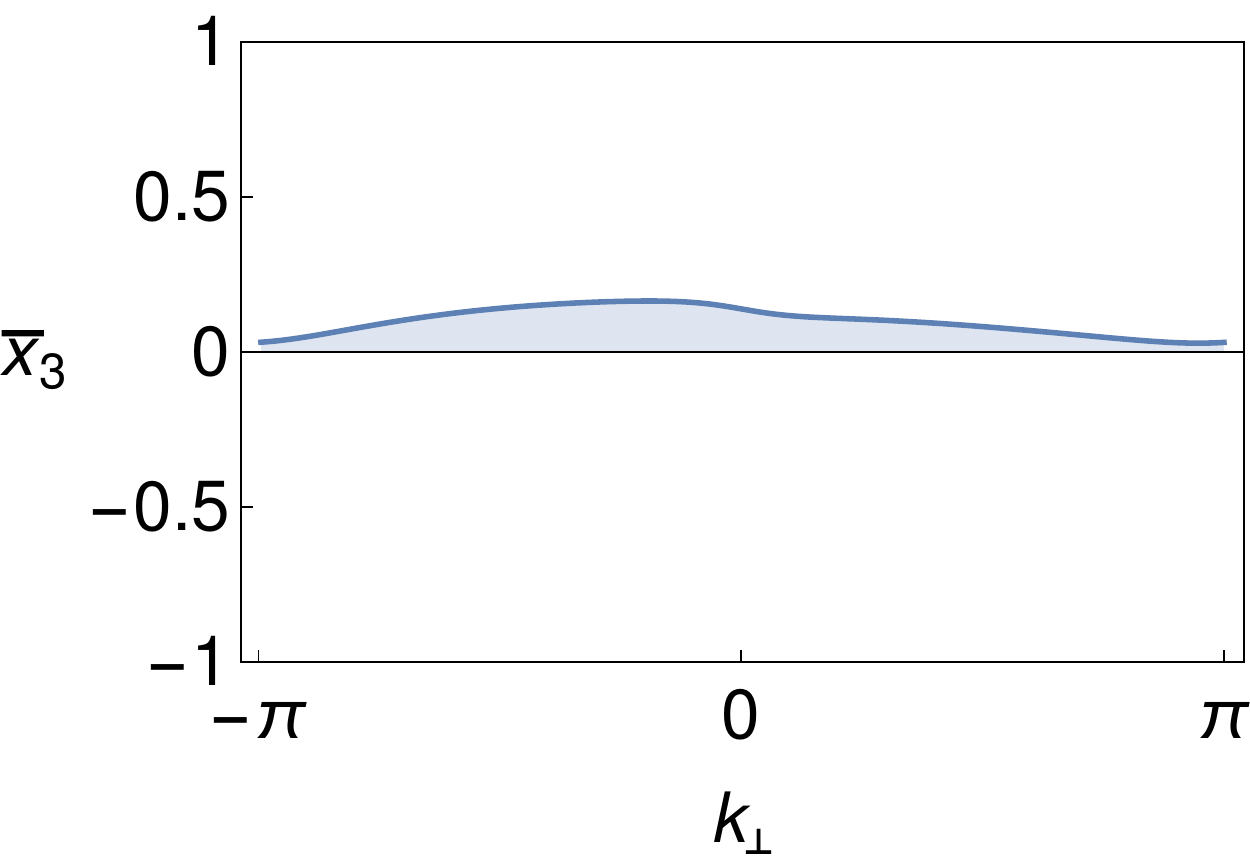} 
\includegraphics[height=3cm]{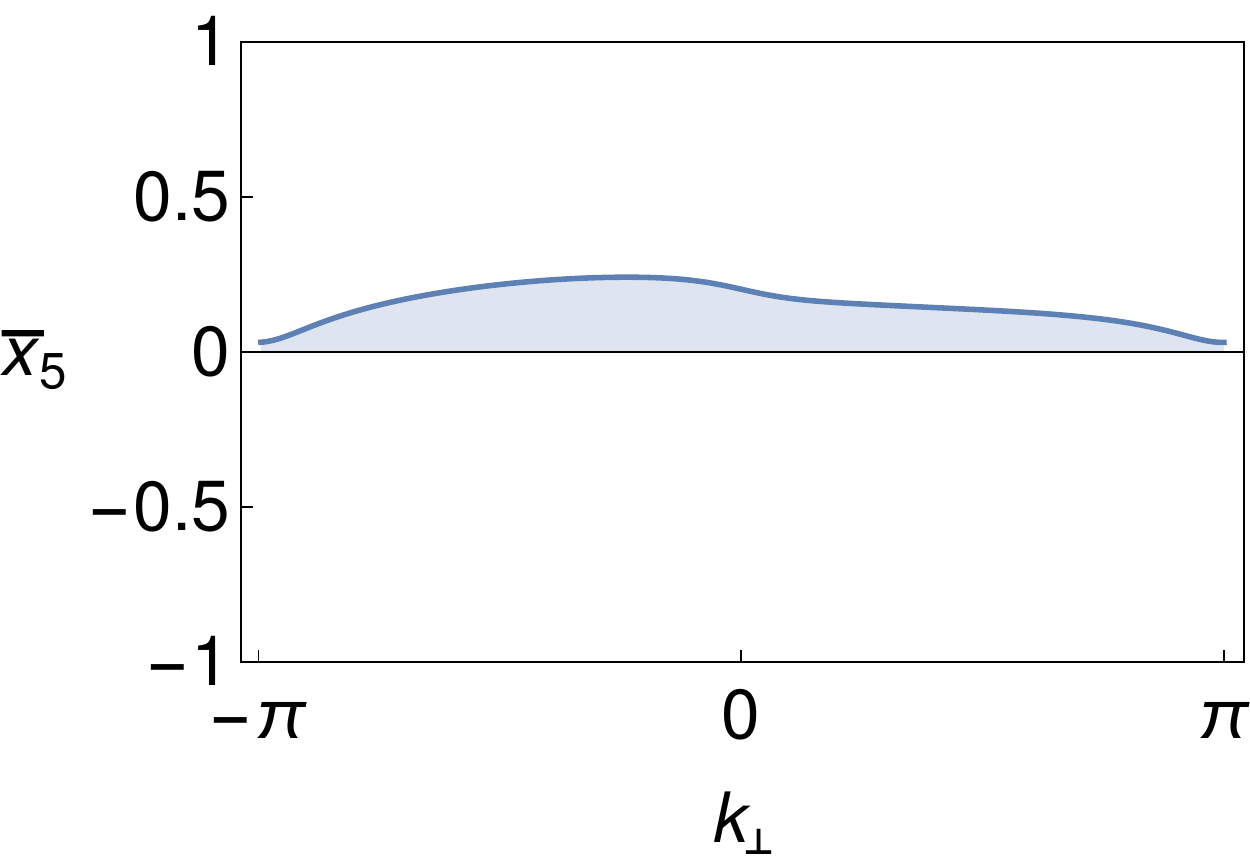} 
\includegraphics[height=3cm]{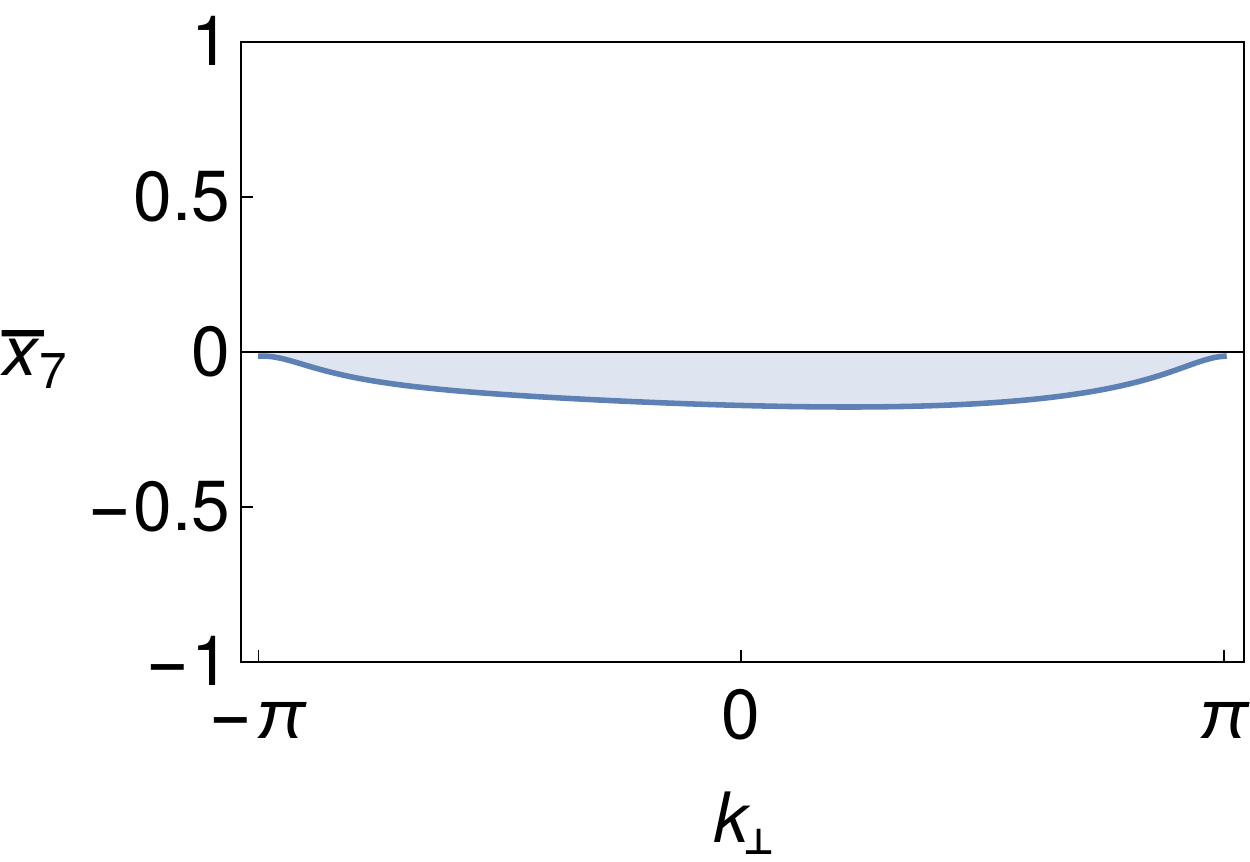}

\includegraphics[height=3cm]{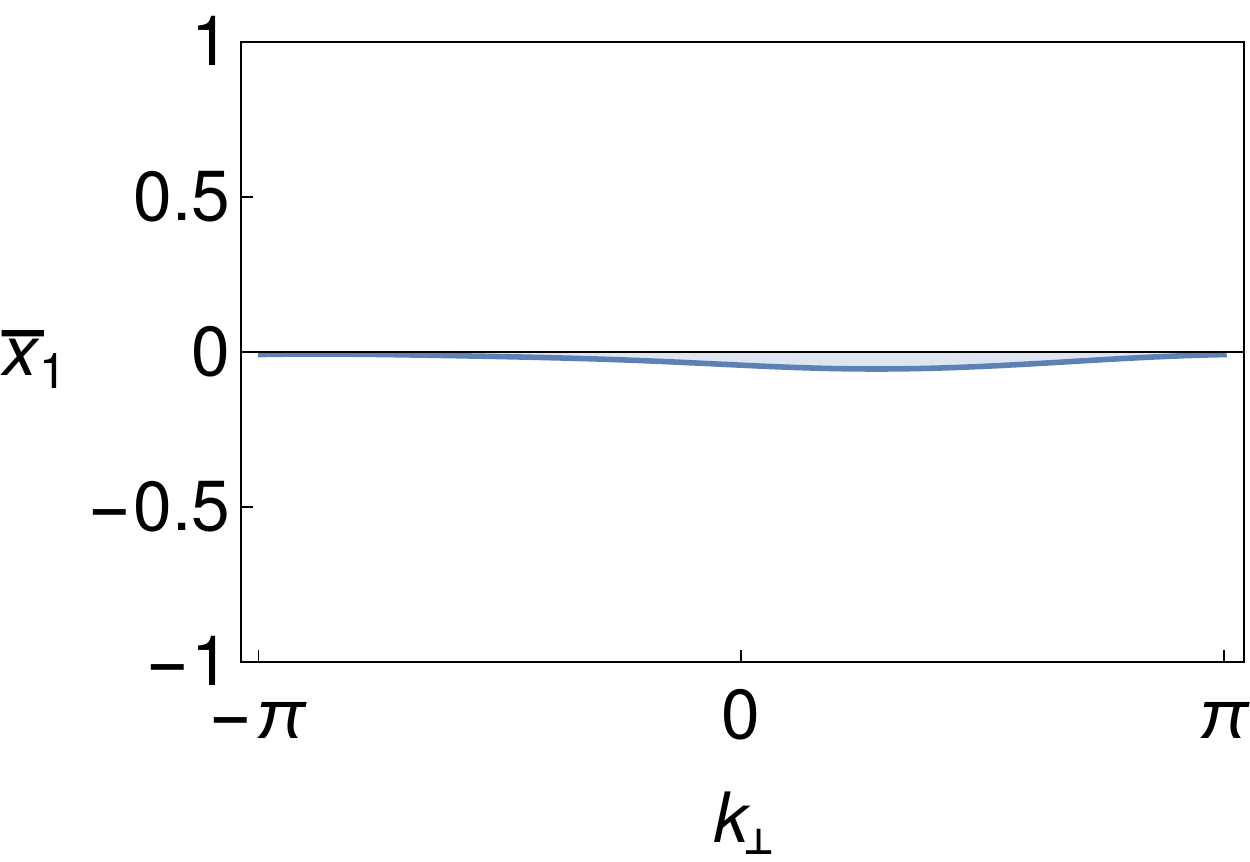} 
\includegraphics[height=3cm]{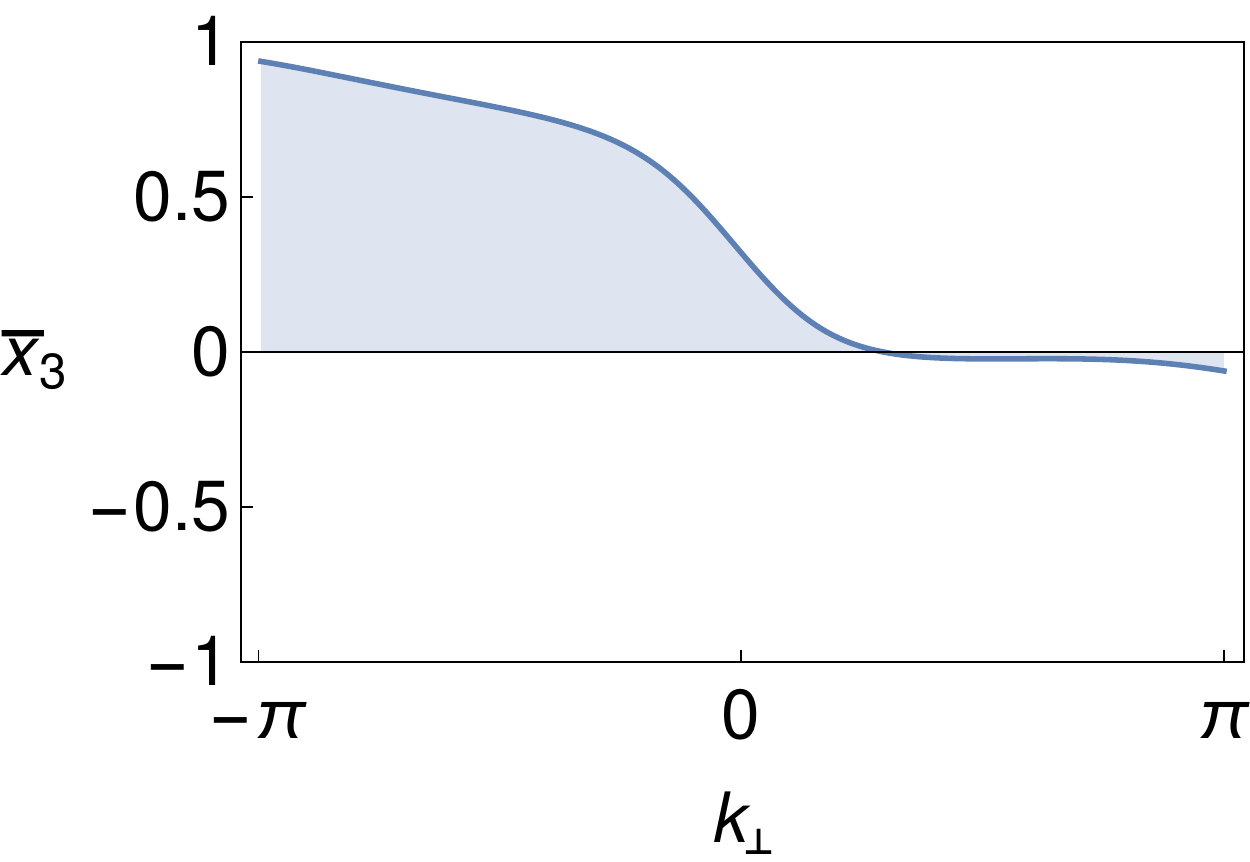} 
\includegraphics[height=3cm]{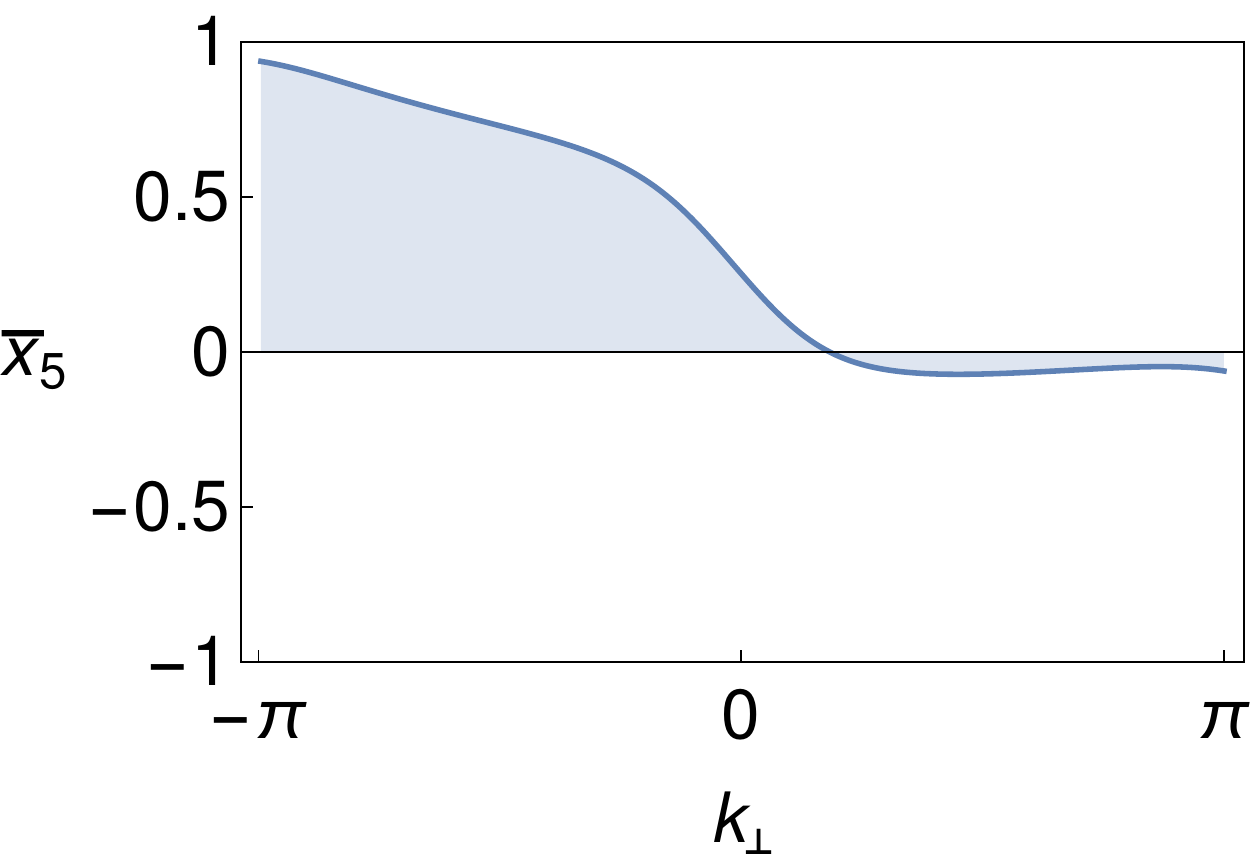} 
\includegraphics[height=3cm]{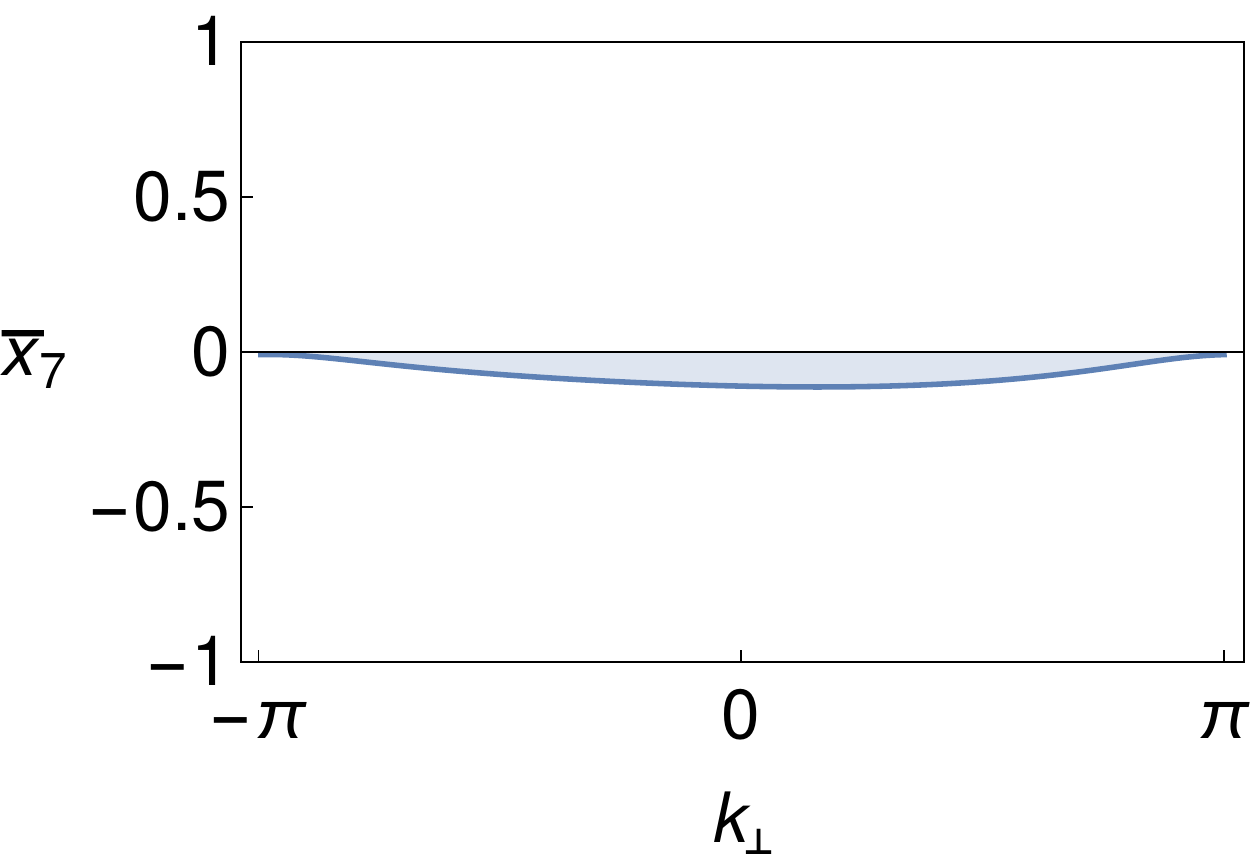}

\caption{(color online) Position of the WCC as a function of $k_\perp$. The 
different columns correspond to a representative of the first, second, third 
and forth pair of bands in this order. The different lines correspond to 
$\Delta\mu=3$ (line 1 and 2) and $\Delta\mu=9$ (line 3 and 4), and $m=2$ (line 
1 and 3) and $m=5$ (line 2 and 4). }
\label{Fig.WCC.all}
\end{figure*}

Using the parallel transport method of the preceding subsection, we define the 
$\ket{u^{s}_{\alpha,k_x,k_y}}$ that is smooth on a mesh in the BZ. Since  
$\bar{x}^s_{k_y,\alpha}$is related to the Berry connection, we find 
\begin{eqnarray}
\bar{x}^s_{k_y,\alpha}&= & i\frac{1}{2\pi}\int_{-\pi}^{\pi}dk_x 
\bra{u^{s}_{\alpha,k_x,k_y}}\nabla_{k_x}\ket{u^{s}_{\alpha,k_x,k_y}}\nn\\
 &=& \frac{1}{2\pi} \sum_{j=1}^{\mathcal{N}} Im \bra{u^{s}_{\alpha,j+1,k_y}} 
 u^{s}_{\alpha,j,k_y}\rangle .
\end{eqnarray}
We observe that the topological invariant differs in sign when computed with 
one 
or the other Kramers partner (see the Fig. \ref{Fig.WCC.ex}). However, as 
$\Delta$ 
is defined modulo 2, this sign has no physical meaning, and either of the two 
Kramers partners may be chosen. 

The position of the WCC for one of the two Kramers partners of each pair of 
bands is plotted as a function of $k_y=k_\perp$ in the Fig. \ref{Fig.WCC.all} 
for the same set of parameters as in the previous section. The results are 
once again in perfect agreement with the expected phase diagram of Fig. 
\ref{PhaseDiagram}.

\subsection{Explicit Construction of the Edge States}
\label{sec.EtatDeBord}

A defining property of $Z_2$ topological insulators is the existence of
topologically-protected edge states. In time reversal invariant systems, a 
non-zero
value of the topological invariant defined above corresponds to systems with
an odd number of Kramers-degenerate pairs of edge states. Here, by explicitly 
constructing the edge states (adapting a method used for the BHZ Hamiltonian 
in [\onlinecite{konig2008quantum}]) we show that this bulk-edge correspondence 
holds 
in the AF case. It corroborates the importance of the time reversal 
polarization 
of the Eq.(\ref{eq.def inv topo with xi AF}) also in the present case.

\begin{figure}

\subfigure[]{\includegraphics[width=4cm]{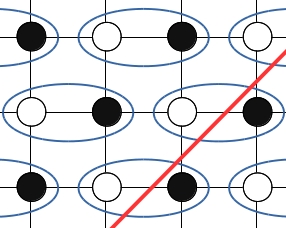}}
\label{LatticeWrong}
~~
\subfigure[]{\includegraphics[width=4cm]{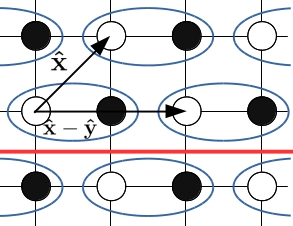}}

\label{LatticeGood}

\caption{(color online) Examples of cuts (red lines) on the dimerised square 
lattice. (a) An example of an edge that does not respect the bulk symmetries 
(the $\ta$ symmetry is broken). (b) The edge considered in this article, and 
that respects the bulk symmetries. In this case,  $k_{x-y}$ (noted 
$k_\parallel$) remains a good quantum number. }
\label{Fig.Edges}
\end{figure}

We consider a system defined on a semi-infinite plane and, hence, with a 
single edge.
The direction of the cut will not be chosen arbitrarily, but in a way that 
respects the symmetries of the bulk, as shown in the Fig. \ref{Fig.Edges}. To 
be more explicit, 
a ``good edge" would involve alternating A and B sites and is symmetric under
$\ta=T\Theta$. By contrast, we will not consider here edges with only the A 
sites,
as such an edge would manifest ferromagnetic order and would explicitly break
the bulk symmetry  $\ta$.
Therefore, we choose the primitive lattice vectors to be  
$\op{x}$ and $\op{x}-\op{y}$, and we choose a cut along $\op{x}-\op{y}$, at 
$\op{x}=0$. 
This breaks the translation invariance along $\op{x}$ while preserving it 
along 
$\op{x}-\op{y}$. Thus $k_{x-y}$ (denoted as $k_\parallel$) remains a good 
quantum 
number, but not the $k_x$.

Hence, we write the Hamiltonian in terms of 
$k_\parallel$ and $-i\partial_x$ and look for the spatial extent of wave 
functions 
in the $x$ direction. In our special case, the Hamiltonian does not mix spin-
up 
with spin-down, allowing us to consider only the 
$H^\uparrow_{k_\parallel}(-i\partial_x)$ acting on the reduced Hilbert space 
made of spin-up states (the spin-down edge states can then be constructed 
using the $\ta$ operation). We now seek the solutions to 
\begin{equation}\label{eq.Equation_sur_psi}
H^\uparrow_{k_\parallel}(-i\partial_x)\Psi_{k_\parallel}
(x)=E\Psi_{k_\parallel}(x)
\end{equation}
with  $E$ in the gap of the bulk spectrum.
As we are looking for states exponentially localized at the edge, 
we use the ansatz $\Psi_{k_\parallel}(x)=e^{\lambda x}\Phi_{k_\parallel}$, 
with $\Phi$ independent of $x$, in Eq.(\ref{eq.Equation_sur_psi}).
Cancelling the exponentials, we obtain the eigenvalue equation:
\begin{equation} \label{eq.Equation avec lambda}
H^\uparrow_{k_\parallel}(-i\lambda)\Phi_{k_\parallel}=E\Phi_{k_\parallel}
\end{equation}
that we will have to solve in $\lambda$ and $\Phi_{k_\parallel}$ 
simultaneously.
Moreover, we will only keep solutions with $Re(\lambda)<0$, as we
wish to obtain normalizable surface states.
For a given value of $E$ and $k_\parallel$, we get several pairs of 
normalizable solutions $(\lambda^i, \Phi^i_{k_\parallel})$ (generally, 
$i=1..4$). We could then build an edge state of the form:
\begin{equation}
\Psi_{edge}(x)=\sum_i C_i e^{\lambda^i x}\Phi^i_{k_\parallel}.
\end{equation}
Given that the wave function must vanish at large negative $x$, we impose
on the $\Psi_{edge}$ the boundary condition $\Psi_{edge}(0)=0$. This can 
only be done if the $\Phi^i_{k_\parallel}$ are linearly dependent, implying 
an existence condition for the edge state.

\begin{figure}[b]

\subfigure[]{\includegraphics[width=0.23\textwidth]{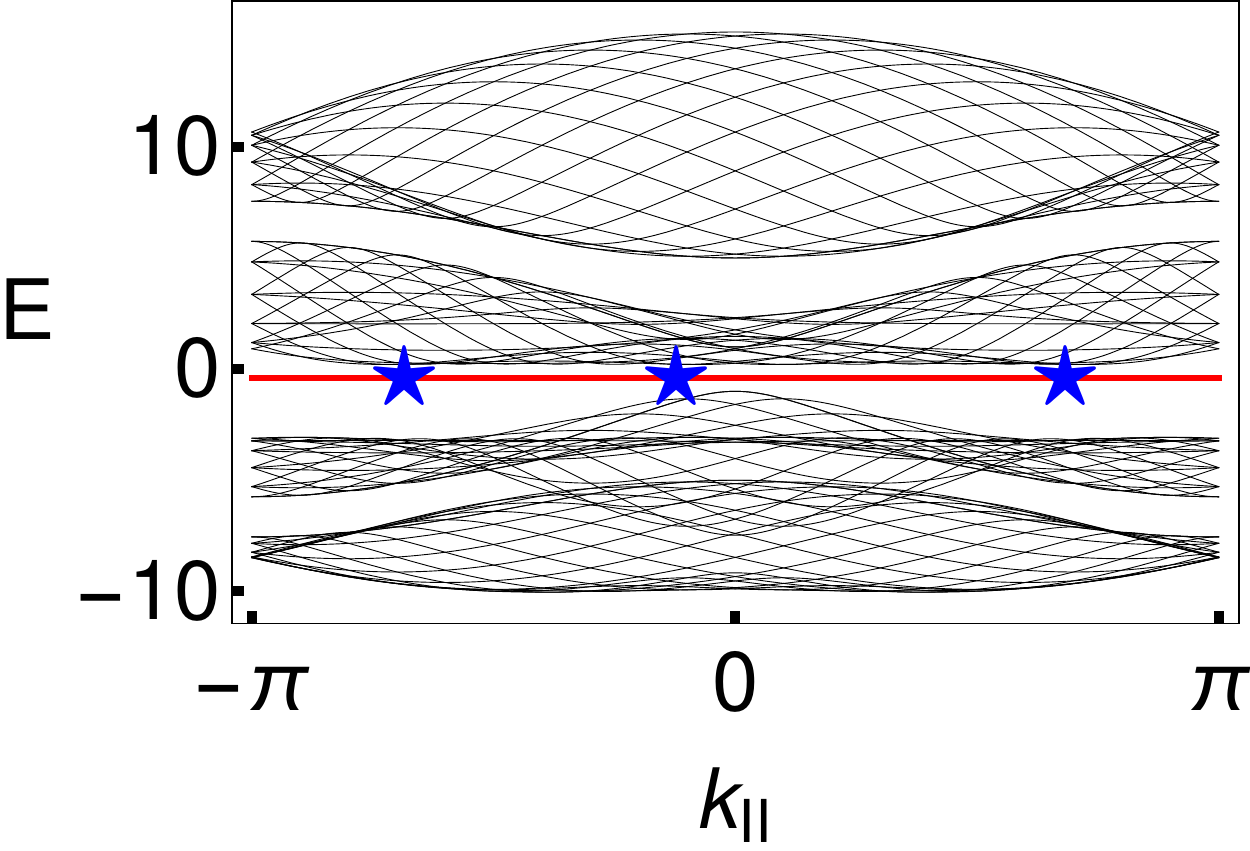}}
\subfigure[]{\includegraphics[width=0.23\textwidth]{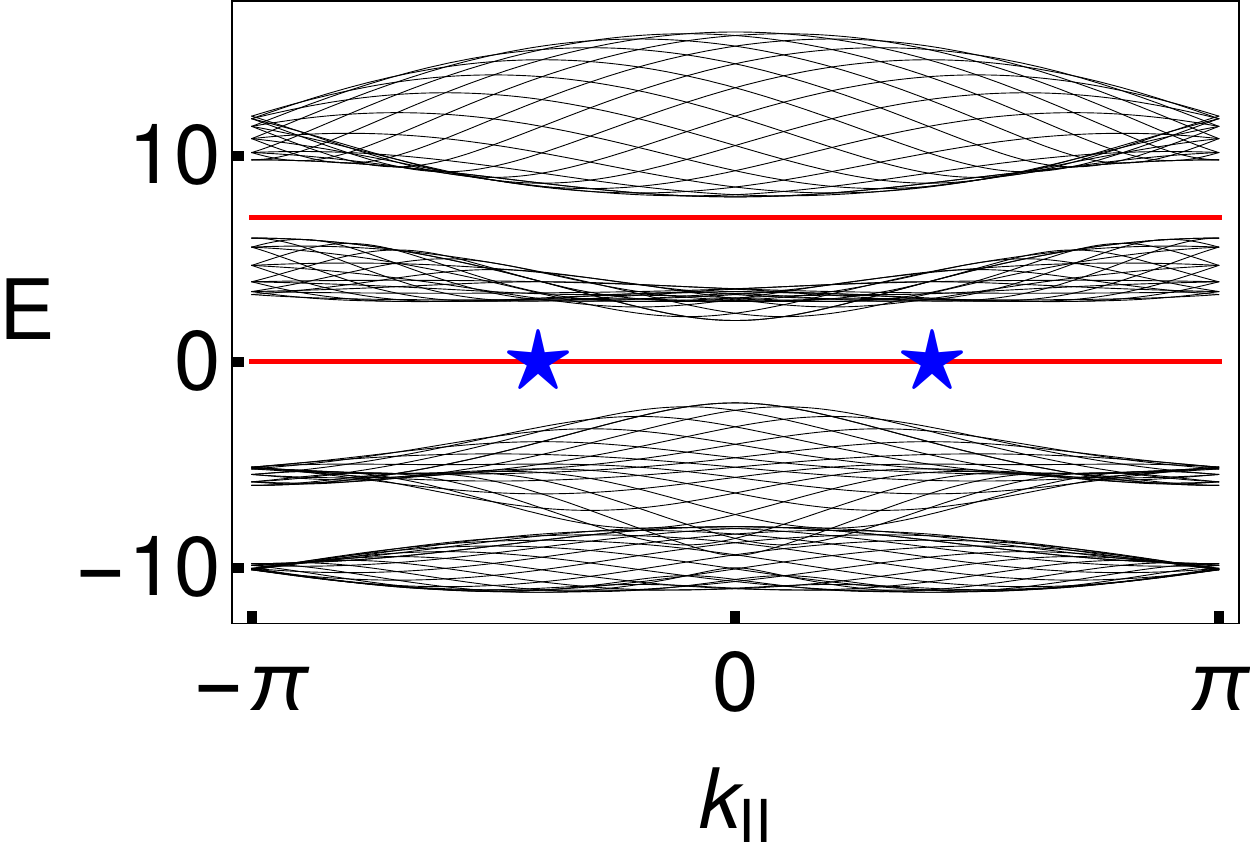}}

\subfigure[]{\includegraphics[width=0.23\textwidth]{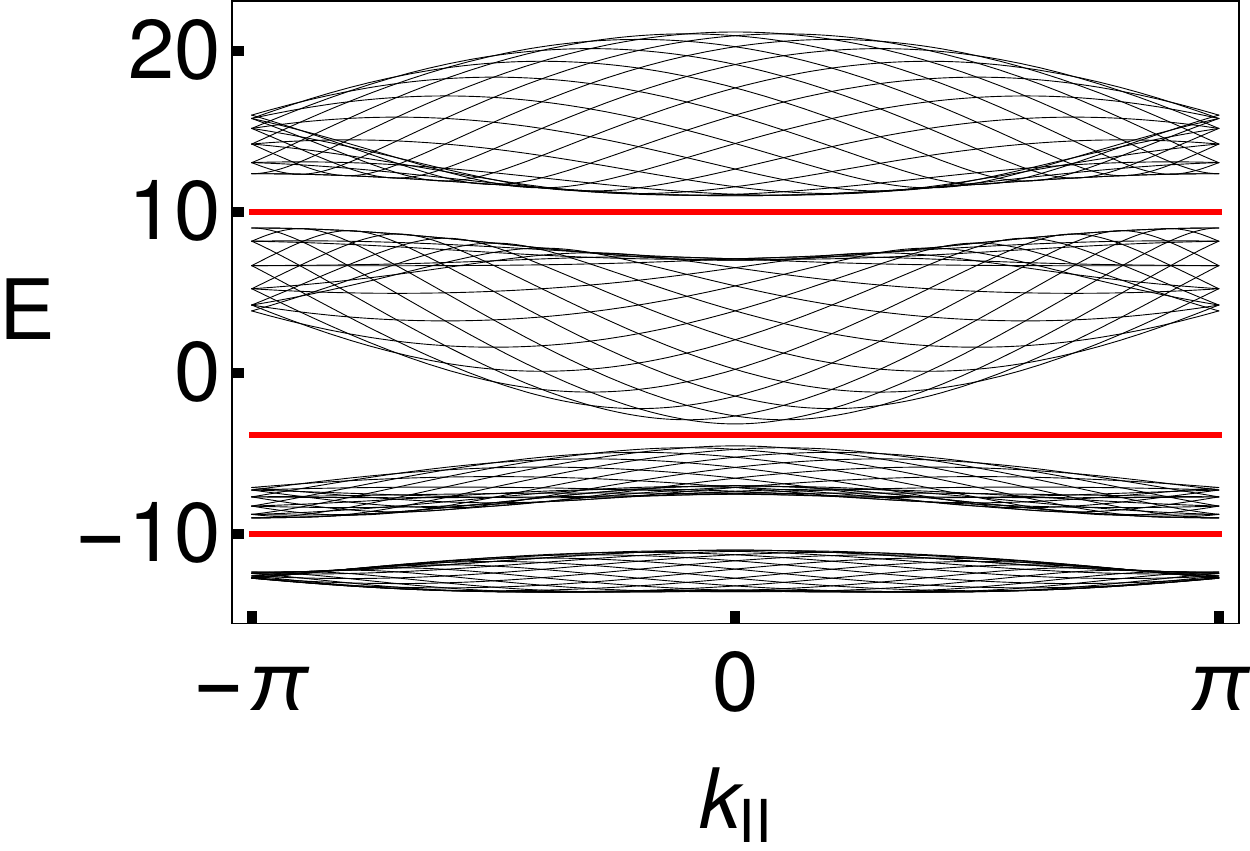}}
\subfigure[]{\includegraphics[width=0.23\textwidth]{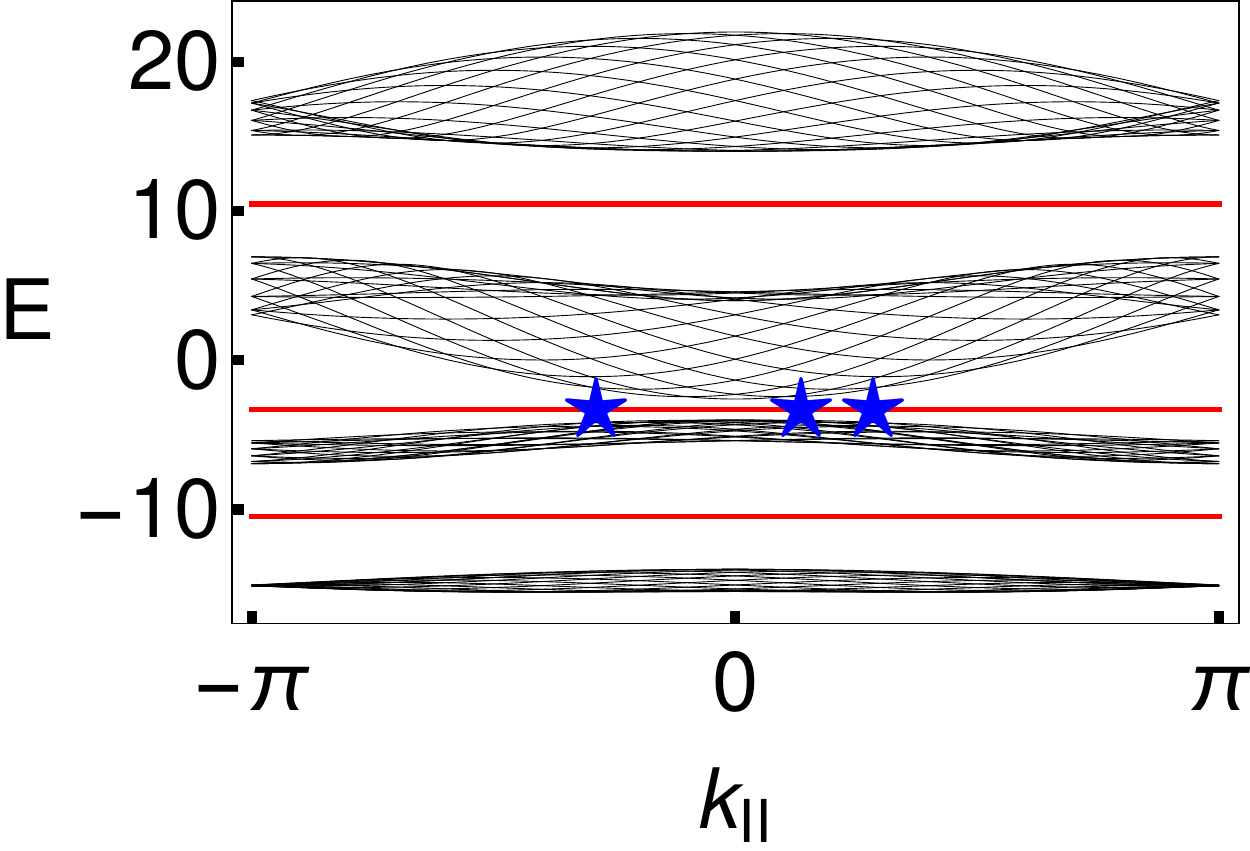}}
\caption{(color online) The projected dispersion relation on $k_\parallel$ is  
plotted in black for several values of the parameters. The red lines 
corresponds to the energy at which we looked 
for edge states using the technique described in \ref{sec.EtatDeBord}. The 
blue stars  
corresponds to the found spin-up edge states. An even number of spin-up edge 
states 
on a given line means that, at the given filling, the system is in the trivial 
phase, 
while an odd number of edge states corresponds to a topological insulator (a) 
$\Delta\mu=3$, $m=2$. (b) $\Delta\mu=3$, $m=5$.  (c) $\Delta\mu=9$, $m=2$.  
(d) $\Delta\mu=9$, $m=2$.  The results are in perfect agreement with the phase 
diagram of the 
Fig. \ref{PhaseDiagram}. In (a) and (b), certain values of filling could not 
be studied 
because of semi-metallic character of the spectrum. As a result, the 
topological nature 
of the first pair of  bands in (a) and (b) could not be verified with the 
present technique.}.
\label{Fig.EdgeStates}
\end{figure}

In practice, for an energy within the gap, we get 4 solutions for Eq.
(\ref{eq.Equation
avec lambda}) and, as we work in the spin-up subspace, the $\Phi^i$ have four 
components. So, to analyse the linear dependence of the obtained solutions 
$\Phi^i$, 
we compute the determinant of the matrix formed by the four vectors, as a 
function of 
$k_\parallel$. We then extract the zeroes of this function. The number of 
zeros 
between $-\pi$ and $\pi$ corresponds to the number of  spin-up edge states.

This number may vary depending on the parameter values. However, provided a 
gap present throughout the BZ, the parity of this number is directly related 
to the value 
of the topological invariant computed in the previous Section: $\Delta=0$ 
corresponds to 
an even number of spin-up edge states, while this number is odd when 
$\Delta=1$. 
This is true for the gap at half filing, which corresponds to the gap of the 
BHZ model, 
but also for the intermediate filling $1/4$ and $3/4$, as we can see, for some 
sets of 
parameters, in the Fig. \ref{Fig.EdgeStates}.

For some values of the parameters, at 1/4 and 3/4 filling the band structure 
does 
not correspond to an insulator, as it has no forbidden energy band, as one may 
observe in the Figs. \ref{Fig.EdgeStates} (a,b): at such a filling the system 
is 
rather a semimetal (an optical insulator). Because of this,  the edge state 
construction above cannot be applied directly: this method explicitly relies 
on the sought edge state having the energy where no bulk states exist. 
However, with the present Hamiltonian, we were unable to obtain both 
fully-insulating and topologically non-trivial behavior at 1/4 and 3/4 
filling. 

\section{Application to non-centrosymmetric systems}
\label{sec.P_broken}

So far, we considered the cases where numerical methods such as tracing 
either the Berry phase or the Wannier charge center trajectory could be tested 
against the product of parity eigenvalues at the TRIM, elegantly encapsulating 
the band topology. Now we would like to show that the former two methods 
allow one to study the band topology even when inversion is no longer a 
symmetry. 
In such a case parity is not a quantum number, and the $Z_2$ invariant cannot 
be 
related to the product of the parity eigenvalues at the TRIM. 

To this end, we consider a perturbation to the Hamiltonian 
(\ref{eq.BHZ+mag k space}) that explicitly breaks the inversion symmetry, such 
as 
\begin{equation}
\label{pert_break_P}
H_{P} (\delta_1, \delta_2) = \delta_1 \tau^x +2 
\delta_2(\sin(k_x)+\sin(k_y))\sigma^z
\end{equation}
where, as before $\tau$ matrices act in the orbital space and $\sigma$ 
matrices in the sublatice space. 
The first term corresponds to an on-site hybridization between the $p$ and $s$ 
orbitals, while the second term gives an inter-sublattice hoping. Each of the 
two 
breaks inversion symmetry and has the same qualitative effect as the 
combination 
we are considering here. We consider a combination of the two in order to 
illustrate 
a general case, where the inversion symmetry is lost, but the $\ta$ remains 
a symmetry. 

\begin{figure}
\includegraphics[width=0.4\textwidth]{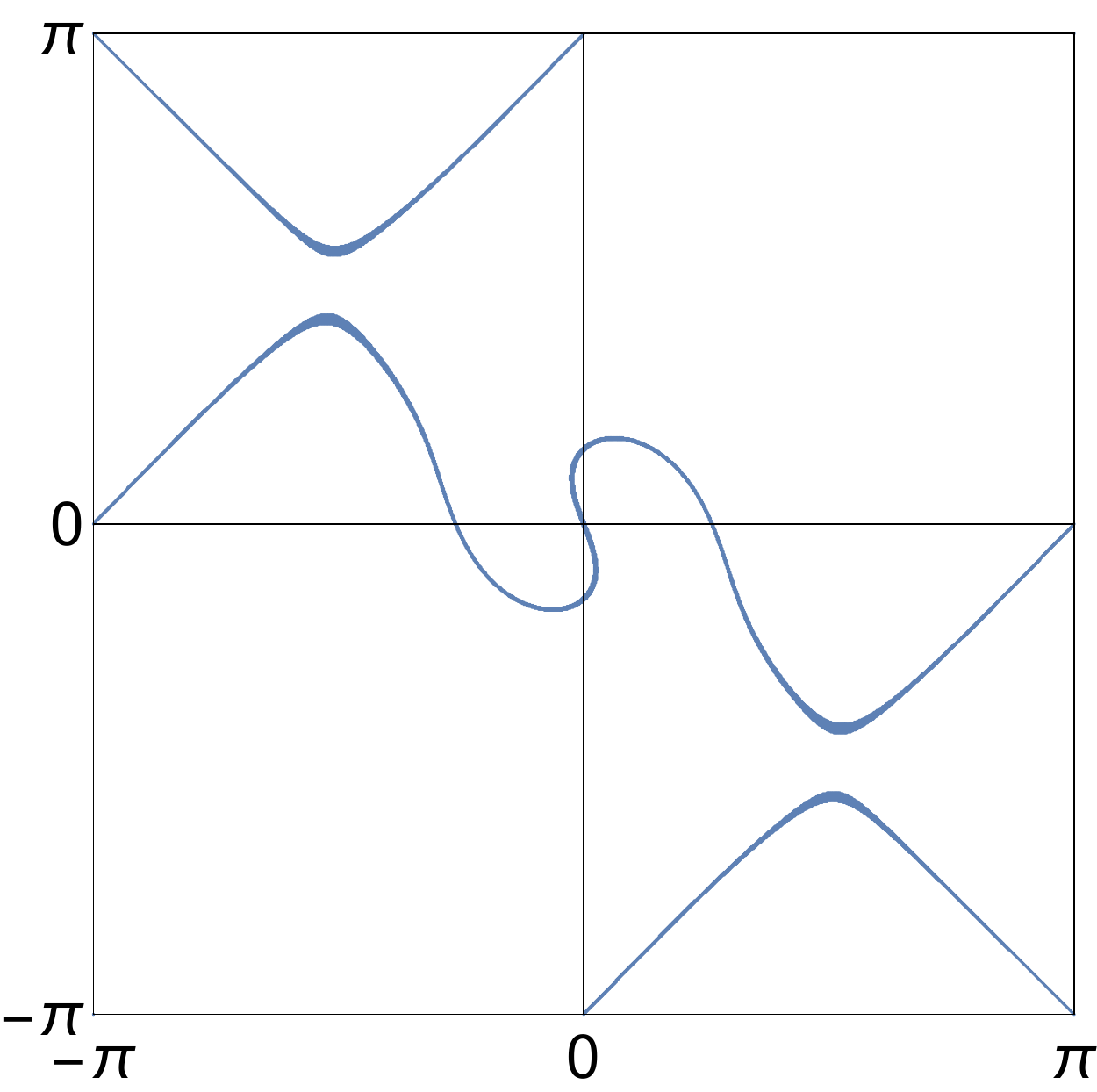}

\caption{(color online) Degeneracy sub-manofild in the BZ for the third and 
fourth bands when the perturbation  (\ref{pert_break_P}) is included. The 
values of the parameters here are  $\delta_1=1.5$ and  $\delta_2=0.5$}
\label{Fig.Deg_lines}
\end{figure}

The perturbation above lifts the double degeneracy, formerly protected by the 
combination of the $\ta$ and $\pa$, throughout the BZ -- except for a line of 
points pinned at the TRIM. The Fig. \ref{Fig.Deg_lines} shows the degeneracy 
line between the third and fourth bands, {\it i.e.} the top two filled bands 
at half filling. 
 
\begin{figure}

\subfigure[]{\includegraphics[width=0.23\textwidth]{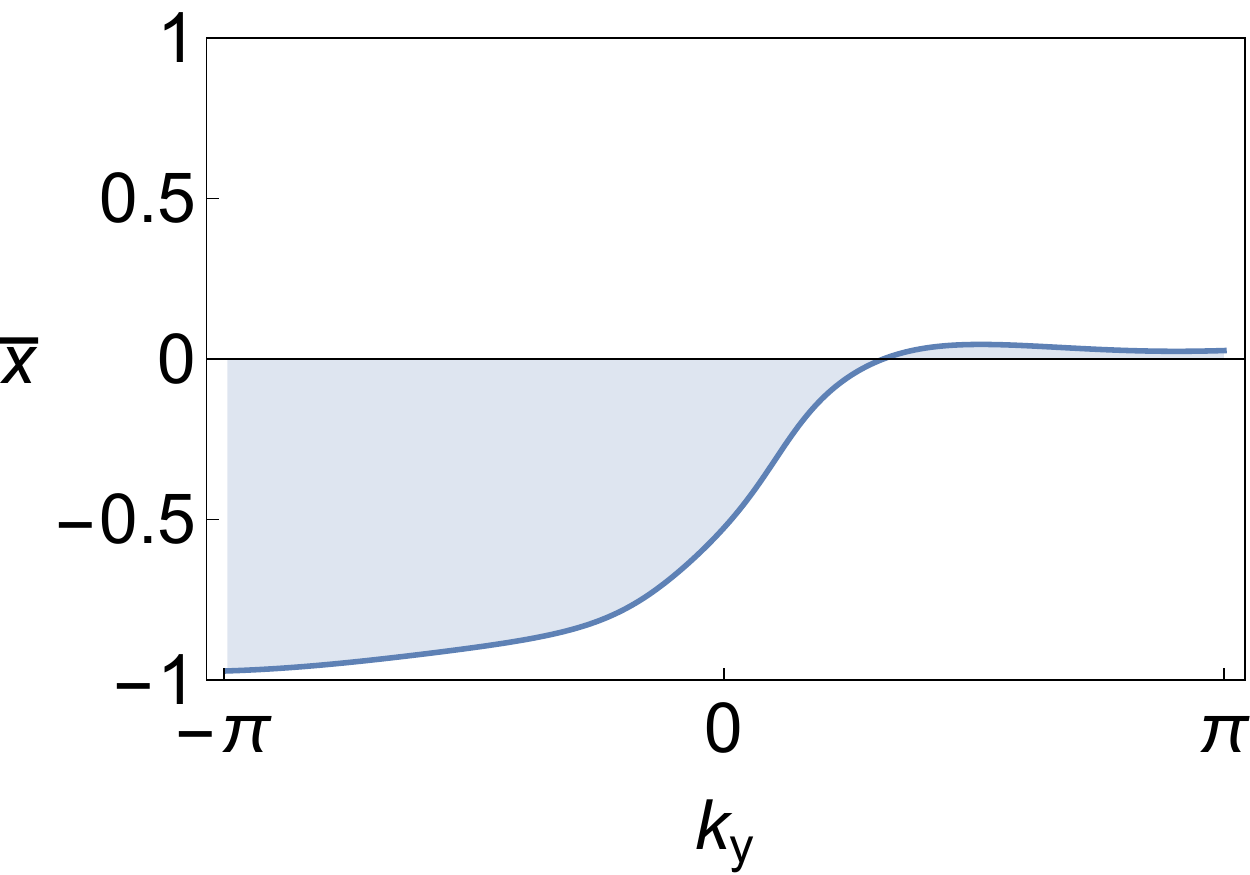}}
\subfigure[]{\includegraphics[width=0.23\textwidth]{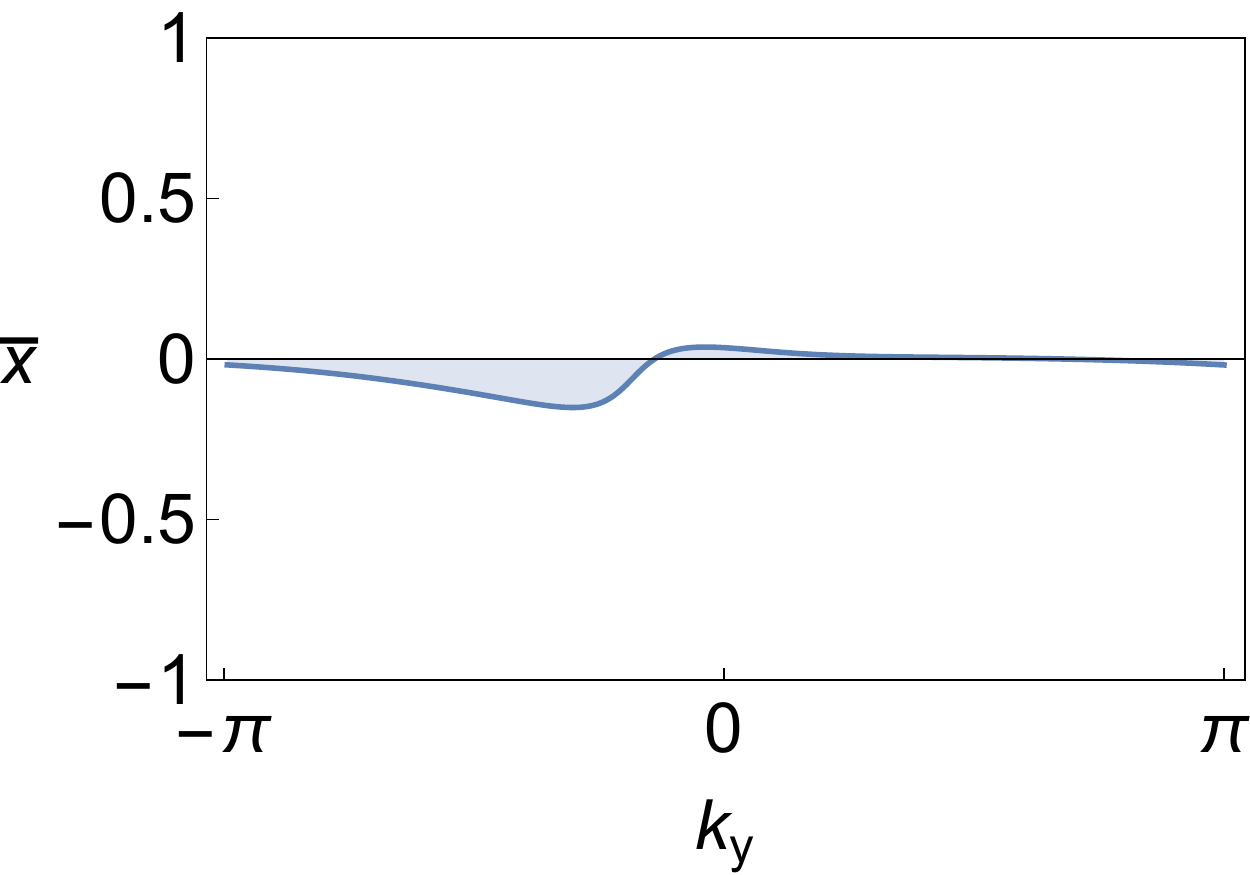}}

\caption{(color online) Analysis of the WCC cumulated on the two pairs of 
filled bands for the system with broken inversion symmetry. The values for 
$t_+,t_-$ and $\alpha$ are chosen as before, and $\Delta\mu=9$, $m=5$ and 
$\delta_2=0.5$. The transition is obtained by varying  $\delta_1$ and the 
critical point is found to be at $\delta^*_1\simeq 2.3$. The left side 
corresponds to  $\delta_1=1.5$ and right to $\delta_1=3$.}
\label{Fig.WCC_P_broken}
\end{figure}

We concentrate on half-filling, and study the insulating gap between the 
(filled) bands 3 and 4 and the (empty) bands 5 and 6. For simplicity we fix 
all the parameters of the unperturbed Hamiltonian, as well as $\delta_2$.  We 
then vary the $\delta_1$ to trigger 
the transition between the topological and topologically trivial phases. By 
choosing the 
same $t_+$ and $t_-$ as above, as well as $\alpha=1$, $\Delta\mu=9$, $m=5$ and 
$\delta_2=0.5$, we observe that the gap closes at $\delta^*_1\simeq 2.3$. For 
the sake 
of simplicity, we show here only the results obtained from the study of the 
WCC, 
but of course the study of the "reconnection phase" gives the same result.
The results are plotted in the Fig. \ref{Fig.WCC_P_broken} on the two sides of 
the transition. They clearly show that for  $\delta_1 < \delta^*_1$ the system 
is topologically 
non-trivial, while  it becomes a trivial insulator for 
$\delta_1 > \delta^*_1$. 

Of course one can consider a wider variety of perturbation terms that break 
inversion symmetry. While for systems with genuine TR symmetry there is still 
the topological invariant (\ref{eq.def inv topo with sew}) proposed by Fu and 
Kane, 
for system with an AF background no simple tool is available to predict 
topological 
phases. However, the extension of both the parallel transport and the WCC 
methods as presented in this article should allow a complete numerical 
identification of topological phases.  

\section{conclusions, discussion and outlook}
\label{sec.CDO}
In the preceding chapters we presented two different methods to test the 
topology of an antiferromagnetic insulator. The first method is based on a 
parallel transport construction of Bloch eigenstates, while the second one 
traces the charge centers of mass of hybrid Wannier functions. The two 
methods were developed by adapting the techniques that have proven 
useful in diagnosing paramagnetic $Z_2$ topological insulators. Thus we 
provided a complete picture and phase diagram of the work initiated by 
Guo {\em et al.} in [\onlinecite{Guo_Feng_Shen}]. The methods we developed 
apply to any system with an anti-unitary momentum-inverting symmetry. 

For centrosymmetric systems, defining a simple criterion of non-trivial 
topology 
was addressed in the Ref.[\onlinecite{Fang_Gilbert_Bernevig}]. This work 
proposed 
to consider the parity eigenvalues at the B-TRIM points only, while entirely 
disregarding their A-TRIM counterparts. In the case at hand, with the help 
of the methods developed above, we showed that this criterion does capture 
the topological phases  correctly. However, for this criterion to hold, it is 
vital 
that there be no band inversion at the A-TRIM points. 
One may further argue that, in a more general model, any band inversion 
at one of the two A-TRIM points should occur at the other point as well, 
as long as there is a symmetry relating these two points. In the present case, 
the spatial ${\pi \over 2}$ rotation symmetry of the Hamiltonian ensures that 
the two A-TRIM points behave identically. As long as this is true, the 
criterion 
due to Fang {\em et al.}~\cite{Fang_Gilbert_Bernevig} does capture a 
topological phase. However, if the two A-TRIM points are not symmetry-related, 
we see no reason that would protect the criterion due to Fang {\em et al.} 

In the latter case, we expect the two methods developed above to prove 
useful for detecting topological phases. To further illustrate this point, 
we studied a generalization of our model to a non-centrosymmetric system, 
where we are no longer aware of a simple expression for a topological 
invariant, that would identify a topological phase. In spite of this, 
the methods developed above allowed us to pinpoint the transition 
between the trivial and topological phases. 

\section{Acknowledgments}

It is our pleasure to acknowledge multiple discussions with Alexey 
A. Soluyanov, whose suggestions helped to greatly improve this work. 

\appendix
\section{Fu and Kane computation of the time reversal 
$Z_2$ topological invariant}
\label{app.Fu and Kane}

In this Appendix, we give the details of the computation of the $Z_2$ 
topological invariant derived by Fu and Kane~\cite{Fu_Kane2006} in the time 
reversal symmetric case.

Here, we choose our primitive lattice vectors to be $X$ and $Y$. We define the 
reciprocal lattice vectors in a standard way, but we treat $k_X$ and $k_Y$  
asymmetrically, 
and define the hybrid Bloch functions $\ket{u^{s}_{\alpha,k_X,k_Y}}$ as 
\begin{equation}
\ket{\Psi^{s}_{\alpha,\op{k}}}=\frac{1}{\sqrt{N_X}}
e^{ik_X \op{\hat{X}} }\ket{u^{s}_{\alpha,k_X,k_Y}} , 
\end{equation}
where $\op{\hat{X}}$ is the position operator in the $\op{X}$ direction, and 
$\ket{\Psi^{s}_{\alpha,\op{k}}}$ is defined in Eq.(\ref{eq.defchi})
In the same spirit, we define the hybrid Wannier functions as:
\begin{equation}
\ket{X,s,\alpha,k_Y}=\frac{a}{2\pi}\int_{-\pi/a}^{\pi/a}dk_X
e^{-ik_X( X-\op{\hat{X}} )}\ket{u^{s}_{\alpha,k_X,k_Y}}
\end{equation}
(in the same way, we define $\ket{u_{n,k_X,k_Y}}$ and $\ket{X,n,k_Y}$ from the 
${\ket{\Psi_{n,\op{k}}}}$ that are used in Eq.(\ref{eq.lien psi et u}).)

Following Fu and Kane,~\cite{Fu_Kane2006} we define the partial polarization 
$P^s_{k_Y}$, for a given value of $k_Y$ as:
\begin{equation}\label{eq.defPs}
P^s_{k_Y}=\sum_\alpha\bra{0,s,\alpha,k_Y}\op{X}\ket{0,s,\alpha,k_Y}
\end{equation}
for $s=\text{I}$ or $\text{II}$.

The total polarization is then equal to:
\begin{equation}
P^{\rho}_{k_Y}=P^\text{I}_{k_Y}+P^\text{II}_{k_Y},
\end{equation}
while the time reversal polarization is defined as:
\begin{equation}
P^{\Theta}_{k_Y}=P^\text{I}_{k_Y}-P^\text{II}_{k_Y},
\end{equation}

One may show that:
\begin{equation}\label{eq.PsAs}
P^s_{k_Y}=\frac{a}{2\pi}\int_{-\pi/a}^{\pi/a}dk_XA^s_{k_Y}(k_X)
\end{equation}
with
\begin{equation}\label{eq.defA}
A^s_{k_Y}(k_X)=i\sum_\alpha 
\bra{u^{s}_{\alpha,k_X,k_Y}}\nabla_{k_X}\ket{u^{s}_{\alpha,k_X,k_Y}}
\end{equation}

After a computation described in the reference~[\onlinecite{Fu_Kane2006}], for 
a gauge continuous over the BZ torus, we obtain:
\begin{align}
P^\Theta_{\Gamma^Y} =\frac{a}{2 \pi i}\bigg[ & \int_{0}^{\pi/a}dk_X 
\nabla_{k_X}\text{log\ det}[\sew(k_X,\Gamma^Y)] \nn \\
&   - 2\log\left(\frac{Pf[\sew(\pi/a,\Gamma^Y))]}{Pf[\sew(0,
\Gamma^Y))}]\right)\bigg],
\end{align}
for $\Gamma^Y\in \{0,\pi/a\}$.

Using this expression, one can go from the definition in (\ref{eq.def inv topo 
with P}) to Eq.(\ref{eq.def inv topo with sew}) in a straightforward way.

\section{Kramers degeneracy in the AF case}\label{app.degeneracy}

In this Appendix, we show that in the presence of commensurate staggered 
magnetization, and thus with broken time reversal symmetry, the inversion 
symmetry protects the Kramers degeneracy (see also the 
Refs.~[\onlinecite{Revaz:PRL.2008,Revaz:PRB.2009}]). This is important for the 
definition of the eigenstates 
in the Eq.(\ref{eq.defchiAF}).

The system at hand possesses two important symmetries.
The first one is associated with the operator

\begin{align}
\ta & =\sum_\op{k} \ta(\op{k})\ket{-\op{k}}\bra{\op{k}}K  \nn\\
 & =\sum_\op{k} ie^{i\frac{k_x-k_y}{2}}s^y\otimes(C_{-}
 \sigma^x-S_{-}\sigma^y)\ket{-\op{k}}\bra{\op{k}}K
\end{align}
which is anti-unitary and such that
\begin{align}
\ta^2 & =-e^{i\Phi_{\op{k}}} =-e^{-i2 a \op{k}\cdot \hat{\op{X}}} .
\end{align}
The second symmetry is associated with
\begin{align}
\pa & =\sum_\op{k} \pa(\op{k})\ket{-\op{k}}\bra{\op{k}}\nn \\
 & =\sum_\op{k}  e^{i\frac{k_x-k_y}{2}}\tau^z\otimes(C_-Id +iS_- \sigma^z) 
 \ket{-\op{k}}\bra{\op{k}}
\end{align}
which is unitary and squares to $1$.

The Kramers theorem arguments goes as follows:
let $A$ be an anti-unitary operator that commutes with the Hamiltonian of the 
system and which square is
$A^2 = e^{i \theta}$ with $\theta \neq 0~mod~2 \pi$.
Let then $\ket{\Psi}$ be an eigenstate of $H$. Then $A\ket{\Psi}$ is also an 
eigenstate of $H$ with the same eigenvalue and
\begin{equation} \label{eq.ortho_A}
\bra{\Psi} A \ket{\Psi} =0
\end{equation}

The fact that $A\ket{\Psi}$ is also an eigenstate of $H$ is strait-forward for 
the commutativity of $A$ and $H$. The 
orthogonality (\ref{eq.ortho_A}) can be shown from the following equalities:
\begin{equation}
\langle{\Psi} \ket{A \Psi} = \langle{A\Psi} \ket{AA \Psi}^* = e^{-i \theta} 
\langle{A\Psi} \ket{ \Psi}^* = e^{-i \theta} \langle{\Psi} \ket{A \Psi}
\end{equation}
where the first equality follows from the fact that $A$ is anti-unitary, and 
the
second from the property of $A^2$. Hence, if $\theta \neq 0~mod~2 \pi$, 
$\bra{\Psi} A \ket{\Psi} =0$.

In the case at hand, this result applies to the operator $\ta$ in the B-TRIM, 
and also to the operator
$\ta \pa$ which is anti-unitary, commutes with the Hamiltonian, and
squares to $-1$ in the entire BZ.

Then, the states $\ket{\Psi_{n,\op{k}}}$ and $\ta\pa\ket{\Psi_{n,\op{k}}}$ are 
degenerate
and orthogonal, while carrying the same momentum label $\op{k}$. Each band is 
thus doubly degenerate. Moreover, the states $\ta\ket{\Psi_{n,\op{k}}}$ and 
$\pa\ket{\Psi_{n,\op{k}}}$ are also degenerate and orthogonal, but carry the 
momentum label $\op{-k}$.

So, we have Kramers pairs of states everywhere in the Brillouin Zone, that we 
can separate into $\ket{\Psi^I_{\alpha,\op{k}}}$ and 
$\ket{\Psi^{II}_{\alpha,\op{k}}}$. The problem is then to find a continuous 
definition of the $\ket{\Psi^{I/II}}$ on the BZ that respect Eq.
(\ref{eq.defchiAF}).
This is done in two steps.

The first step consists in finding a continuous definition of the 
$\ket{\Psi^I_{\alpha,\op{k}}}$ over the BZ, that respects
\begin{equation}\label{eq.choice of u I}
\pa\ket{\Psi^I_{\alpha,\op{k}}}=e^{i\phi(\op{k})}\ket{\Psi^I_{\alpha,\op{-k}}}
\end{equation}
Then, we just need to construct 
$\ket{\Psi^{II}_{\alpha,\op{k}}}=\ta\pa\ket{\Psi^I_{\alpha,\op{k}}}$. The 
continuity of the $\ket{\Psi^{II}}$ states is a direct consequence of the 
continuity of the $\ket{\Psi^{I}}$, and Eq.(\ref{eq.defchiAF}) is a 
consequence of Eq.(\ref{eq.choice of u I}).

To properly define the $\ket{\Psi^I_{\alpha,\op{k}}}$, we first choose a 
continuous definition of it for $k_y=0$ and $k_x\in \left[ 0,\pi \right] $ 
such that $\pa\ket{\Psi^I_{\alpha,\op{\Gamma}}}$ is proportional to 
$\ket{\Psi^I_{\alpha,\op{\Gamma}}}$ for $\op{\Gamma}=(0,0)$ and $(\pi,0)$. 
Then, we construct the state at $k_y=0$ and $k_x\in \left] -\pi ,0 \right[ $ 
as $\ket{\Psi^I_{\alpha,\op{k}}}=\pa \ket{\Psi^I_{\alpha,\op{-k}}}$. This 
definition may be discontinuous at $k_x=0$ and $k_x=\pi$, but we have:
\begin{align}
\lim_{k_x\rightarrow 0^+} \ket{\Psi^I_{\alpha,-k_x,k_y=0}} &= 
\lim_{k_x\rightarrow 0^+} \pa \ket{\Psi^I_{\alpha,k_x,k_y=0}} \nn \\
&= \pa \ket{\Psi^I_{\alpha,k_x=0,k_y=0}}\nn\\
&=e^{i\phi(0)}\ket{\Psi^I_{\alpha,k_x=0,k_y=0}}
\end{align}
and a similar phase appears at $k_x=\pi$.
To get rid of this phase, we multiply all states at $k_y=0$ and 
$k_x\in \left] -\pi ,0 \right[ $ by a continuous phase $e^{-i\phi(k_x)}$ 
such that 
$\lim_{k_x\rightarrow 0^-}e^{i\phi(k_x)}=e^{i\phi(0)}$ and 
$\lim_{k_x\rightarrow -\pi}e^{i\phi(k_x)}=e^{i\phi(\pi)}$. We end up with a 
definition of $\ket{\Psi^I_{\alpha,\op{k}}}$ that is continuous along the 
circle $k_y=0$ and that respects Eq.(\ref{eq.choice of u I}).
This set of states can now be continuously extended to the entire upper half 
of the Brillouin zone ($k_y\in \left] 0 ,\pi \right]$) in a way described in 
[\onlinecite{Soluyanov_Vanderbilt_2012}].
We apply $\pa$ to construct the states in the lower half of the Brillouin 
zone.
Once again, we may have a discontinuity while crossing the line $k_y=0$,
but we recover continuity by multiplying all the states with $k_y<0$ by the
phase $e^{-i\phi{|k_x|}}$.

As was said before, we can now apply $\ta\pa$ to construct the 
$\ket{\Psi^{II}_{\alpha,\op{k}}}$ states, and get a continuous definition over 
the cylindrical BZ with edges $k_y=\pi$ and $k_y=-\pi$, that respects Eq.
(\ref{eq.defchiAF}).

\section{Computation of the $Z_2$ invariant in the AF case}
\label{app.Fu and Kane AF}

In this Appendix, we define the hybrid Bloch states and the hybrid Wannier 
functions as in Appendix \ref{app.Fu and Kane}, the goal being once again to 
compute the time reversal polarization, but in the case of broken time 
reversal symmetry. In Appendix \ref{app.Fu and Kane}, we showed how to get an 
expression of the $Z_2$ topological invariant in terms of the determinant and 
the pfaffian of the sewing matrix at the four TRIM (Eq.(\ref{eq.def inv topo 
with sew})). In order to define the Pfaffian, we needed the sewing matrix to 
be anti-symmetric at the four TRIM. However, in the case of the added 
staggered magnetization, the sewing matrix defined with $\ta$ is symmetric at 
the A-TRIM (TRIM where $\ta^2=1$). This is why, in the AF case, we do not get 
an expression (\ref{eq.def inv topo with chi}) 
for the $Z_2$ invariant only in terms of the sewing matrix, but also in terms 
of the phase 
$\chi$.

As in the time reversal symmetric case, we consider two primitive lattice 
vectors $\op{a}$ and $\op{b}$. We make the Wannier transformation along 
$\op{a}$, while keeping plane waves along $\op{b}$. We now wish to compute 
$P^{\ta}_{\Gamma_b}=P^\text{I}_{\Gamma_b}-P^\text{II}_{\Gamma_b}$.

We first concentrate on $P^\text{I}_{\Gamma_b}$. Equation (\ref{eq.PsAs}) for 
$s=\text{I}$ may be rewritten as:
\begin{eqnarray}
P^\text{I}_{\Gamma_b}&=&\frac{1}{2\pi}\int_{0}^{\pi}dk_a[A^\text{I}_{\Gamma_b}
(k_a)+A^\text{I}_{\Gamma_b}(-k_a)]
\end{eqnarray}
Moreover, using Eqs.(\ref{eq.defchiAF}) and (\ref{eq.defA}) and the 
antiunitarity of $\ta$, one may show that:
\begin{eqnarray}
A^\text{I}_{-k_{b}}(-k_a)&=&A^\text{II}_{k_{b}}(k_a) + \sum_\alpha\nabla_{k_a}
(\chi_{\alpha,k_a,k_b} + \frac{1}{2}\Phi_{k_a,k_b})\\
&+&i\sum_\alpha \brab{u}{II}{\alpha,k_a,k_b}
(\nabla_{k_a}\tah(\op{k}))\tah(\op{k})^\dagger\ketb{u}{II}{\alpha,k_a,k_b}\nn
\end{eqnarray}

Because of the derivatives of $\ta$ and $\Phi$, this expression may seem not 
very useful. But $\ta$ and $\Phi$ both
depends only on $k_x-k_y$. So, if we choose $k_a=k_x$ and 
$k_b=k_x-k_y$, the derivative term disappear and we get:
\begin{eqnarray}
A^\text{I}_{-k_{b}}(-k_a)&=&A^\text{II}_{k_{b}}(k_a) + 
\sum_\alpha\nabla_{k_a}\chi_{\alpha,k_a,k_b}
\end{eqnarray}

And so,
\begin{eqnarray}
P^\text{I}_{\Gamma_b}&=&\frac{1}{2\pi}\int_{0}^{\pi}dk_a(A^{I}_{-\Gamma_b}
(k_a)+A^{II}_{\Gamma_b}(k_a)) \nn\\
&+&\frac{1}{2\pi} \sum_\alpha (\chi_{\alpha,
\pi,-\Gamma_b}-\chi_{\alpha,0,-\Gamma_b})
\end{eqnarray}

Finally, using $P^{\ta}_{\Gamma_b} = 2 P^\text{I}_{\Gamma_b}-
P^\rho_{\Gamma_b}$, and 
folowing Fu and Kane,~\cite{Fu_Kane2006} this can be rewritten in a continuous 
gauge as:
\begin{eqnarray}\label{eq.time rev pol app}
P^{\ta}_{\Gamma_b} &=&\frac{1}{2\pi i}\int_{0}^{\pi}dk_a 
\nabla_{k_a}\text{log\ det}[\tilde{\sew}_{\Gamma_b}(k_a)] \nn\\
&&+\frac{1}{\pi} \sum_\alpha (\chi_{\alpha,0,\Gamma_b}+\chi_{\alpha,\pi,
\Gamma_b})
\end{eqnarray}
where $\tilde{\sew}$ is the sewing matrix defined by:
\begin{equation}
\tilde{\sew}_{k_{b}}(k_a)_{mn}=\bra{\Psi_{m,\op{-k}}} 
\ta\ket{\Psi_{n,\op{k}}} .
\end{equation}

%%%%%%%%%%%%%%%%%%%%%%%%%%%%%%%%%%%%%%%%%%%%%%%%%%%%%%%%%%%%%%%%%%%%%%
%%%%%%%%%%%%%%%%%%%%%%%%%%%%%%%%%%%%%%%%%%%%%%%%%%%%%%%%%%%%%%%%%%%%%%
%
%       BIBLIOGRAF\'{I}A
%

%%%%%%%%%%%%%%%%%%%%%%%%%%%%%%%%%%%%%%%%%%%%%%%%%%%%%%%%%%%%%%%%%%%%%%
%%%%%%%%%%%%%%%%%%%%%%%%%%%%%%%%%%%%%%%%%%%%%%%%%%%%%%%%%%%%%%%%%%%%%%
%
\end{document}